\documentclass[11pt,preprintnumbers,nofootinbib,amsmath,amssymb]{revtex4}
\usepackage{float}
\usepackage{indentfirst}
\usepackage{amsmath,amssymb,epsfig,subfigure}
\usepackage{graphicx}
\usepackage[export]{adjustbox}
\usepackage{color}
\usepackage{multirow}
\usepackage{booktabs}
\usepackage{changes}
\usepackage{footnote}
\usepackage{mathrsfs} 
\usepackage[shortcuts]{extdash}

\usepackage{hyperref}
\hypersetup{colorlinks=true,linkcolor=blue,citecolor=magenta}

\begin{document}
\renewcommand{\baselinestretch}{1.15}
\makeatletter
\renewcommand{\thesubfigure}{\alph{subfigure}}
\renewcommand{\@thesubfigure}{(\thesubfigure)\hskip\subfiglabelskip}
\makeatother
\title{Frozen solitonic Hayward-boson stars in Anti-de Sitter Spacetime}
\author{Shu-Cong Liu$^{1,2}$}
\author{Yong-Qiang Wang$^{1,2}$}
\author{Zhen-Hua Zhao$^{3}$\footnote{Electronic address: zhaozhh78@sdust.edu.cn, corresponding author}}

\affiliation{$^{1}$Lanzhou Center for Theoretical Physics,
 Key Laboratory of Theoretical Physics of Gansu Province,
 School of Physical Science and Technology,Lanzhou University, Lanzhou 730000, China\\$^{2}$Institute of Theoretical Physics $\&$ Research Center of Gravitation,
 Lanzhou University, Lanzhou 730000, China\\$^{3}$Department of Applied Physics,
 Shandong University 
 of Science and Technology, Qingdao 266590, China
}

\begin{abstract}
We construct solitonic Hayward-boson stars (SHBSs) in Anti-de Sitter (AdS) spacetime, which consists of the Einstein-Hayward model and a complex scalar field with a soliton potential. Our results reveal a critical magnetic charge $q_c$. For $q\geq q_c$ in the limit of $\omega    \rightarrow 0$, the matter field is primarily distributed within the critical radius $r_c$, beyond which it decays rapidly, while the metric components $-g_{tt}$ and $1/g_{rr}$ become very small at $r_c$. These solutions are termed ``frozen solitonic Hayward-boson stars" (FSHBSs). Continuously decreasing $\Lambda$ disrupts the frozen state. However, we did not find a frozen solution when $q<q_c$. The value of $q_c$ depends both on the cosmological constant $\Lambda$ and the self-interaction coupling $\eta$. We also found that for high frequency solutions, increasing $\eta$ can yield a pure Hayward solution. However, for low frequency solutions, increasing $\eta$ reduces both $1/g_{rr}$ and $-g_{tt}$. Furthermore, we analyzed the effective potential of SHBSs and identified an extra pair of light rings in the second solution branch.
\end{abstract}
\maketitle
\thispagestyle{empty}
\newpage
\section{INTRODUCTION}\label{Sec1}
General relativity predicts that the interior of a black hole contains a spacetime singularity, where the curvature diverges and all known physical laws cease to apply. To resolve this theoretical difficulty, many attempts have been made, one of which is to construct regular black holes~\cite{Lan:2023cvz,Torres:2022twv,Ansoldi:2008jw,Carballo-Rubio:2025fnc}. Shirokov~\cite{Shirokov:1947elm} and Duan~\cite{Duan:1954bms} were the first to attempt to solve the black hole singularity problem. In 1966, A. D. Sakharov~\cite{Sakharov:1966aja} and E. B. Gliner~\cite{Gliner:1966} proposed that the central singularity of classical black hole solutions could be avoided by a de Sitter core. In 1968, Bardeen introduced a radius-dependent mass function to remove the classical central singularity, thus establishing the first model of a regular black hole~\cite{Bardeen:1968}. Three decades later, Ayon-Beato and Garcia demonstrated that the Bardeen black hole is actually a gravitational configuration sourced by magnetic monopoles in nonlinear electrodynamics~\cite{Ayon-Beato:1998hmi,Ayon-Beato:1999kuh,Ayon-Beato:2000mjt}. In 2006, Hayward proposed another model of a regular black hole~\cite{Hayward:2005gi}. In recent years, the application of nonlinear electromagnetic fields in the study of regular black holes has attracted widespread attention~\cite{Bronnikov:2000vy,Dymnikova:2004zc,Lemos:2011dq,Balart:2014cga,Bronnikov:2017sgg,Breton:2004qa,Ayon-Beato:2004ywd,Berej:2006cc,Junior:2023ixh,Cordeiro:2025ydg}. 

However, the event horizon of regular black holes still raises debates, such as the information loss~\cite{Hawking:1974rv,Hawking:1975vcx,Hawking:1976ra} and the firewall paradox~\cite{Almheiri:2013hfa,Almheiri:2012rt}. These challenges have motivated investigations into horizonless compact objects, among which boson stars~\cite{Schunck:2003kk,Liebling:2012fv} have attracted particular attention. The theoretical foundation of boson stars can be traced to Wheeler’s geons~\cite{Wheeler:1955zz,Power:1957zz}, a hypothetical structure formed by the coupling of electromagnetic waves with Einstein's gravitational field. However, subsequent studies showed that such configurations are unstable. Latter, Kaup~\cite{Kaup:1968zz} and Ruffini~\cite{Ruffini:1969qy} replaced the electromagnetic field with a complex scalar field and obtained stable solutions. 

Recent study has coupled a complex scalar field to the Einstein-Hayward theory, resulting in a class of solutions termed Hayward boson stars (HBSs)~\cite{Yue:2023sep}. In particular, HBSs permit extreme solutions with $\omega\rightarrow 0$ when the magnetic charge $q$ exceeds a critical value $q_c$. In this case, the metric component $1/g_{rr}$ becomes very small but non-zero at the critical radius $r_c$. Simultaneously, the metric component $-g_{tt}$ approaches zero within $r_c$. Solutions exhibiting this behavior are referred to as ``frozen stars", and the spherical surface
represented by $r = r_c$ is called the critical horizon. The study of frozen stars can be traced back to the seminal works of Oppenheimer, Snyder~\cite{Oppenheimer:1939ue}, Zel'dovich, and Novikov~\cite{zeldovichbookorpaper}. The term ``frozen" refers to the phenomenon in which, from the perspective of a distant observer, the collapse of an ultra-compact object appears to take an infinite amount of time, as if the object were frozen at the event horizon. Solutions of frozen stars can also be extended to other scenarios~\cite{Mathur:2023uoe,Wang:2023tdz,Huang:2023fnt,Mathur:2024mvo,Ma:2024olw,Huang:2024rbg,Chen:2024bfj,Wang:2024ehd,Sun:2024mke,Zhang:2025nem,Huang:2025cer,Tan:2025jcg}.

The AdS/CFT correspondence~\cite{Maldacena:1997re,Gubser:1998bc,Witten:1998qj} is an important conjecture in string theory and quantum gravity research. Studies in AdS spacetime have significantly advanced our understanding of quantum gravity and have gradually expanded to various theoretical models, including regular black holes~\cite{Fan:2016rih,Guo:2024jhg,Xie:2024xkh}, boson stars~\cite{Astefanesei:2003qy,Astefanesei:2003rw,Buchel:2013uba,Maliborski:2013ula} and proca stars~\cite{Duarte:2016lig}. Ref.~\cite{Zhao:2025yhy} has extended HBSs to AdS spacetime and found that variations in the cosmological constant can significantly affect the ADM mass, the critical magnetic charge, and other properties.

Early mini-boson star models possessed masses well below the Chandrasekhar limit, which limited their astrophysical relevance. By introducing self-interaction into the potential term, the resistance to gravitational collapse becomes more effective, allowing the star to achieve a larger mass while maintaining an equilibrium state. Depending on the scalar potential, boson stars can be categorized into types such as massive boson stars~\cite{Colpi:1986ye}, solitonic boson stars~\cite{Friedberg:1986tq,Lee:1986ts} and axion boson stars~\cite{Guerra:2019srj,Delgado:2020udb}, among which solitonic boson stars have attracted considerable interest due to their ability to form bound states even in the absence of gravity~\cite{Lynn:1988rb,Kleihaus:2005me,Macedo:2013jja,Dzhunushaliev:2014bya,Brihaye:2015veu,Collodel:2017biu,Boskovic:2021nfs,Collodel:2022jly,Siemonsen:2024snb,Ogawa:2024joy}. Therefore, in this work, we replace the matter field in~\cite{Zhao:2025yhy} with a complex scalar field featuring a solitonic potential, aiming to investigate the solutions and spacetime geometry of solitonic Hayward boson stars (SHBSs) in AdS spacetime. Interestingly, we also find frozen solutions, with the value of the critical magnetic charge $q_c$ being jointly determined by the cosmological constant $\Lambda$ and the self-interaction coupling $\eta$. Furthermore, for high frequency solutions, increasing $\eta$ leads to a pure Hayward solution, while for low frequency solutions, increasing $\eta$ decreases both $1/g_{rr}$ and $-g_{tt}$. By analyzing the effective potential of SHBSs, we identify an additional pair of light rings in the second branch of the solutions.

The structure of this paper is as follows. In Sect. \ref{sec2}, we construct solitonic Hayward boson stars in AdS spacetime, consisting of a nonlinear electromagnetic field and a complex scalar field minimally coupled with gravity. In Sect. \ref{sec3}, we present the boundary conditions for the equations of motion. Sect. \ref{sec4} displays the numerical results. Finally, we give a brief conclusion and discussion  in Sect. \ref{sec5}.

\section{THE MODEL}\label{sec2}
\subsection{The framework of SHBSs}
In this section, we present a concise theoretical framework that combines Einstein's gravity with electromagnetic field~\cite{Fan:2016hvf}, further coupled to a complex scalar field~\cite{Friedberg:1986tq}. The action of this model is the following
\begin{equation}\label{equ1}
  S=\int\sqrt{-g}d^4x\left(\frac{R-2\Lambda}{4}+\mathcal{L}^{(1)}+\mathcal{L}^{(2)}\right),
\end{equation}
with
\begin{equation}\label{equ2}
\mathcal{L}^{(1)} = - \frac{ 3}{ 2 s } \frac{ (2 q^2 {\cal F})^{3/2}}{\left(  1 + ( 2 q^2 {\cal F})^{3/4}\right)^2},
\end{equation}
\begin{equation}\label{equ3}
\mathcal{L}^{(2)} = -\nabla_\mu \Psi^* \nabla^ \mu \Psi - U(\Psi,\Psi^*),
\end{equation}
where
 $$U(\Psi,\Psi^*) = \mu^2 \Psi \Psi^* \left(1-2 \eta^2\Psi\Psi^*\right)^2.$$ 
 
In this framework, $R$ represents the scalar curvature, $\Psi$ denotes a complex scalar field. ${\cal F} \equiv \frac{1}{4}F_{\mu\nu}F^{\mu\nu}$ defines the electromagnetic invariant, where the field strength tensor $F_{\mu\nu} = \partial_\mu A_\nu - \partial_\nu A_\mu$. The model incorporates three fundamental parameters: the magnetic charge $q$, scalar field mass $m$, and $\eta$ the coupling parameter controlling self-interaction. Variations of the action \eqref{equ1} with respect to the metric, electromagnetic field, and scalar field yield the corresponding equations of motion
\begin{eqnarray} \label{eq:EKG1}
R_{ab}-\frac{1}{2}g_{ab}R+\Lambda g_{ab}-2 (T^{(1)}_{ab}+T^{(2)}_{ab})&=&0 \ ,  \nonumber\\
\bigtriangledown_{a} \left(\frac{ \partial {\cal L}^{(1)}}{ \partial {\cal F}}  F^{a b}\right) &=& 0,    \\
\Box\Psi- \frac{\partial \, U}{\partial\lvert \Psi \rvert^2}\Psi &=& 0, \nonumber
\end{eqnarray}
with
\begin{equation}\label{equ5}
T^{(1)}_{ab} =- \frac{ \partial {\cal L}^{(1)}}{ \partial {\cal F}} F_{a c} F_{ b }^{\;\;c} + g_{ab} {\cal L}^{(1)},
\end{equation}
\begin{equation}\label{equ6}
T^{(2)}_{a b} = \partial_a \Psi^* \partial_b \Psi + \partial_b \Psi^* \partial_a \Psi - g_{a b} \left[ \frac{1}{2}g^{ab}\left(\partial_a\Psi^*\partial_b\Psi + \partial_b\Psi^*\partial_a\Psi\right) + U(\Psi,\Psi^*)\right]\,.
\end{equation}

Following Noether's theorem, the $U(1)$ gauge invariance of the action under the transformation $\Phi \rightarrow e^{i\alpha}\Phi$ (where $\alpha$ is a constant parameter) generates a conserved current associated with the complex scalar field
\begin{equation}\label{equ7}
	J^{a} = -i\left(\Psi^*\partial^a\Psi - \Psi\partial^a\Psi^*\right).
\end{equation}
By integrating the timelike component of this conserved current over a spacelike hypersurface $\Sigma$, one obtains the Noether charge
\begin{equation}\label{equ8}
	Q = \frac{1}{4\pi}\int_{\varSigma}J^t .
\end{equation}

We consider a general static spherically symmetric spacetime and adopt the following ansatz:
\begin{equation}\label{equ9}
	ds^2 = -N(r)\sigma^2(r)dt^2 + \frac{dr^2}{N(r)} + r^2\left(d\theta^2 + \sin^2\theta d\varphi^2\right),
\end{equation}
where 
$$N(r) = 1- \frac{2 n(r)}{r}-\frac{\Lambda r^2}{3}.$$ 
The metric functions $N(r)$, $n(r)$, and $\sigma(r)$ depend solely on the radial coordinate $r$. For the matter fields, we adopt the following ansatz:
\begin{equation}\label{equ10}
   A= q \cos(\theta)d\varphi,\;\;\; \Psi = \frac{\phi(r)}{\sqrt{2}}e^{-i\omega t},
\end{equation}
where $\phi(r)$ represents the real scalar field amplitude and $\omega$ denotes the oscillation frequency of the complex field $\Psi$. Using the electromagnetic field ansatz specified in Eq.~\eqref{equ10}, we obtain the magnetic field expression:
\begin{equation}\label{equ11}
F_{\theta \varphi } =  q \sin \theta.
\end{equation}
From this solution, the magnetic charge can be determined through the surface integral:
\begin{equation}\label{equ12}
	\frac{1}{4\pi}\oint_{S^\infty}d A =q  .
\end{equation}
Besides, the Noether charge obtained from  Eq.~(\ref{equ8}) is written as
\begin{equation}\label{equ13}
	Q = \int_0^\infty r^2\frac{\omega\phi^2}{N~\sigma}dr\, .
\end{equation} 

Substituting the ansatz  Eq.~(\ref{equ9}) and  Eq.~(\ref{equ10}) into the field equations Eq.~(\ref{eq:EKG1}), we obtain the following radial equations:
\begin{eqnarray}
	 \phi^{\prime\prime}+\left(\frac{2}{r} + \frac{N^\prime}{N} + \frac{\sigma^\prime}{\sigma}\right)\phi^\prime + \left(\frac{\omega^2}{N \sigma^2} - \dot{U}(\phi)\right)\frac{\phi}{N} &=& 0\, ,\label{equ14}\\
 N' +\frac{\omega^2 r \phi^2 }{N \sigma^2}+N r \phi'^2+\frac{N}{r} + r (2 U(\phi))+\frac{3 q^6 r }{s \left(q^3+r^3\right)^2}-\frac{1}{r}&=&0, \label{equ15}
 \\
  	\frac{\sigma^\prime}{\sigma} - r\left(\phi^{\prime2} + \frac{\omega^2\phi^2}{N^2 \sigma^2}\right)&=&0.  \label{equ16}
\end{eqnarray}

The equations of motion (\ref{equ14})-(\ref{equ16}) have solutions in two special cases:
1. When $q = 0$ but $\phi \neq 0$, the action \eqref{equ1} reduces to the Einstein scalar theory, yielding boson star solutions~\cite{Kaup:1968zz,Ruffini:1969qy}.
2. When $\phi = 0$ but $q \neq 0$, the model becomes Hayward's regular black hole~\cite{Hayward:2005gi}, with the static spherically symmetric metric:
\begin{equation}\label{equ17}
ds^2 = -f(r)dt^2 + f(r)^{-1}dr^2 + r^2(d\theta^2 + \sin^2\theta d\varphi^2),
\end{equation}
and
\begin{equation}\label{equ18}
f(r) = 1 - \frac{2Mr^2}{r^3 + q^3},
\end{equation}
where $M = \frac{q^3}{2s}$. The metric function $f(r)$ obtains local minima at $r = 2^{1/3}q$.
When $q = \frac{\sqrt{3s}}{2^{1/3}}$, there is a horizon for the Hayward black hole. When $q < \frac{\sqrt{3s}}{2^{1/3}}$, no horizon is formed and when $q > \frac{\sqrt{3s}}{2^{1/3}}$, there are two horizons.

\subsection{Null geodesics of SHBSs}
The geodesic equation in the corresponding space-time of the ansatz  (\ref{equ9}) can be derived from the Lagrangian
\begin{equation}\label{equ19}
2\mathcal{L} = g_{\mu\nu} \frac{dx^\mu}{d\tau} \frac{dx^\nu}{d\tau} = -N(r) \, \sigma^2(r) \, \dot{t}^2 + \frac{1}{N(r)} \, \dot{r}^2 + r^2 \, \dot{\theta}^2 + r^2 \sin^2\theta \, \dot{\varphi}^2 ,
\end{equation}
where $\tau$ is an affine parameter along the geodesic and the dot indicates the derivative with respect to $\tau$. The corresponding regular momentum is
\begin{equation}\label{equ20}
p_t = -\frac{\partial\mathcal{L}}{\partial \, \dot{t}} = N \, \sigma^2\,\dot{t},\ p_r = \frac{\partial\mathcal{L}}{\partial \, \dot{r}}= \frac{\dot{r}}{N},\ p_\varphi = \frac{\partial\mathcal{L}}{\partial \, {\varphi}} = r^2\sin^2\theta \, \dot{\varphi} .
\end{equation}

For null geodesics, equation (\ref{equ19}) equals zero. We consider particles confined to the equatorial plane with $\theta = \pi/2$ and we have
\begin{equation}\label{equ21}
\frac{d \, p_t}{d \, \tau} = -\frac{\partial \mathcal{L}}{\partial \, {t}} = 0 ,\ \frac{d \, p_\varphi}{d \, t}=\frac{\partial \mathcal{L}}{\partial \, {\varphi}} = 0.
\end{equation}
Therefore, we can define the conserved quantities for null particles: energy $E=N(r) \, \sigma^2(r) \, \dot{t}$ and angular momentum $L = r^2\dot{\varphi}$. By substituting into Eq. (\ref{equ19}), we derive
\begin{equation}\label{equ22}
\dot{r}^2=\frac{{L}^2}{\sigma^2} \ (\frac{E^2}{{L}^2}-\frac{N \, \sigma^2}{r^2}).
\end{equation}
Given that $E$ and $L$ are constants, it is convenient to define the effective potential~\cite{Alcubierre:2021psa}
\begin{equation}\label{equ23}
V_{eff} = \frac{N \, \sigma^2}{r^2}.
\end{equation}
The position of the light ring $R_{LR}$ is determined by the extrema of the effective potential:
\begin{equation}\label{equ24}
\left. \frac{d V_{\text{eff}}}{dr} \right|_{r = R_{\text{LR}}} = 0.
\end{equation}
Stable light rings occur where the second derivative $V'' > 0$, whereas unstable ones appear where $V'' < 0$. Further studies on the stability of light rings can be found in~\cite{Cunha:2017qtt}.

\section{BOUNDARY CONDITIONS AND NUMERICAL METHODS}\label{sec3}
To solve this system of coupled ordinary differential equations, we must impose appropriate boundary conditions for each function. For asymptotically AdS solutions, the metric functions $n(r)$ and $\sigma(r)$ must satisfy the following boundary conditions:
\begin{equation}\label{equ25}
n(0) = 0, \qquad \sigma(0) = \sigma_0, \qquad n(\infty) = M, \qquad \sigma(\infty) = 1.
\end{equation}

The asymptotic value $\sigma_0$ and the ADM mass $M$ of the solution are determined through numerical integration of the coupled differential equations. For the complex scalar field, we impose the following boundary conditions to ensure regularity and finite energy:
\begin{equation}\label{equ26}
\phi(\infty) = 0,\;\;\;\left. \frac{d\phi(r)}{dr}\right|_{r = 0} = 0.
\end{equation}

To perform numerical calculations, we adopt the following scaling transformations to obtain the dimensionless variables below:
\begin{equation}\label{equ27}
r\to r\mu\, ,\ \omega\to\omega/\mu \, ,\ \Lambda \to\Lambda/\mu^2 \, ,\ \Psi\to\sqrt{4\pi G}\,\Psi \,,\ M \to \mu GM \, , \  Q \to \mu^2 GQ.
\end{equation}

Furthermore, without loss of generality, we set $\mu=1, s = 0.2$ and $4\pi G = 1$. To solve the coupled system of Eqs. (\ref{equ14}), (\ref{equ15}), and (\ref{equ16}) subject to the boundary conditions (\ref{equ25}) and (\ref{equ26}) , we introduce a compactified radial coordinate $x=\frac{r}{1+r}$. Map the computational domain to $x \in [0,1]$. The nonlinear ODEs were solved numerically via the finite element method and the numbers of grid points were set to 1000 spanning $0 \leq x \leq 1$. We employ the Newton-Raphson method as our iterative scheme.  The relative error of our numerical solutions is controlled below $10^{-5}$.
\section{NUMERICAL RESULTS}\label{sec4}

In this section, we present the numerical results. The solutions are divided into two categories based on the value of the magnetic charge. When $q < q_c$, we do not find extreme solutions where the frequency approaches zero. When $q \geq q_c$, the frequency can converge to zero. Therefore, we discuss case by case.
\subsection{Small magnetic charge $q < q_c$ } 

\begin{figure}[!htbp]
\centering
\subfigure{\includegraphics[width=0.45\textwidth]{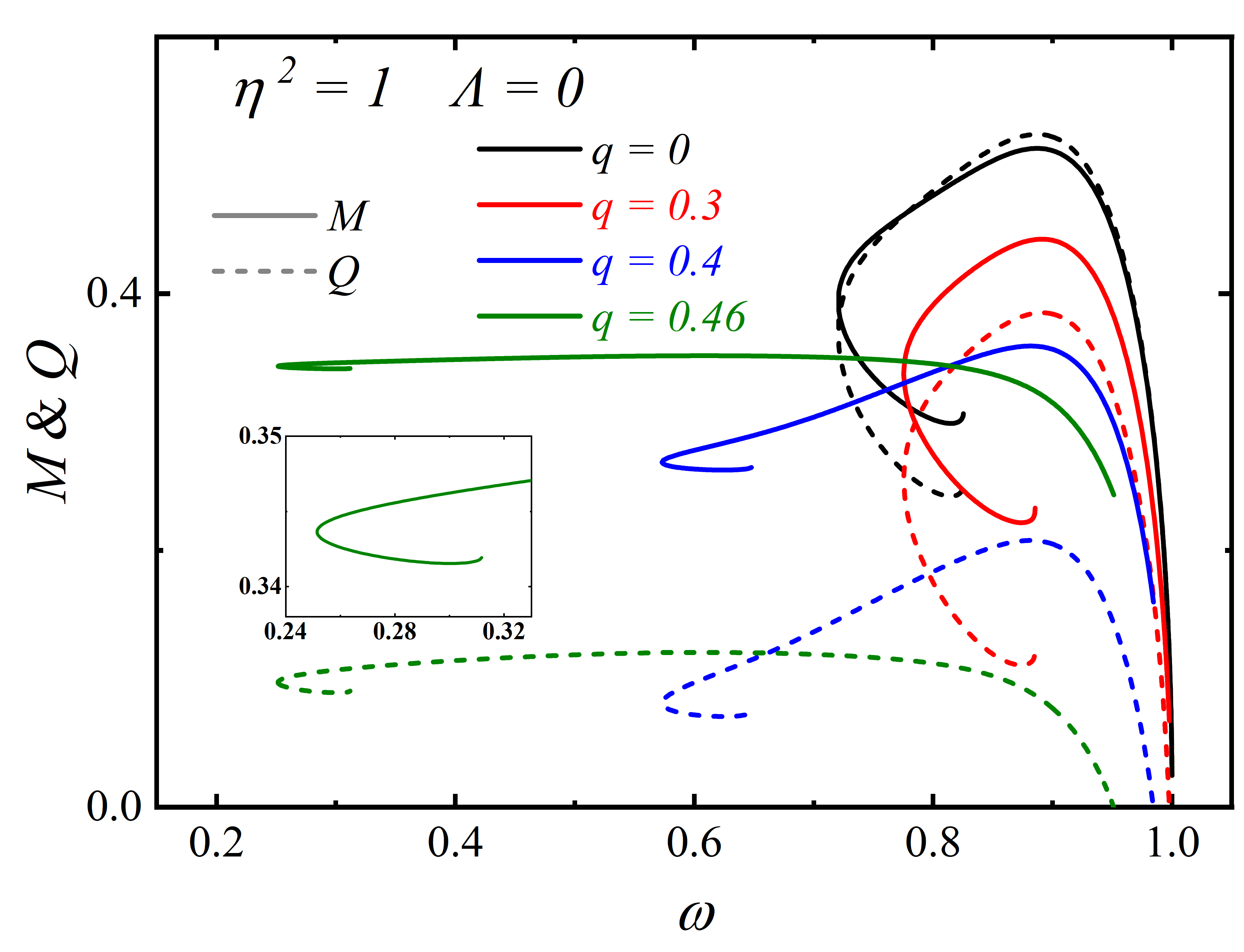}}
\subfigure{\includegraphics[width=0.45\textwidth]{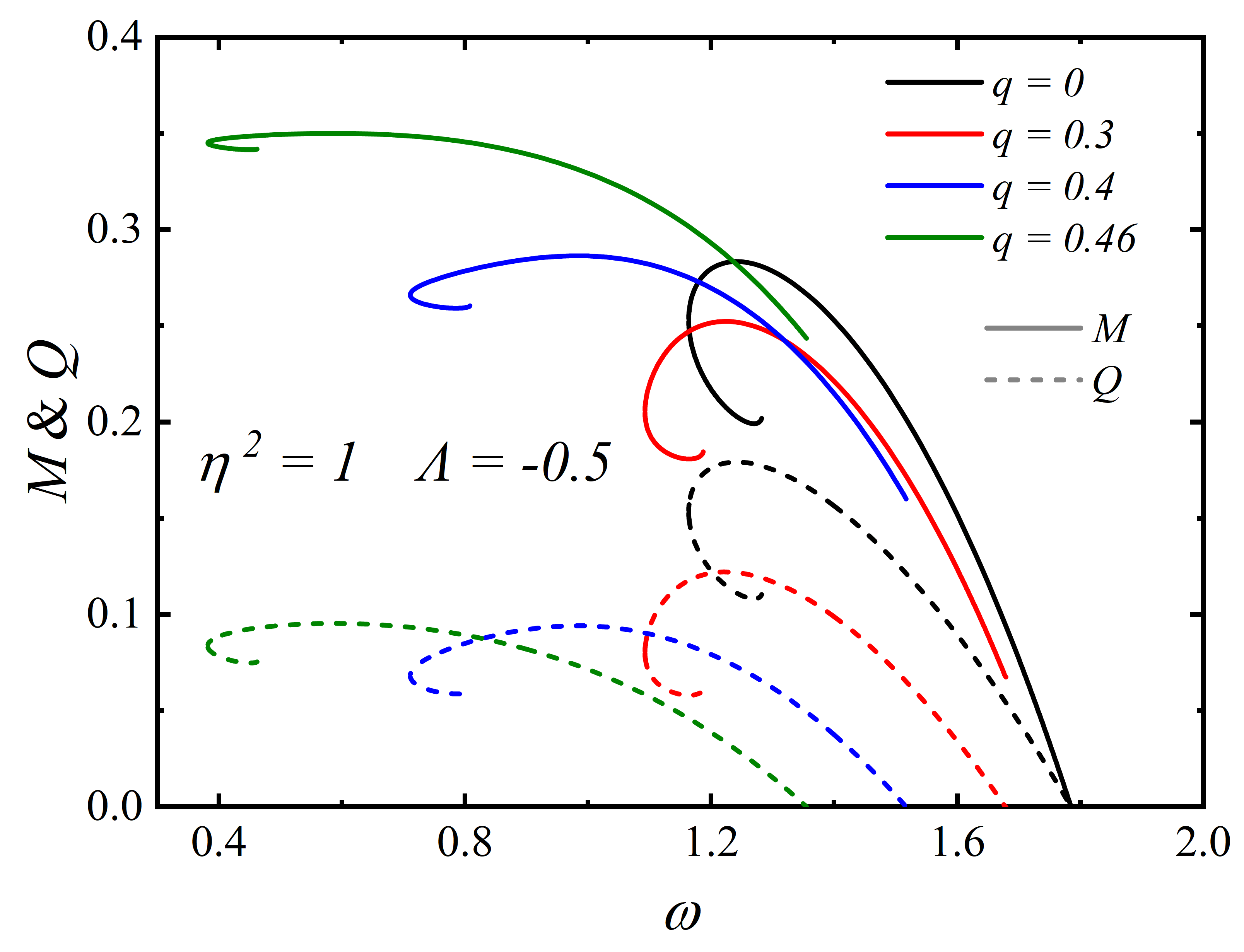}}
\subfigure{\includegraphics[width=0.45\textwidth]{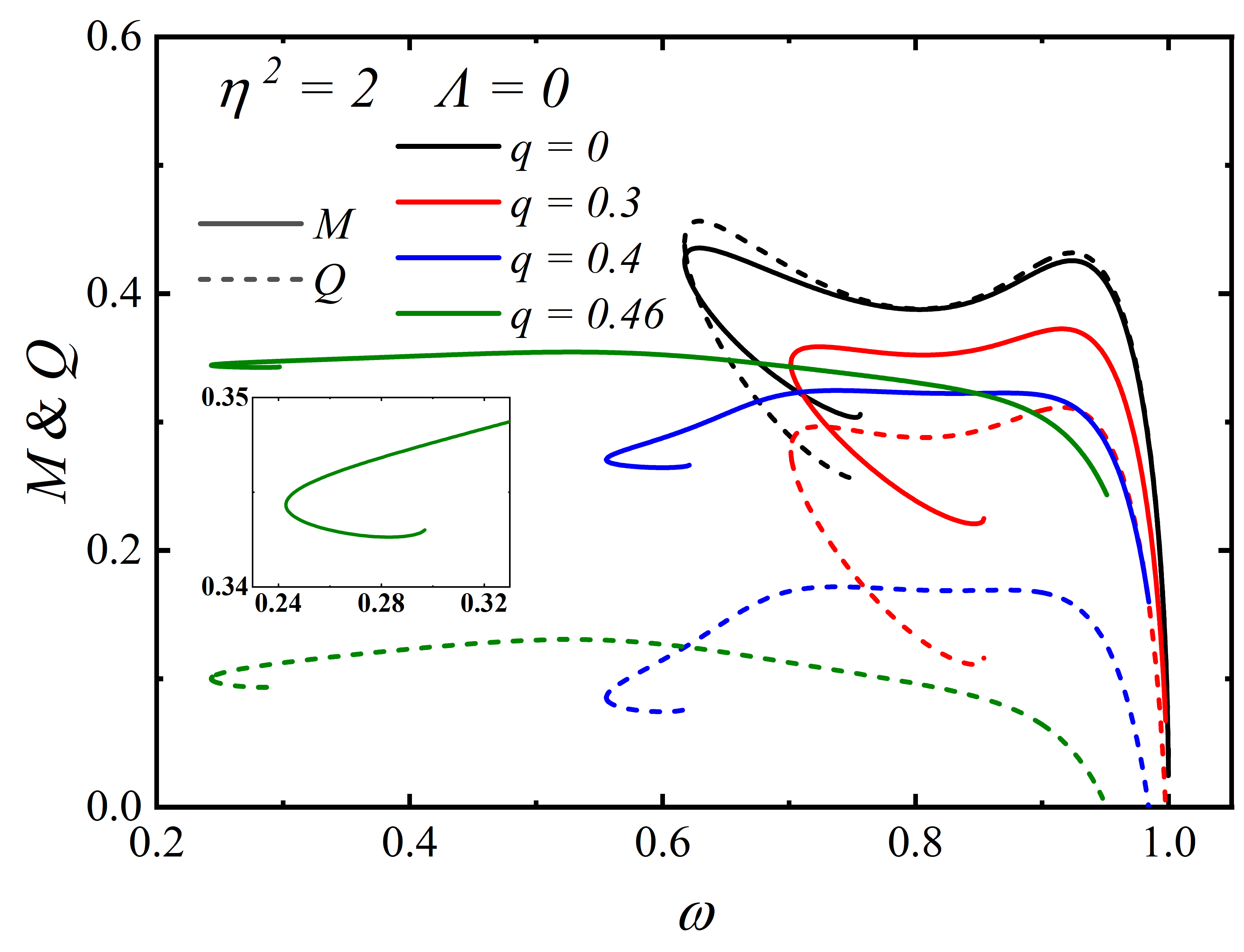}}
\subfigure{\includegraphics[width=0.45\textwidth]{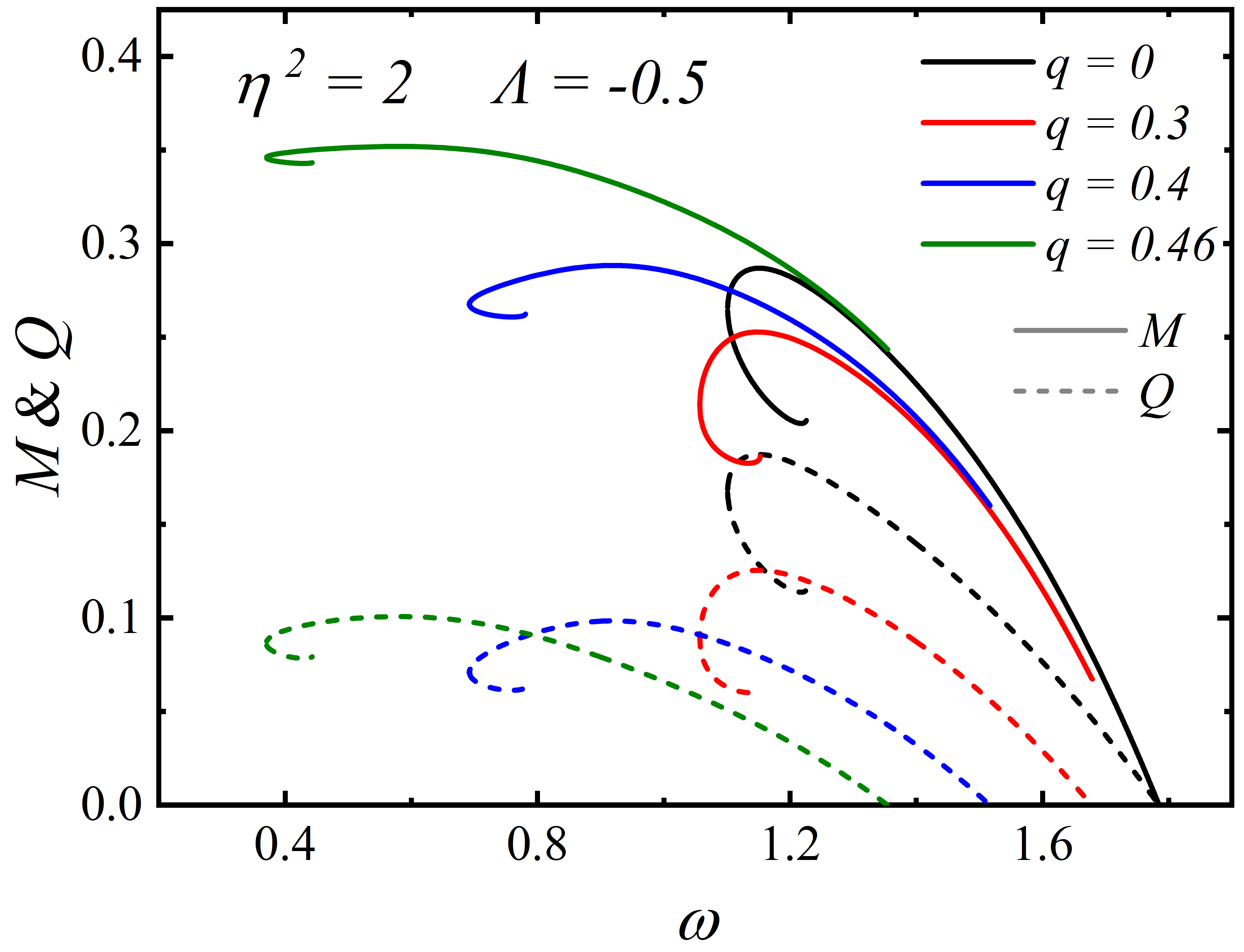}}
\subfigure{\includegraphics[width=0.43\textwidth]{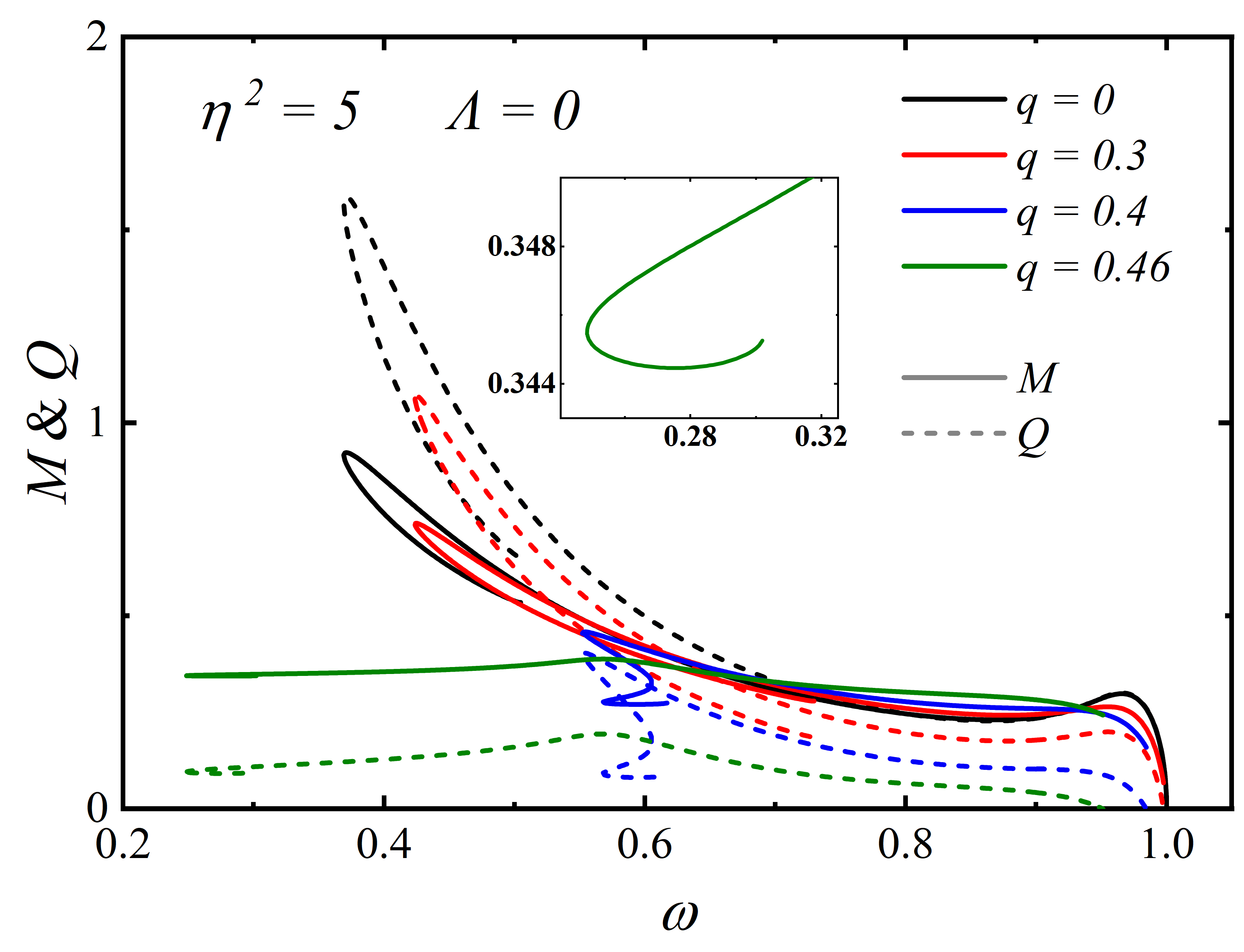}}
\subfigure{\includegraphics[width=0.45\textwidth]{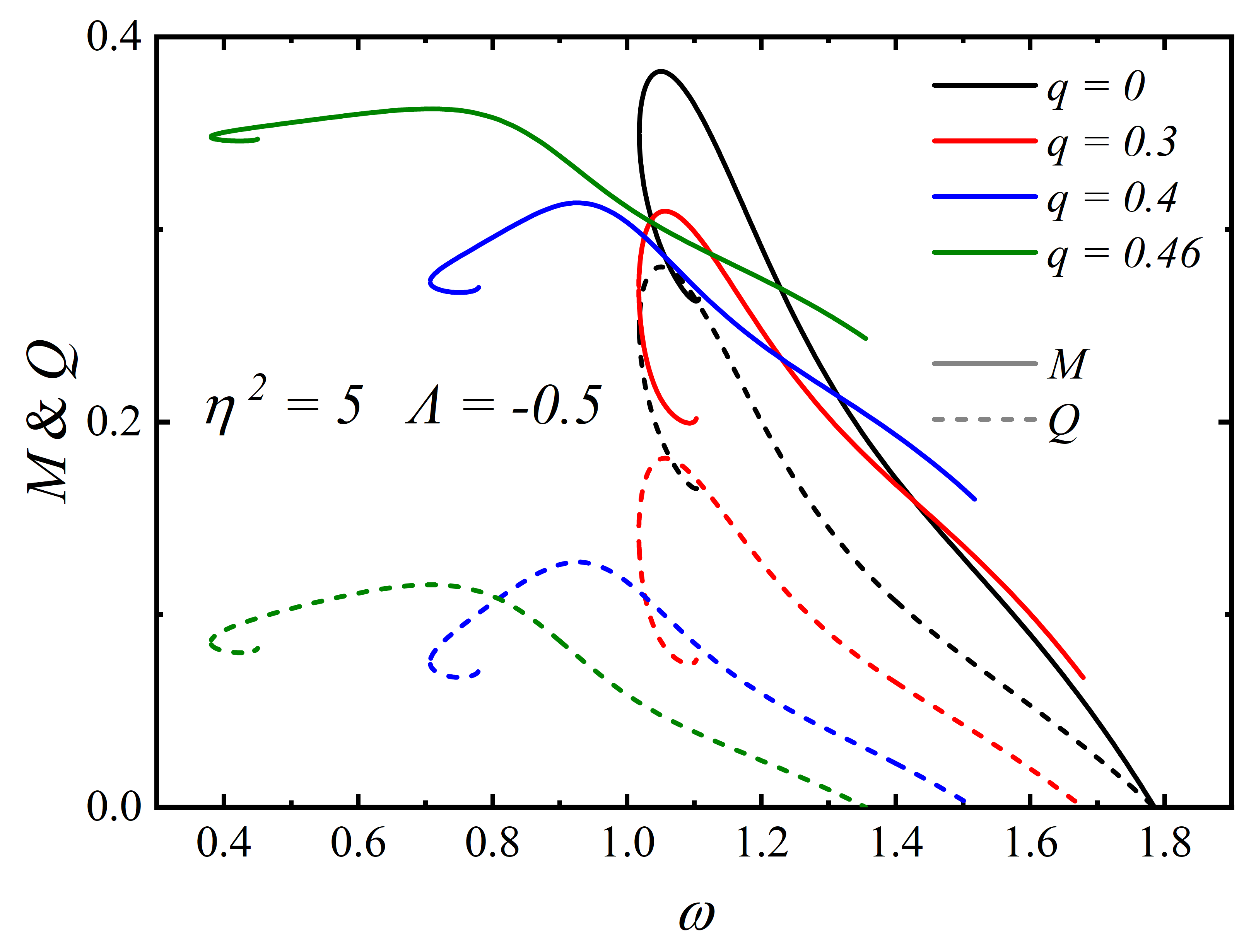}}
\caption{The ADM mass $M$ (solid line) and Noether charge $Q$ (dashed line) as a function of frequency $\omega$ with $q<q_c$ at $\eta^2 = \{1,2,5\}$. The left panels corresponds to $\Lambda=0$, the right panels corresponds to $\Lambda=-0.5$.}\label{fig.1}
\end{figure}

First, we give the solitonic Hayward boson star solutions for $q < q_c$. In Fig. \ref{fig.1}, we show the function of the ADM mass $M$ (solid line) and Noether charge $Q$ (dashed line) with frequency $\omega$ for different magnetic charges $q$ = \{0, 0.3, 0.4, 0.46\} and the coupling parameter $\eta^2$ = \{1, 2, 5\}. The left panels correspond to the asymptotic Minkowski spacetime with $\Lambda = 0$, and the right panels correspond to the asymptotic Ads spacetime with $\Lambda = -0.5$. When $\Lambda = 0$, the maximum frequency satisfies $\omega_{\max} < 1$. The curve unfolds gradually with increasing magnetic charge $q$ for a fixed $\eta^2$. As $\eta^2$ increases, the mass $M$ and the Noether charge $Q$ exhibit two local maxima and become sharper. However, at $\Lambda = -0.5$, this phenomenon is not observed and the maximum frequency $\omega_{\max}$ can exceed 1, which is related to the effect of the cosmological constant. Furthermore, due to the contribution of the Hayward term, the minimum of the ADM mass is not zero when $q>0$. Notice that unlike the ADM mass $M$, the Noether charge $Q$ represents only the number of scalar field particles, so its minimum value remains zero.

\begin{figure}[!htbp]
\centering
\subfigure{\includegraphics[width=0.49\textwidth]{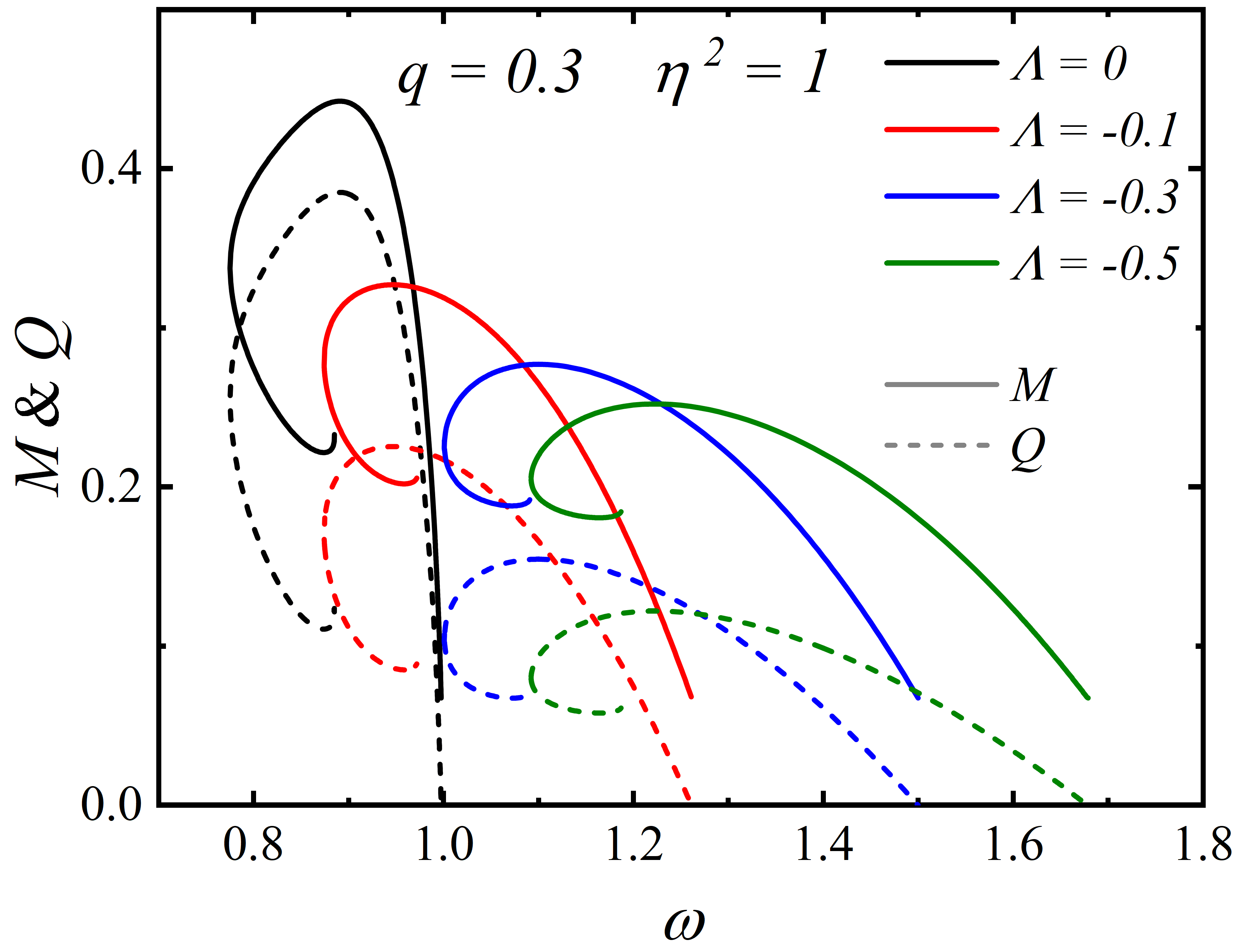}}
\caption{The ADM mass $M$ (solid line) and Noether charge $Q$ (dashed line) as a function of $\omega$ with different $\Lambda$ at $q=0.3,\ \eta^2 = 1$.
  }\label{fig.2}
 \end{figure}
 
We demonstrate the influence of the cosmological constant on the $M(Q)-\omega$ curve in Fig. \ref{fig.2}. It can be observed that as $\Lambda$ decreases, the maximum mass $M_{\max}$ decreases, while the minimum mass $M_{\min}$ remains almost unchanged. The minimum frequency $\omega_{\min}$ and the maximum frequency $\omega_{\max}$ increase. It appears as if the curve is being stretched horizontally. The curve $Q$ - $\omega$ exhibits similar properties.
\begin{figure}[!htbp]
\centering
\subfigure{\includegraphics[width=0.49\textwidth]{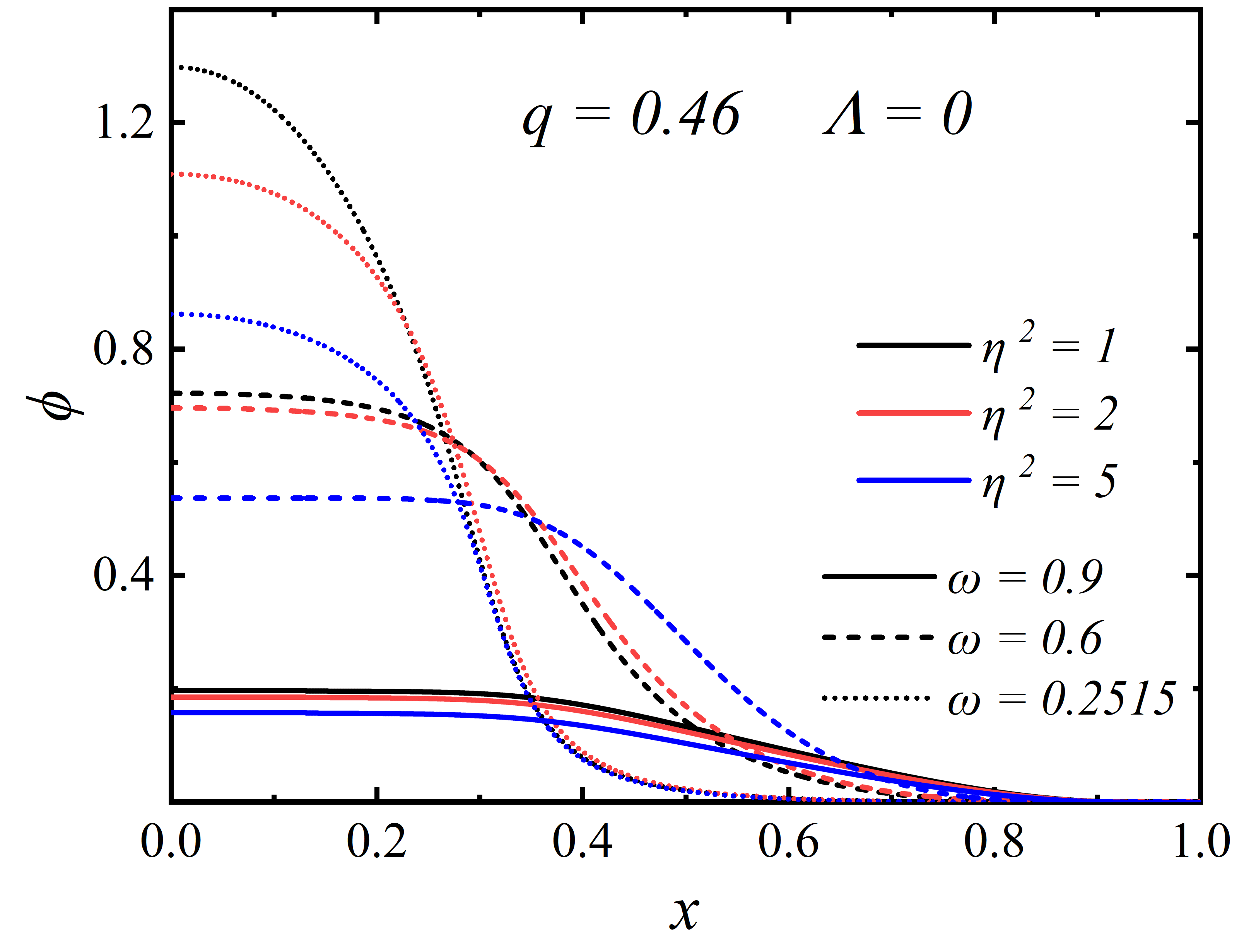}}
\subfigure{\includegraphics[width=0.49\textwidth]{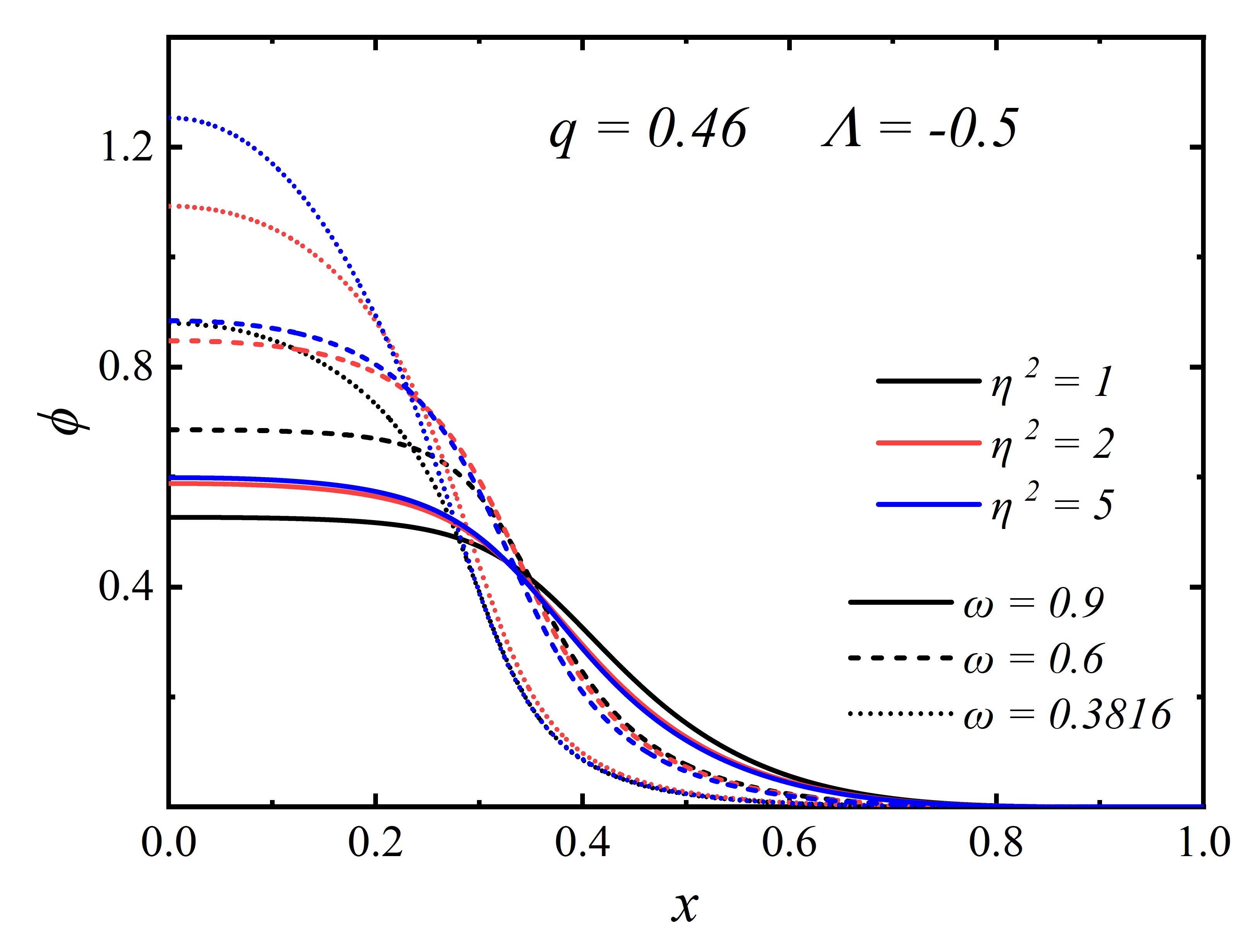}}
\subfigure{\includegraphics[width=0.49\textwidth]{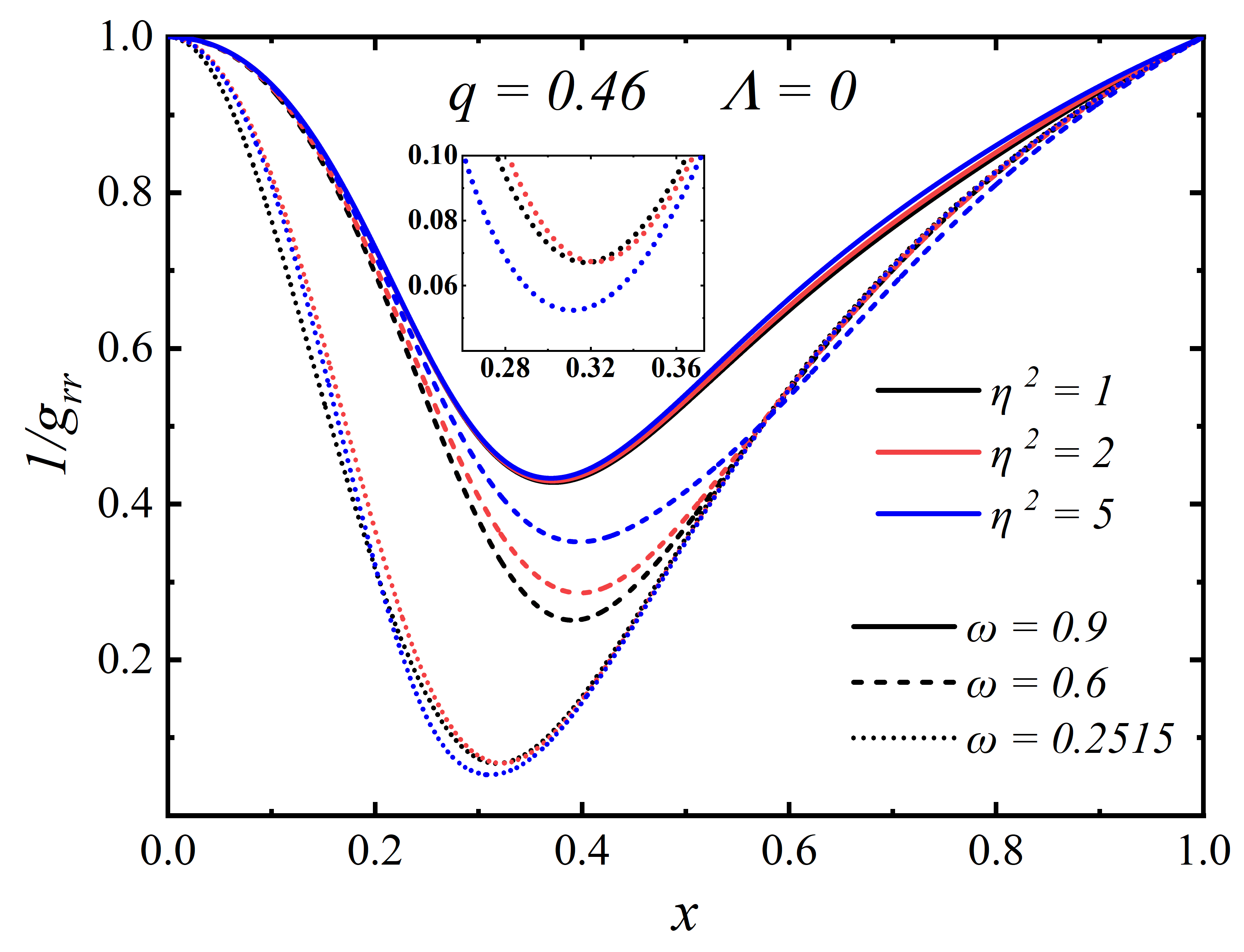}}
\subfigure{\includegraphics[width=0.49\textwidth]{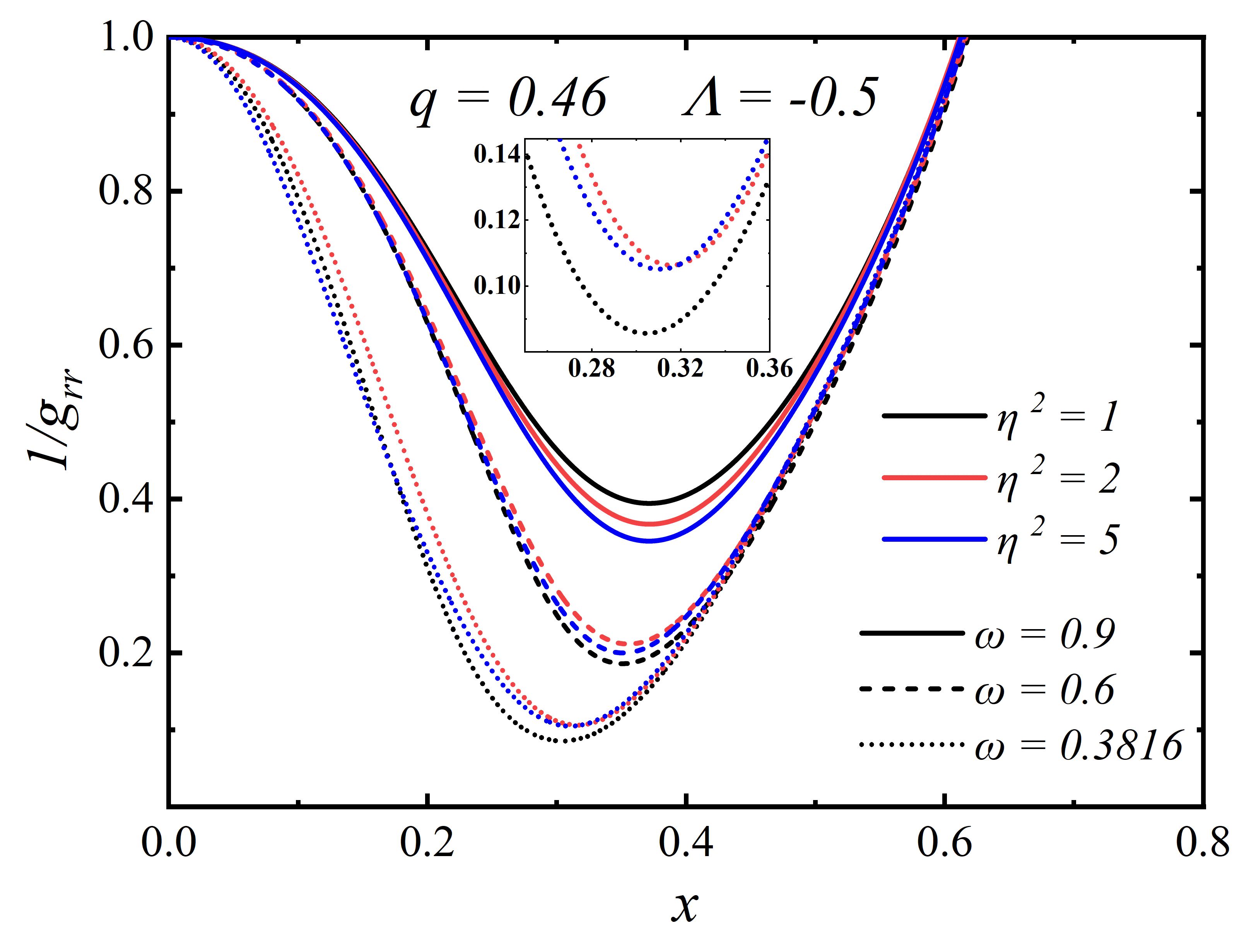}}
\subfigure{\includegraphics[width=0.49\textwidth]{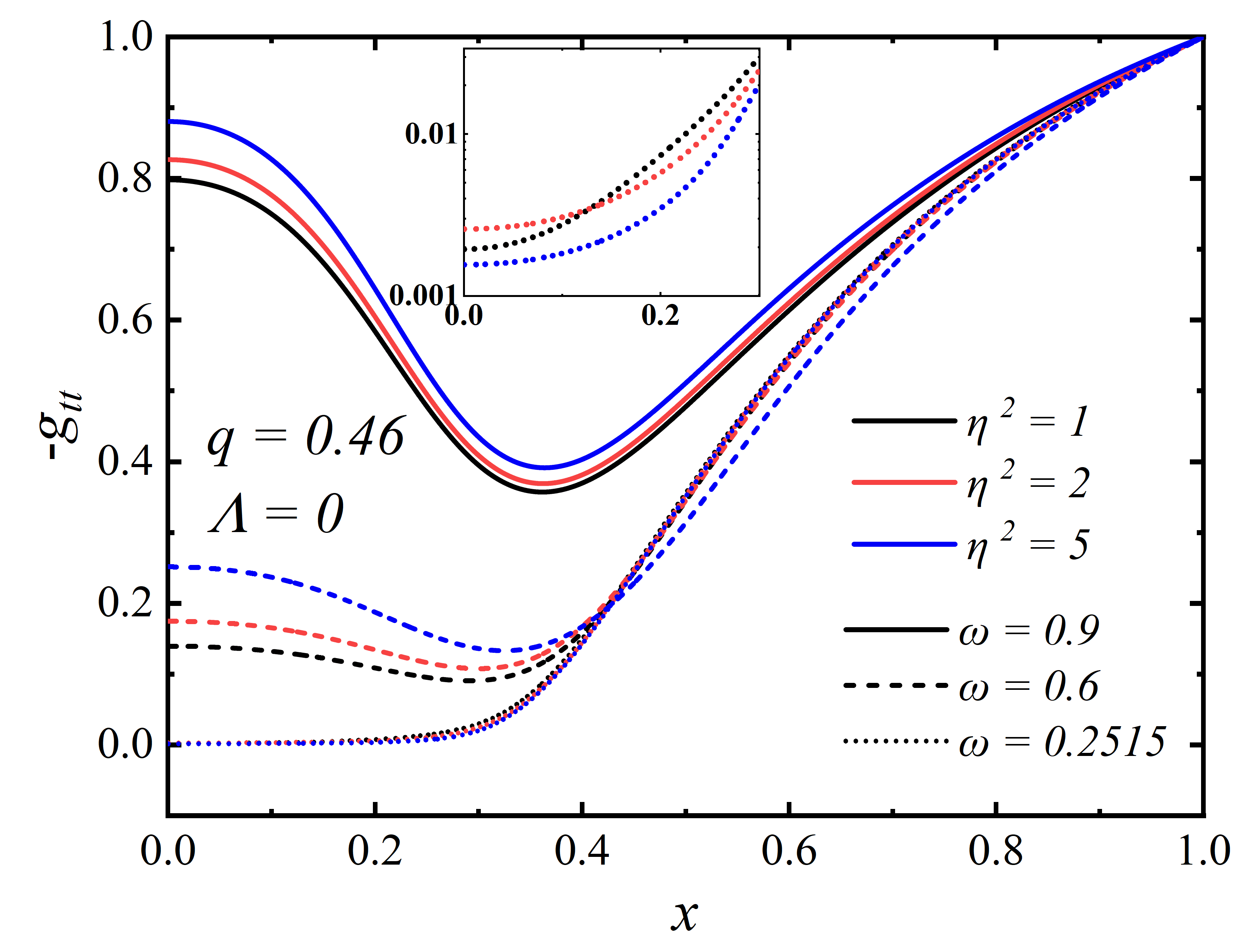}}
\subfigure{\includegraphics[width=0.49\textwidth]{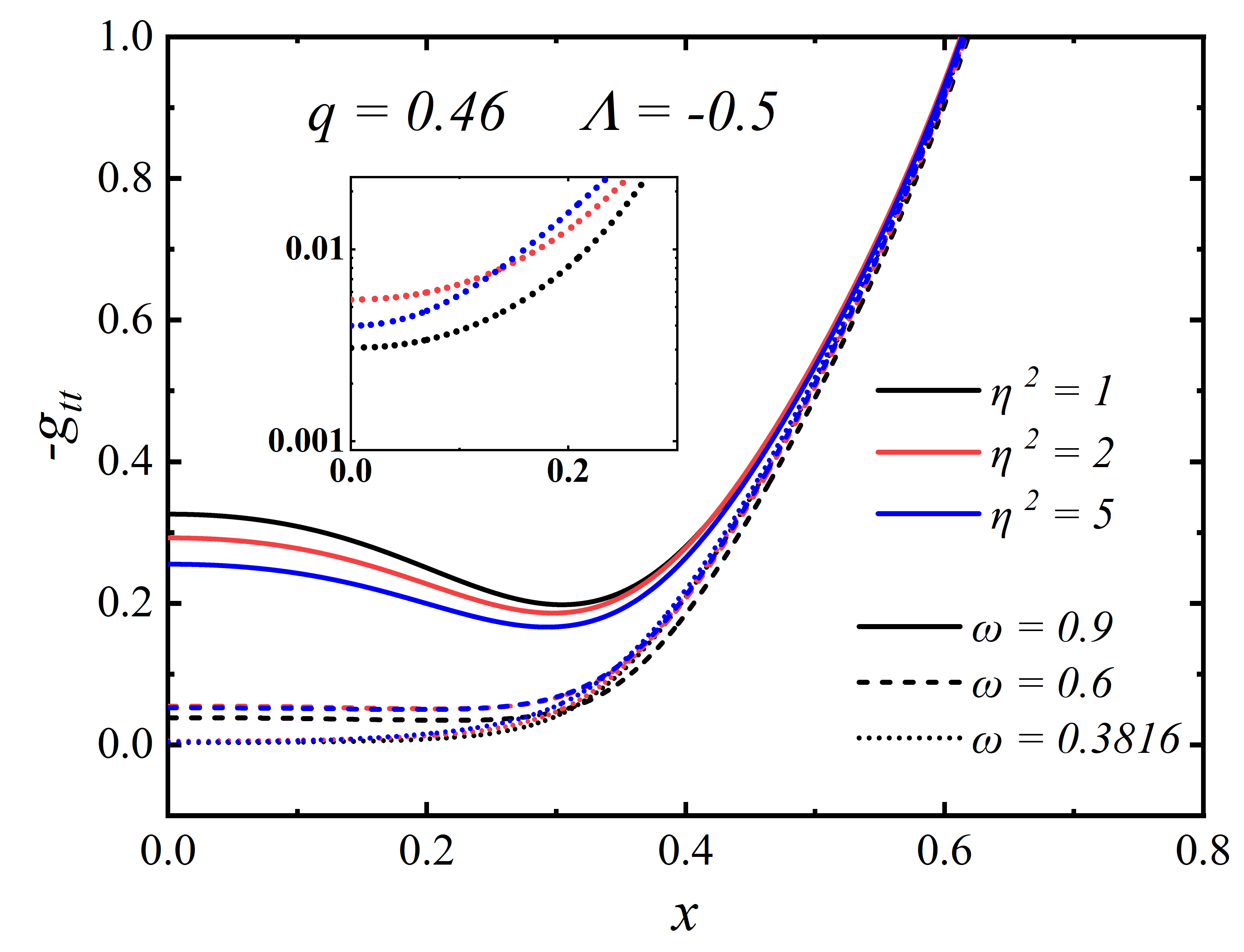}}
\caption{The scalar field $\phi$, metric functions $-g_{tt}=N\,\sigma^2$ and $1/g_{rr}=N$ as functions of $x$ with $q=0.46$. The left (right) panels corresponds to $\Lambda = 0$ ($\Lambda = -0.5)$.
  }\label{fig.3}
 \end{figure}

In Fig. \ref{fig.3}, we analyze the radial distributions of the scalar field $\phi$, the metric functions $-g_{tt}=N\,\sigma^2$ and $1/g_{rr}=N$ with $q = 0.46$. The left panels corresponds to $\Lambda = 0$, and the right panels corresponds to $\Lambda = -0.5$. It can be observed that as $\omega$ decreases, the radial distribution of $\phi$ becomes more compact and the maximum value increases. For the configuration with $\eta^2=1$ and $\Lambda=0 \ (\Lambda=-0.5)$, the minimum achievable frequency is $\omega_{\min}=0.2515$ ($\omega_{\min}=0.3816$), corresponding to the leftmost end of the first branch of the $M-\omega$ curve (see Fig. \ref{fig.1}). Simultaneously, the minimum values of $1/g_{rr}$ and $-g_{tt}$ are approximately on the order of $10^{-1}$ and $10^{-2}$, respectively. By comparing the curves of different colors, it can be observed that under different parameters, the influence of $\eta^2$ on the same function can be entirely different: when $\Lambda=0$, $\phi_{{\max}}$ decreases as $\eta^2$ increases; for $\Lambda=-0.5$, the opposite behavior is observed. Furthermore, it is noteworthy that $-g_{tt}$ and $1/g_{rr}$ with $\Lambda=-0.5$ diverge at infinity, which is the effect of the cosmological constant. 

\begin{figure}[!htbp]
\centering
\subfigure{\includegraphics[width=0.49\textwidth]{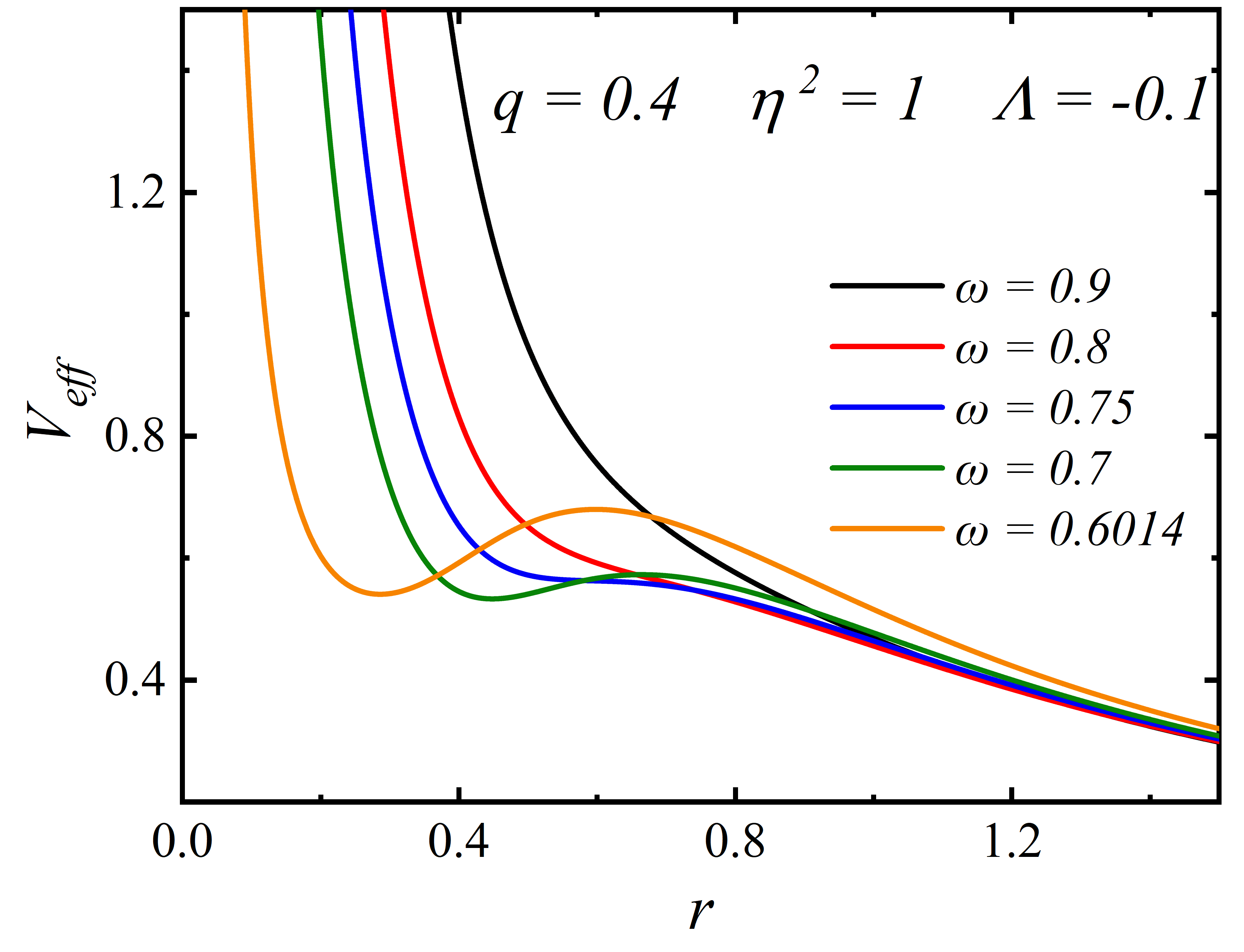}}
\subfigure{\includegraphics[width=0.49\textwidth]{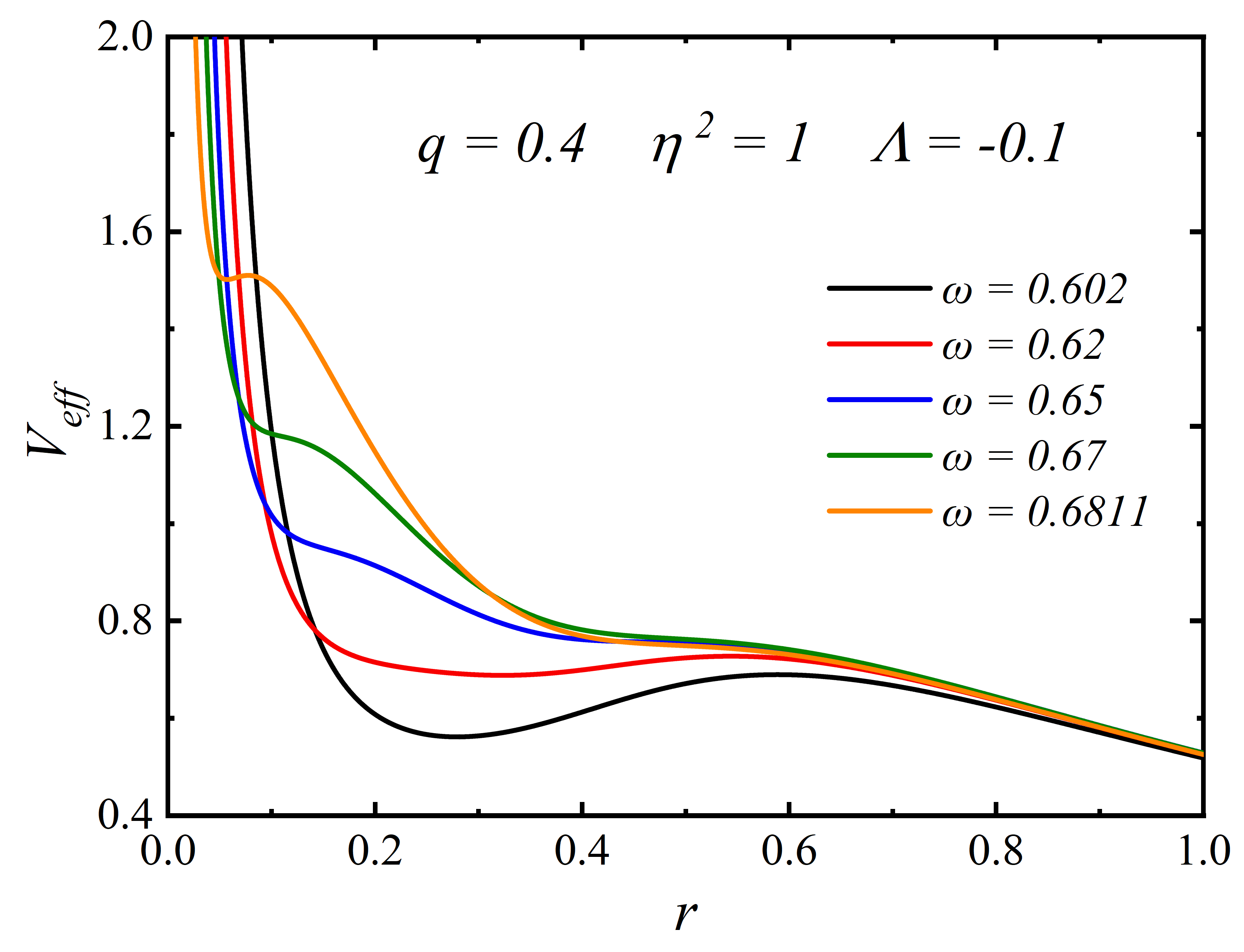}}
\caption{The effective potential $V_{eff}$ as a function of $r$ with different $\omega$ at $q=0.3,\ \eta^2 = 1$ and $\Lambda=-0.1$. The left and right panels correspond to the first and second branch solutions of SHBSs, respectively.
  }\label{fig.4} 
 \end{figure}
 
Next, we discuss the light rings of SHBSs for $q<q_c$. In Fig. \ref{fig.4}, we present the effective potential $V_{eff}$ as a function of the radius $r$ with different frequencies $\omega$ for $q = 0.4,\, \eta^2 = 1 $ and $\Lambda=-0.1$. The left and right panels correspond to the first and second branch solutions of SHBSs, respectively. As explained in Sect. \ref{sec2}, the positions of the light rings can be determined by the $V_{eff}$. As can be seen from the left panel, for the first branch solution: as the frequency $\omega$ decreases, the number of light rings in this configuration increases from zero to two. Based on their radial positions, we classify them as the inner light ring and outer light ring, where the inner one is stable and the outer one is unstable. However, for the second branch solution (see the right panel), as $\omega$ increases, the number of light rings decreases from two to zero, and then increases to two again. 

\begin{figure}[!htbp]
\centering
\subfigure{\includegraphics[width=0.49\textwidth]{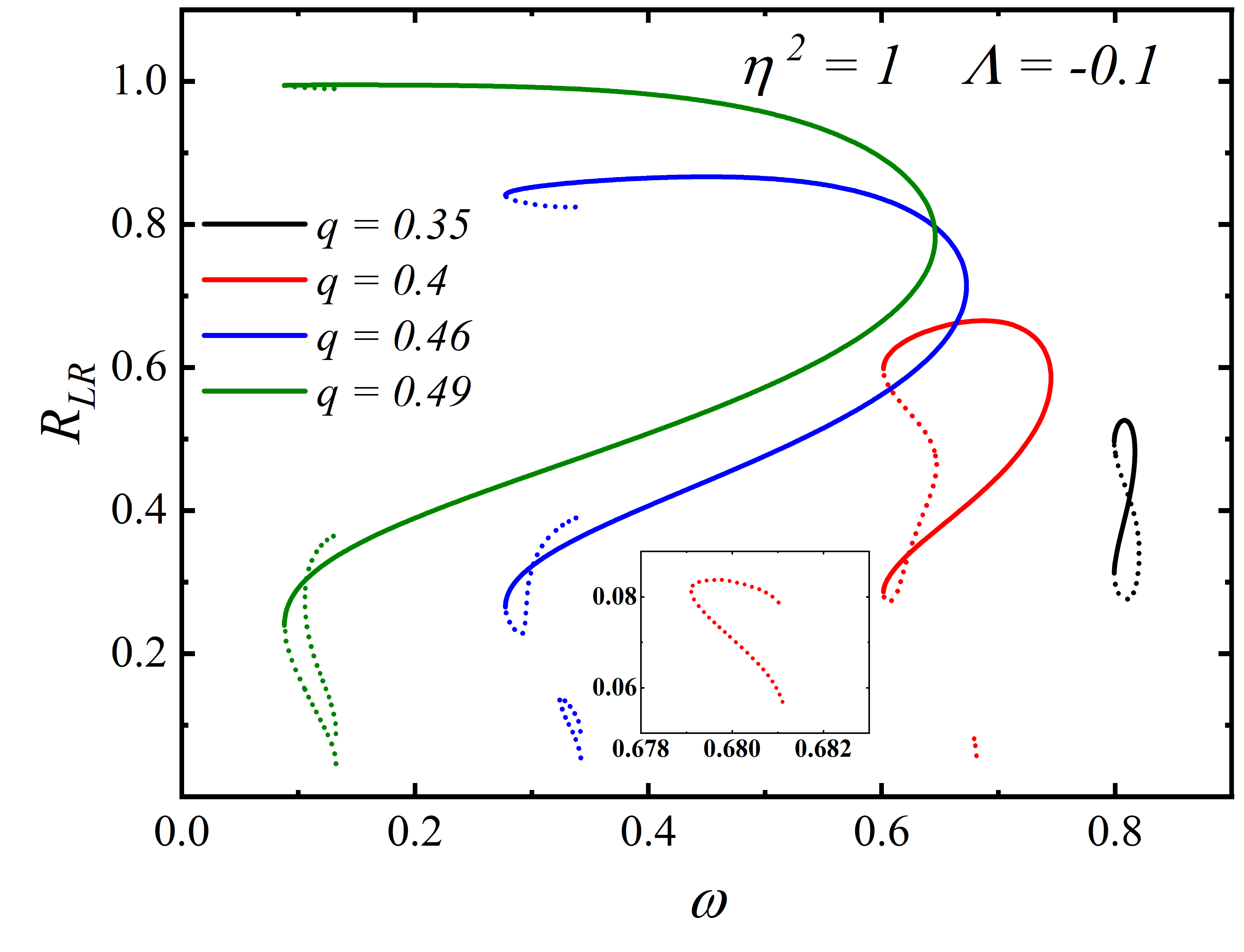}}
\subfigure{\includegraphics[width=0.49\textwidth]{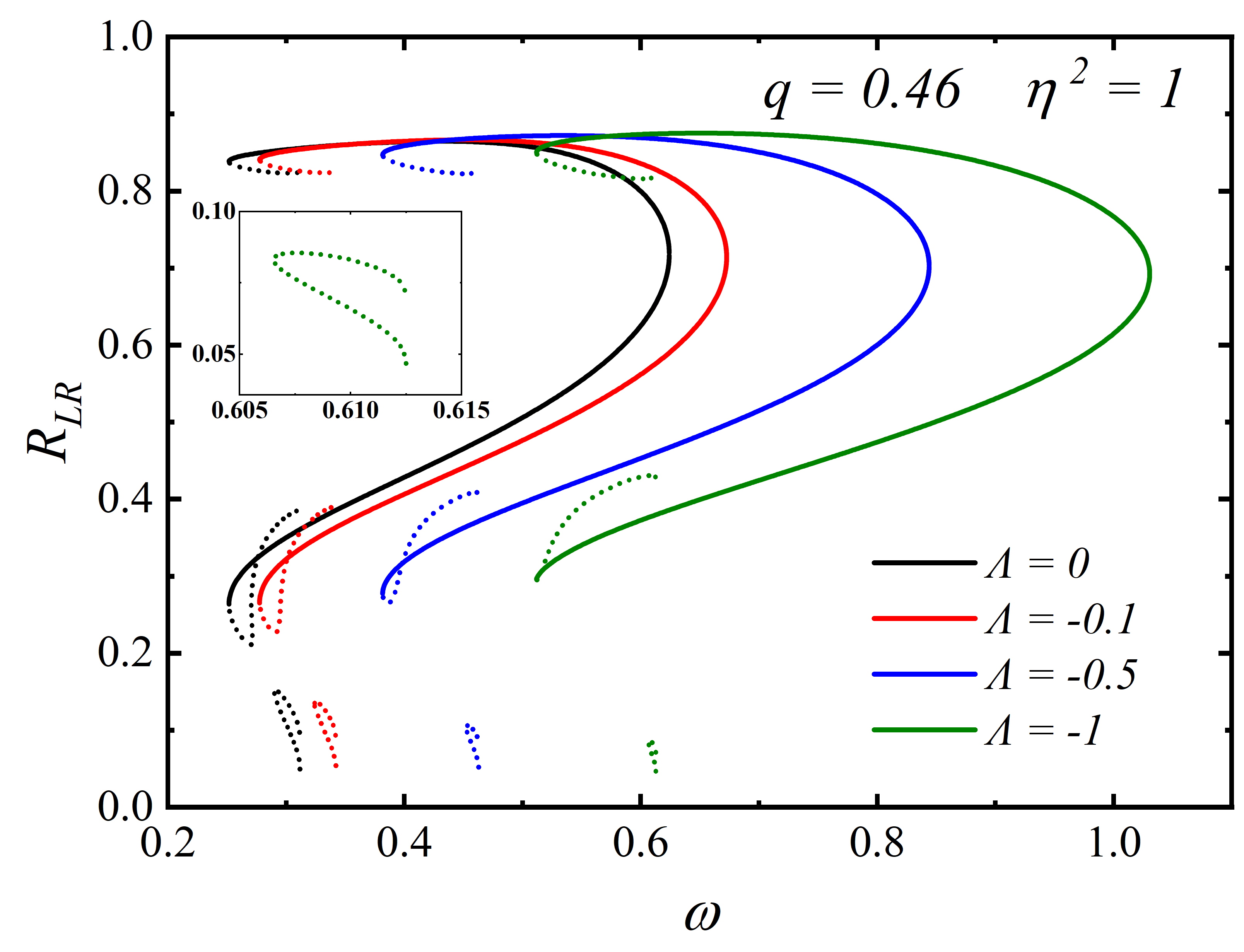}}
\subfigure{\includegraphics[width=0.49\textwidth]{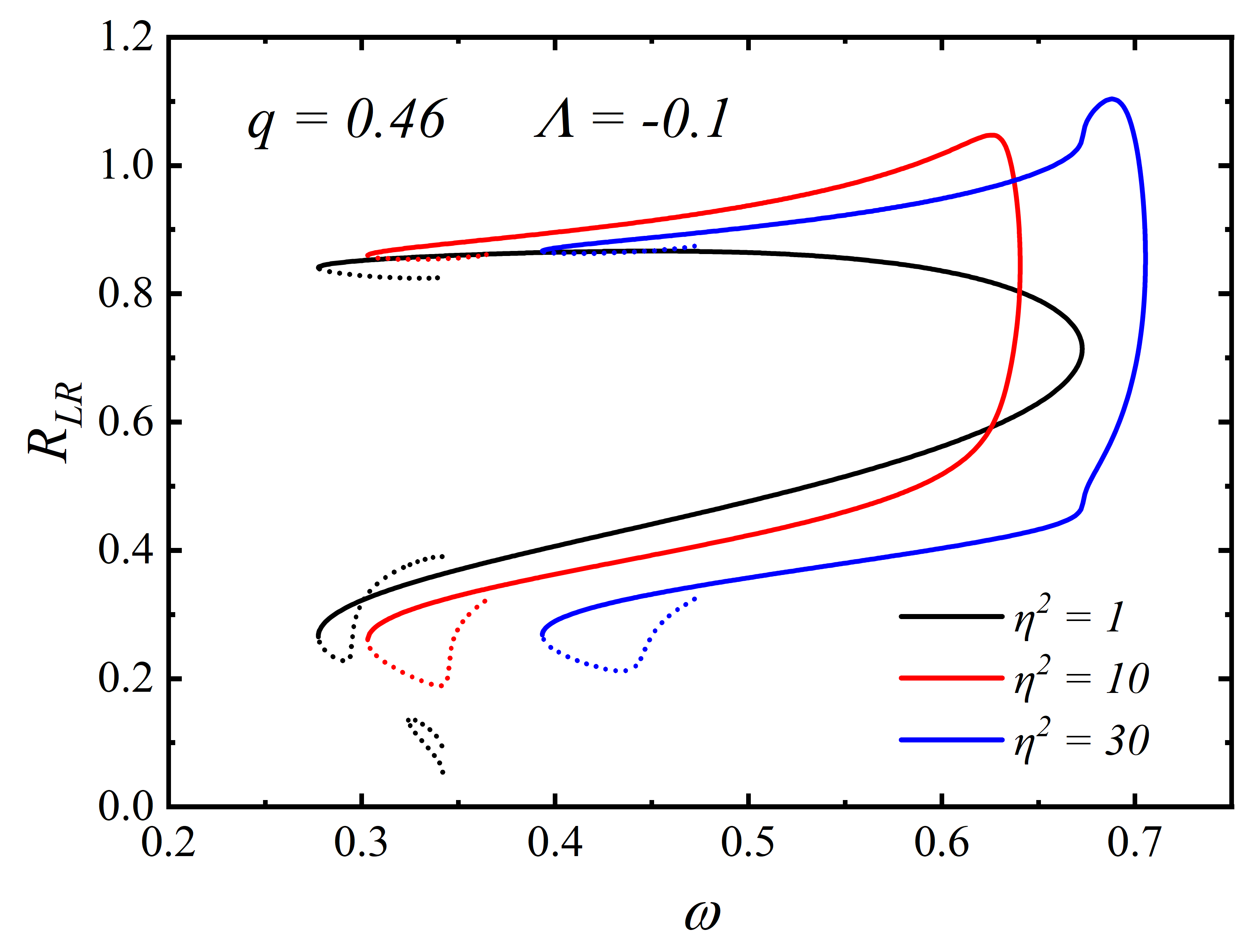}}
\caption{The position of light ring $R_{LR}$ as a function of frequency $\omega$ for different parameters. The solid and dashed lines correspond to the first and second branch solutions, respectively.
  }\label{fig.5}
 \end{figure}

In Fig. \ref{fig.5}, we present the variation of light ring positions $R_{LR}$ with $\omega$ for different parameters $q$, $\Lambda$, and $\eta$, where solid and dotted lines correspond to the first and second branch solutions, respectively. For the first branch, we find that as $q$ increases, both the maximum and minimum frequencies ($\omega_{\text{maxl}}$ and $\omega_{\text{minl}}$) of the existence of light ring solutions decrease, but $\Delta\omega = \omega_{\text{maxl}} - \omega_{\text{minl}}$ increases. Moreover, for fixed $q$, reducing $\Lambda$ leads to an increase in both $\omega_{\text{maxl}}$ and $\omega_{\text{minl}}$. For the second branch, when $q=0.35$, the number of light rings decreases to zero with increasing $\omega$. However, as $q$ increases to 0.4, the red curve in the top left panel shows that the number of light rings first decreases from two to zero with increasing frequency, and then increases back to two (see the inset). This behavior is consistent with the analysis of the radial effective potential $V_{eff}$ in Fig. \ref{fig.4}. When $q$ increases to 0.46, four light rings emerge within a specific frequency range. Further increasing $q$ enlarges the frequency range of the four light ring solutions. Conversely, decreasing $\Lambda$ or increasing $\eta^2$ (as shown in the top right and lower panels) causes this frequency range to shrink and eventually vanish.
\subsection{Large magnetic charge $q \geq q_c$}

\begin{figure}[!htbp]
\centering
\subfigure{\includegraphics[width=0.45\textwidth]{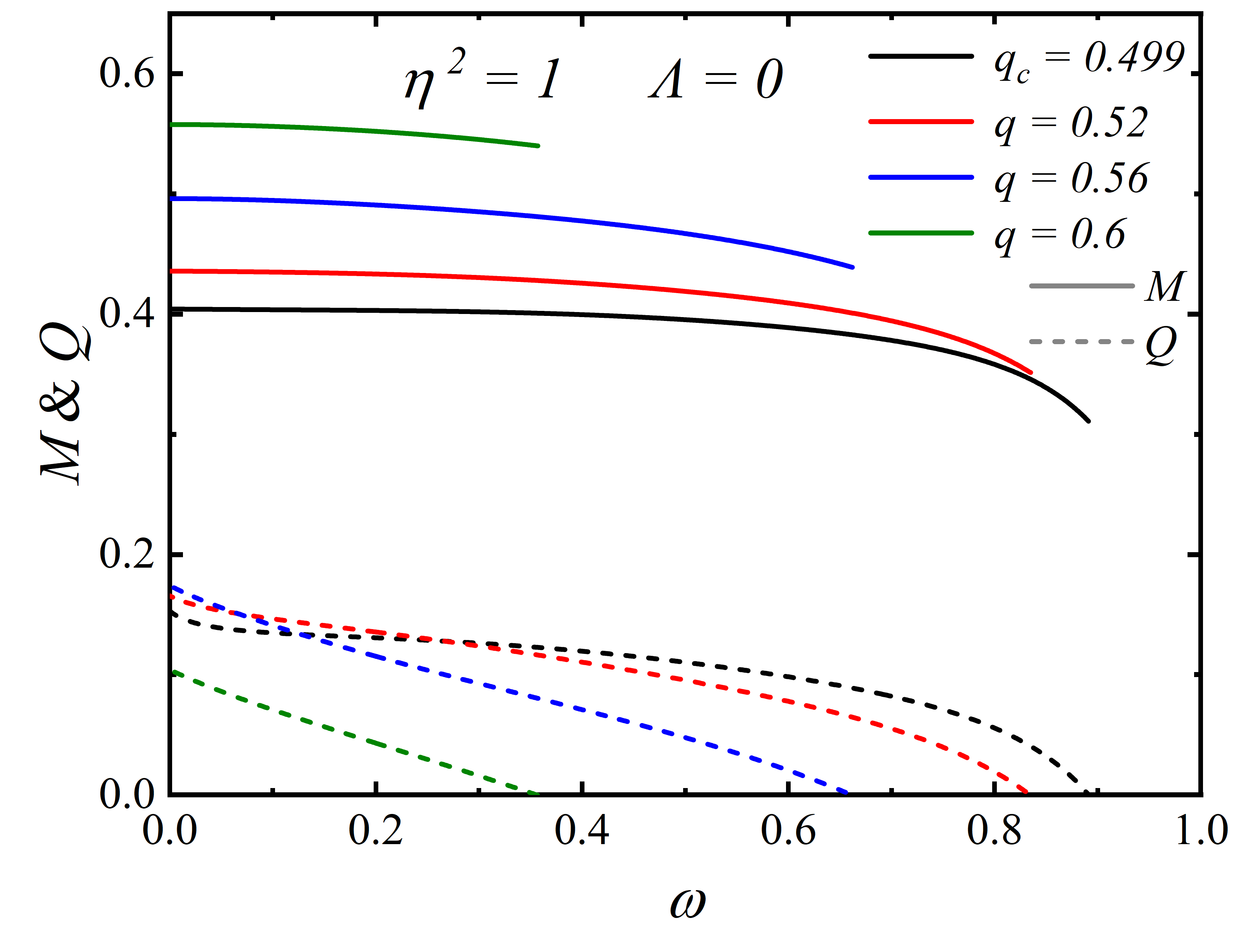}}
\subfigure{\includegraphics[width=0.45\textwidth]{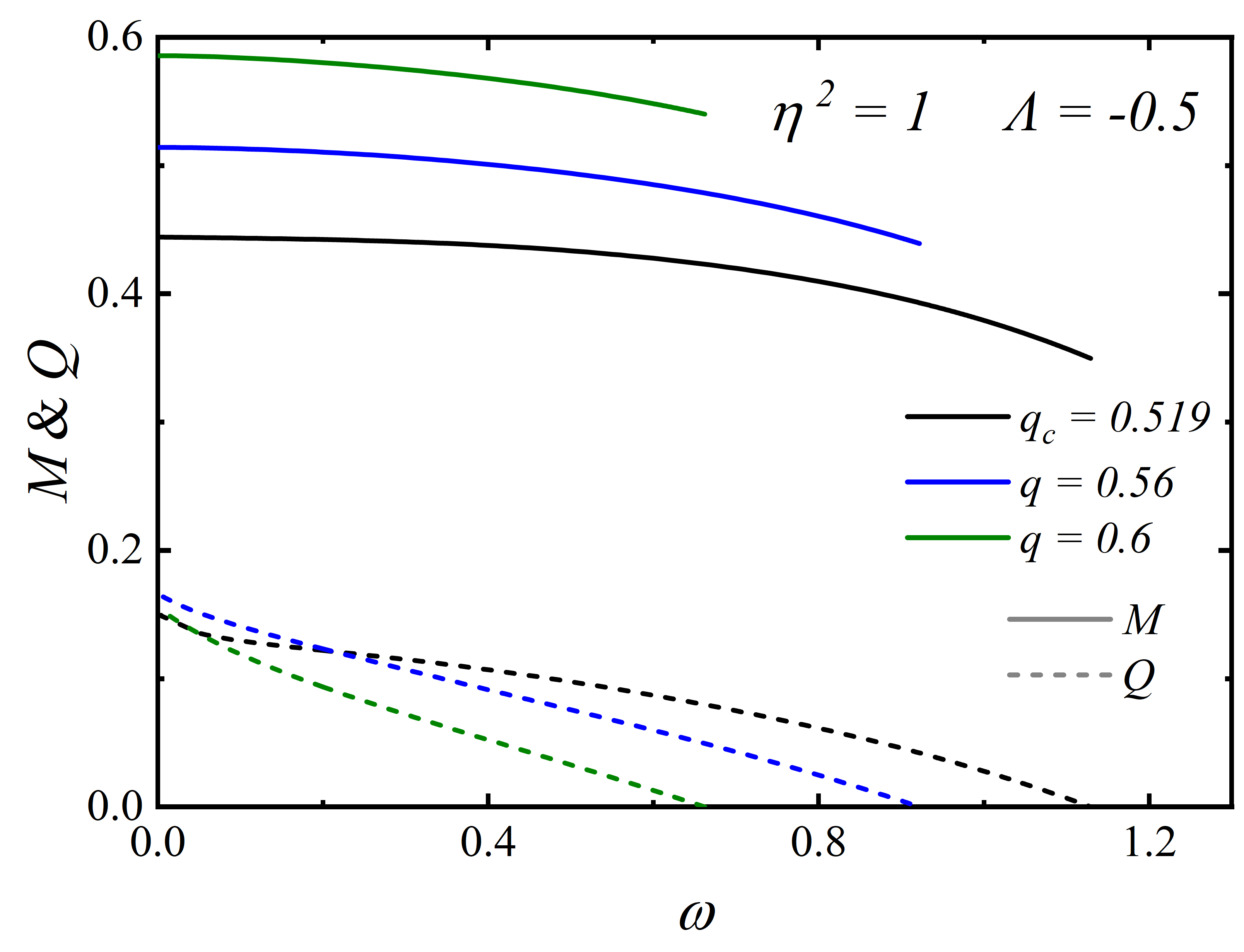}}
\subfigure{\includegraphics[width=0.45\textwidth]{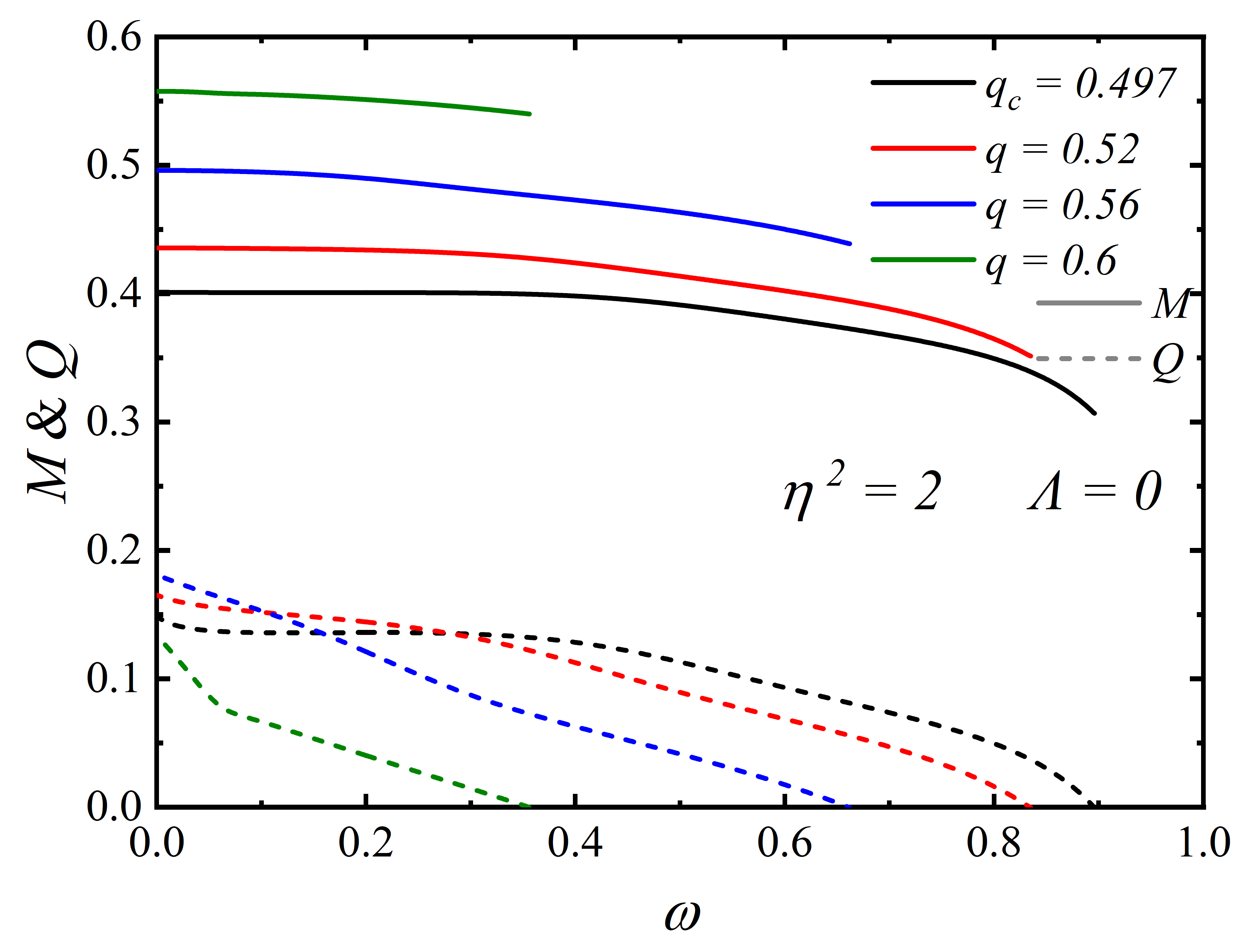}}
\subfigure{\includegraphics[width=0.45\textwidth]{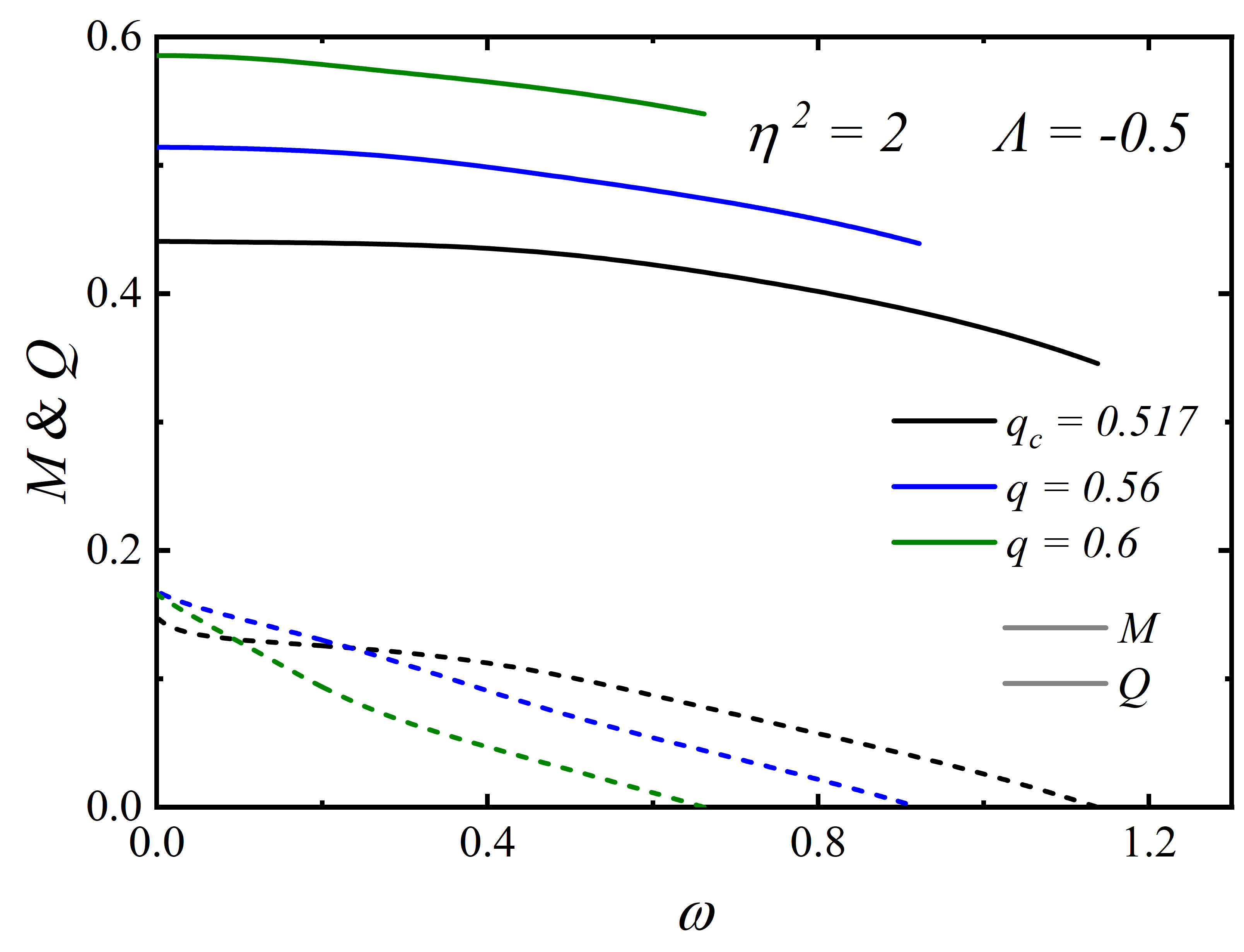}}
\subfigure{\includegraphics[width=0.45\textwidth]{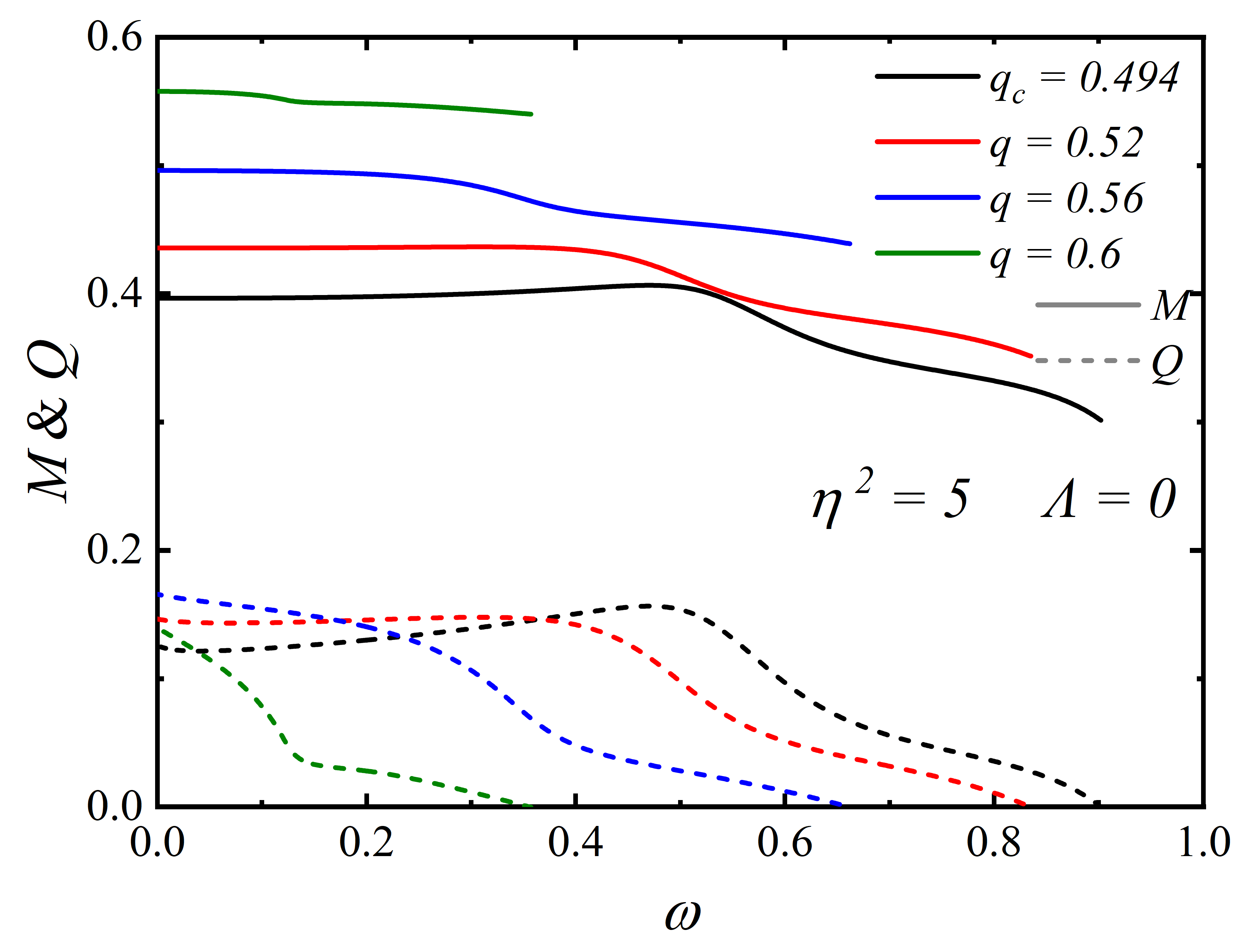}}
\subfigure{\includegraphics[width=0.45\textwidth]{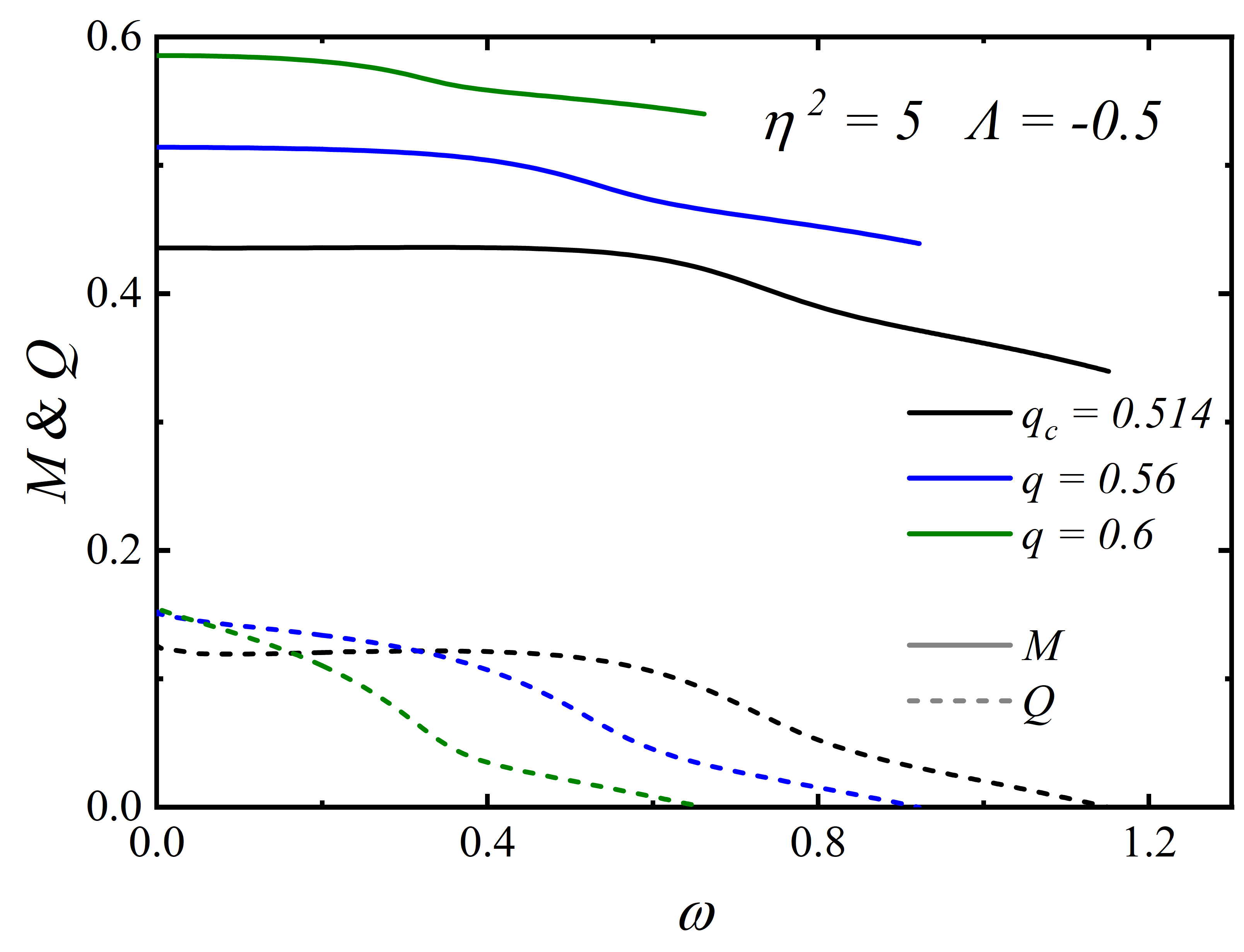}}
\caption{The ADM mass $M$ (solid line) and Noether charge $Q$ (dashed line) versus the frequency $\omega$ with large magnetic charge $q>q_c$ for $\eta^2 = \{1,2,5,10\}$. The left panels corresponds to $\Lambda=0$, the right panels corresponds to $\Lambda=-0.5$.
  }\label{fig.6}
\end{figure}

Next, we give the mass $M$ and charge $Q$ of the solitonic Hayward boson stars versus frequency $\omega$ for $q \geq q_c$ in Fig. \ref{fig.6}. The black line corresponds to the Hayward boson stars with the critical magnetic charge. It can be observed that unlike the case $q < q_c$, extreme solutions with $\omega \to 0$ exist. For both $\Lambda=0$ and $\Lambda=-0.5$, the critical magnetic charge becomes slightly smaller as $\eta^2$ increases. For a fixed $\eta^2$, the negative cosmological constant requires a larger $q_c$. We calculated the values of $q_c$ corresponding to $\eta^2=\{0.01,0.1,0.5,1,2,5,10\}$ with $\Lambda = \{0,-0.1,-0.5,-1,-2\}$ and summarized in Table. \ref{table1}. From this table, we observe that for a fixed $\Lambda$, the influence of $\eta^2$ on $q_c$ is non monotonic, with $q_c$ reaching an extreme value around $\eta^2=1$. This behavior is consistent with the case in the asymptotically flat $\Lambda=0$ Bardeen spacetime~\cite{Zhao:2025hdg}.

Additionally, it can be observed that when $\eta^2$ and $\Lambda$ are fixed, the ADM mass of the extreme solution increases with magnetic charge $q$, while it remains nearly unchanged with variations in $\eta^2$ when $q$ is fixed. This suggests that the mass in the extreme solution is almost independent of $\eta^2$. However, by comparing the left and right panes, it is found that the cosmological constant $\Lambda$ affects the mass of the extreme solution. Unlike mass, the Noether charge $Q$ of the extreme solution is significantly affected by $\eta^2$ and $\Lambda$.

\begin{table}[H]
  \centering
  \begin{tabular}{|c|c|c|c|c|c|c|c|}
    \hline
   $\eta^2$ & 0.01 & 0.1 & 0.5 & 1 & 2 & 5 & 10\\
    \hline
    $q_c\,(\Lambda = 0)$ & 0.4932 & 0.4938 & 0.4968 & 0.4982 & 0.4967 & 0.4939 &0.4926\\
    \hline
    $q_c\,(\Lambda = -0.1)$ & 0.4972 & 0.4978 & 0.5008 & 0.5023 & 0.5007 & 0.4979 &0.4965\\
    \hline
   $q_c\,(\Lambda=-0.5)$ & 0.5131 & 0.5137 & 0.5170 & 0.5186 & 0.5169 & 0.5137 & 0.5122\\
    \hline
    $q_c\,(\Lambda = -1)$ & 0.5337 & 0.5344 & 0.5380 & 0.5397 & 0.5379 & 0.5343 &0.5325\\
    \hline
    $q_c\,(\Lambda = -2)$ & 0.5780 & 0.5787 & 0.5831 & 0.5851 & 0.5830 & 0.5786 &0.5765\\
    \hline
  \end{tabular}
  \caption{The critical charge $q_c$ of SHBSs under different values of parameters $\eta^2$ and $\Lambda$.}
    \label{table1}
\end{table}

\begin{figure}[!htbp]
\centering
\subfigure{\includegraphics[width=0.49\textwidth]{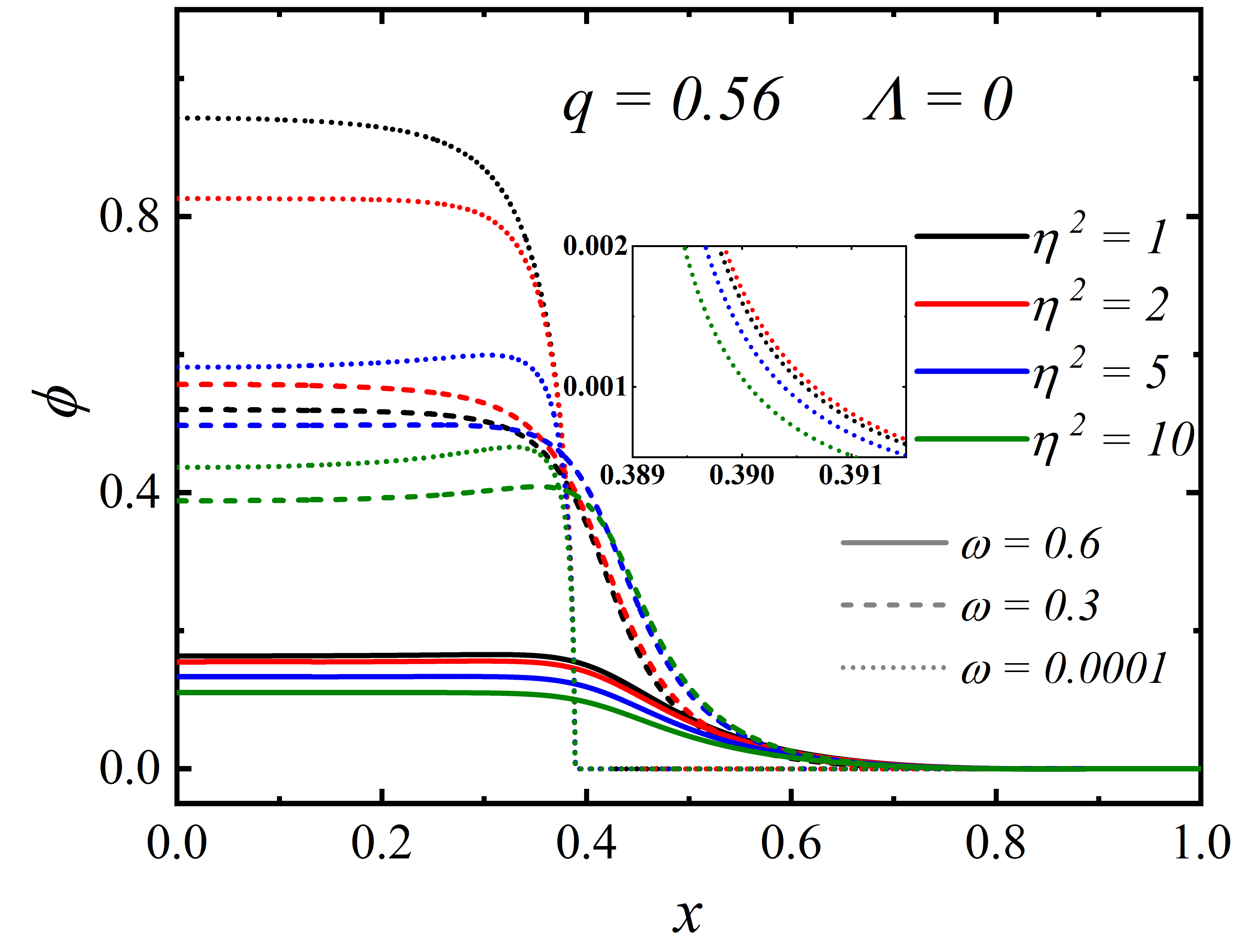}}
\subfigure{\includegraphics[width=0.49\textwidth]{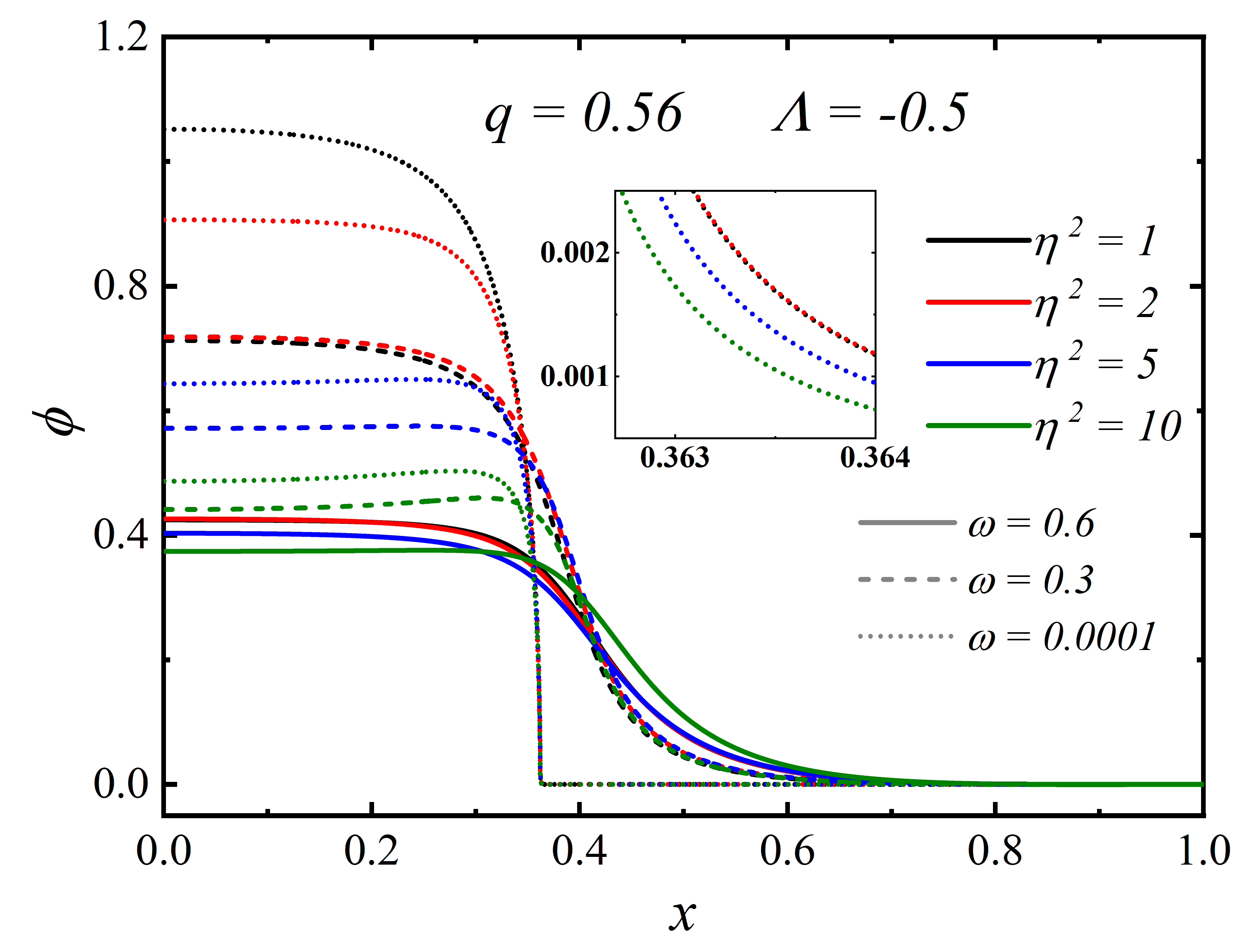}}
\subfigure{\includegraphics[width=0.49\textwidth]{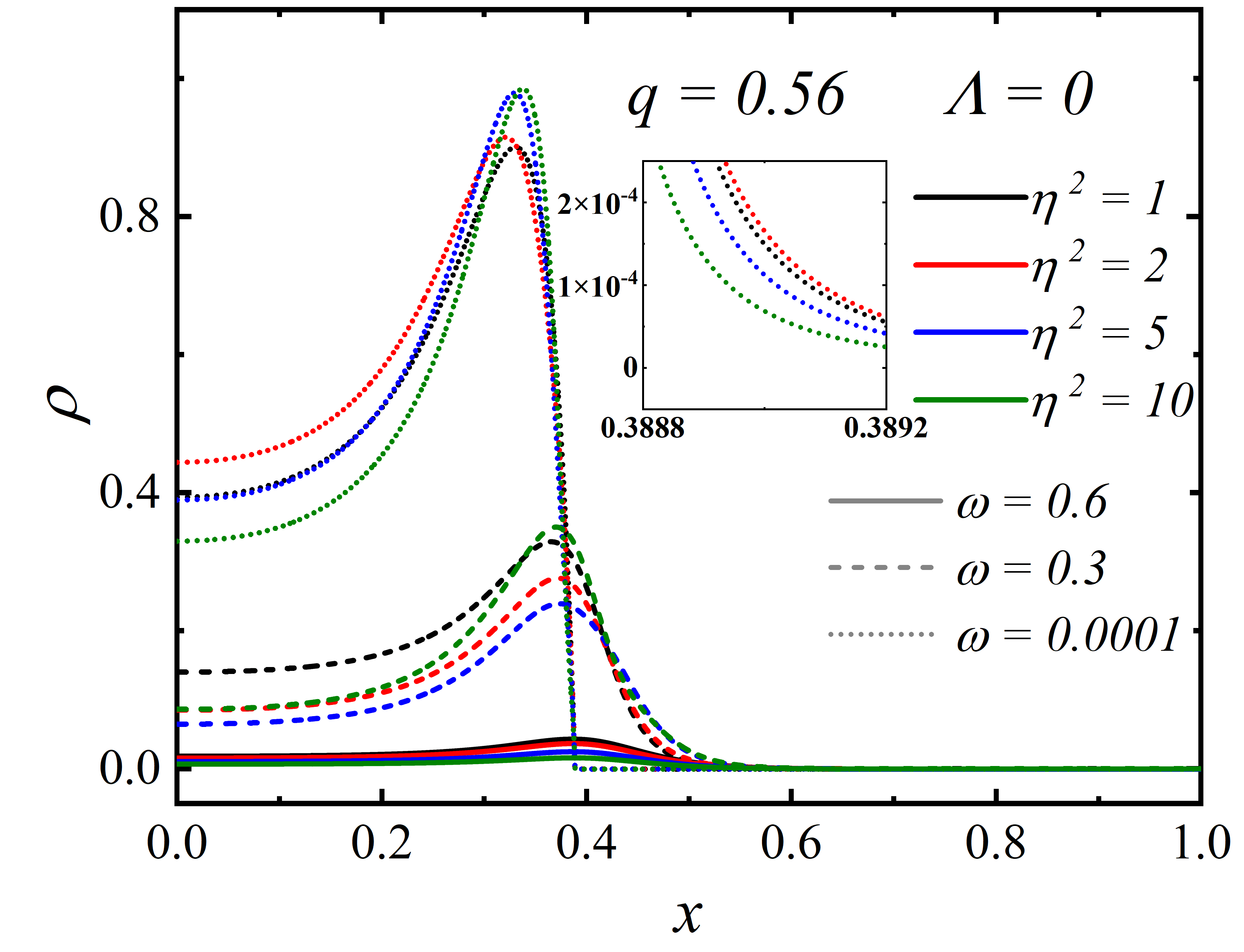}}
\subfigure{\includegraphics[width=0.49\textwidth]{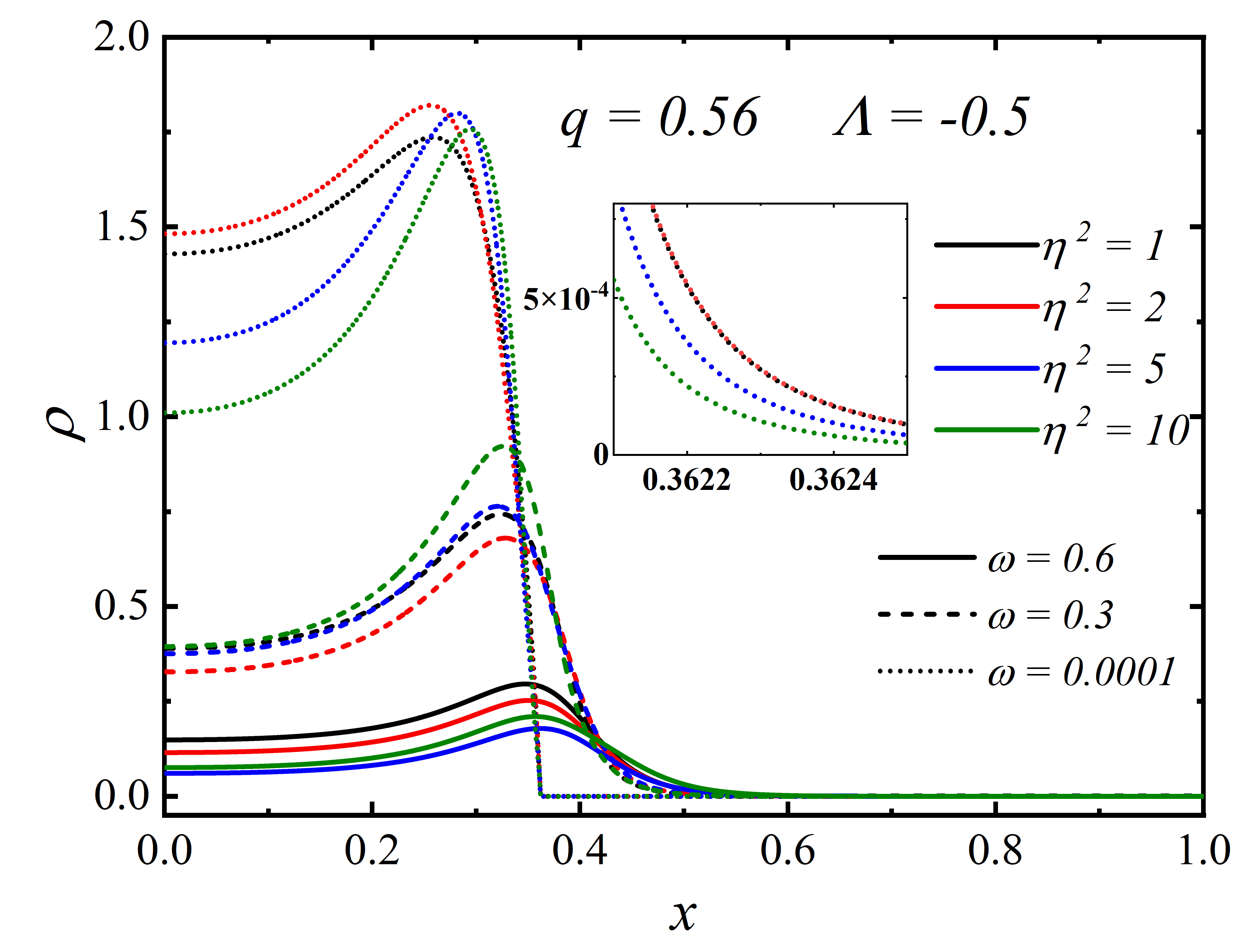}}
\caption{The radial distribution of the scalar field $\phi$, the energy density $\rho$ of the SHBSs with $\eta^2 = \{1,2,5,10\}$ and $\omega =\{ 0.6, 0.3, 0.0001\}$ at $q = 0.56$. The left panels corresponds to $\Lambda = 0$ , the right graphs corresponds to $\Lambda = -0.5$.
  }\label{fig.7}
\end{figure}

In order to explore the behavior of extreme solutions. In Fig. \ref{fig.7} we analyze the radial distributions of the scalar field $\phi$ and the energy density $\rho=-T^{(2)0}_{0}$ of the scalar field with $\eta^2 = \{1,2,5,10\}$ and $\omega = \{0.6, 0.3, 0.0001\}$ for $q = 0.56$. The left panels correspond to $\Lambda = 0$, and the right panels correspond to $\Lambda = -0.5$. The inset shows the details of the curves. From the top panels we can see that when $\omega$ decreases to 0.0001\footnote{In practical numerical computations, if we further refine the step size, $\omega$ can continue to decrease. However, this requires significantly longer computation time without substantially altering the properties of the solution. Therefore, we adopt $\omega = 0.0001$ as an illustrative example.}, the scalar field and its energy density are almost entirely confined within a specific coordinate $x_c$. Beyond this coordinate, they decay rapidly, forming a very steep wall. By comparing the curves of different colors, we observe that variations in $\eta^2$ only affect the distribution of the function within $x_c$. Furthermore, for a fixed $\eta^2$, the case of $\Lambda=-0.5$ yields higher values of $\phi$ and $\rho$ within $x_c$.    

\begin{figure}[!htbp]
\centering
\subfigure{\includegraphics[width=0.49\textwidth]{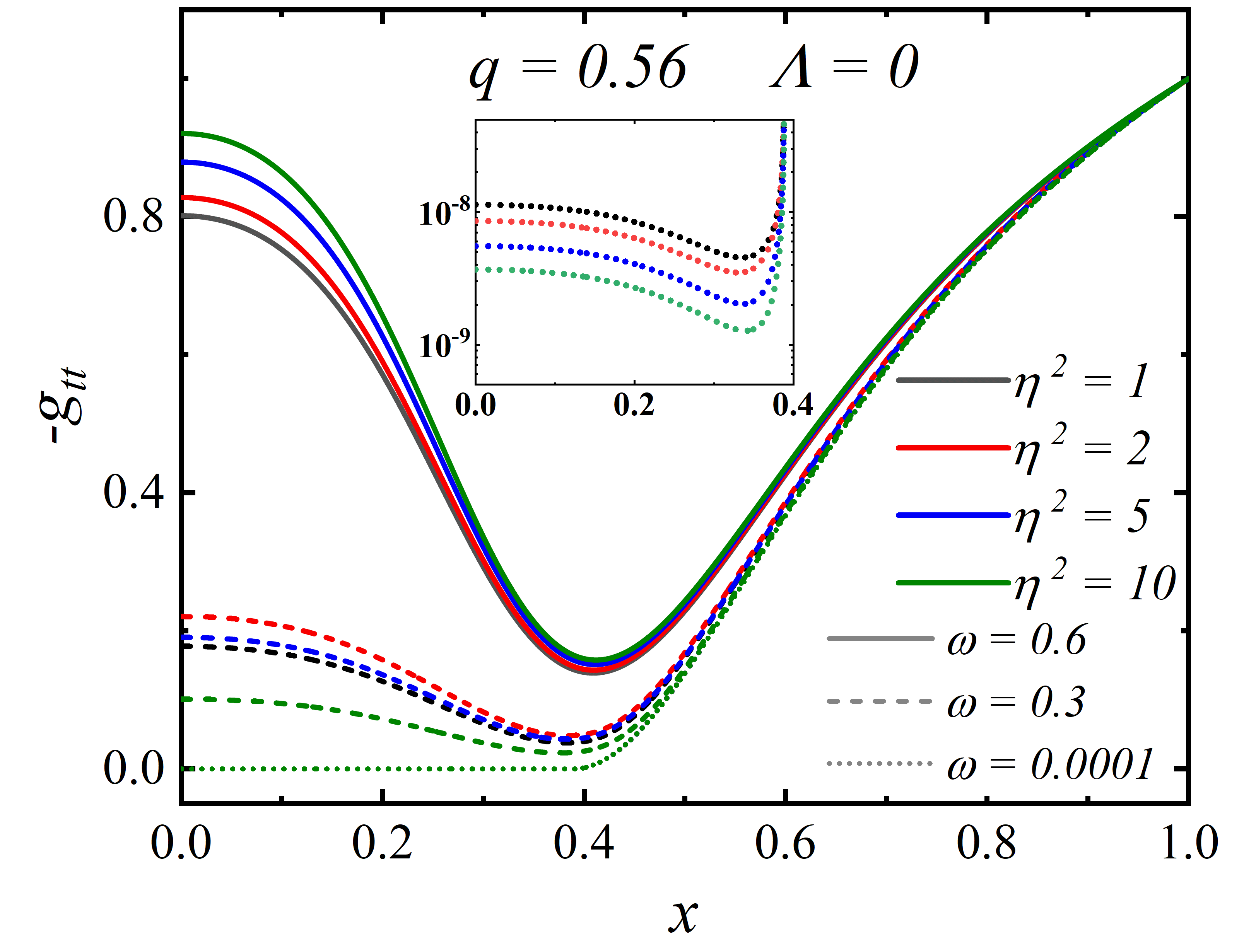}}
\subfigure{\includegraphics[width=0.49\textwidth]{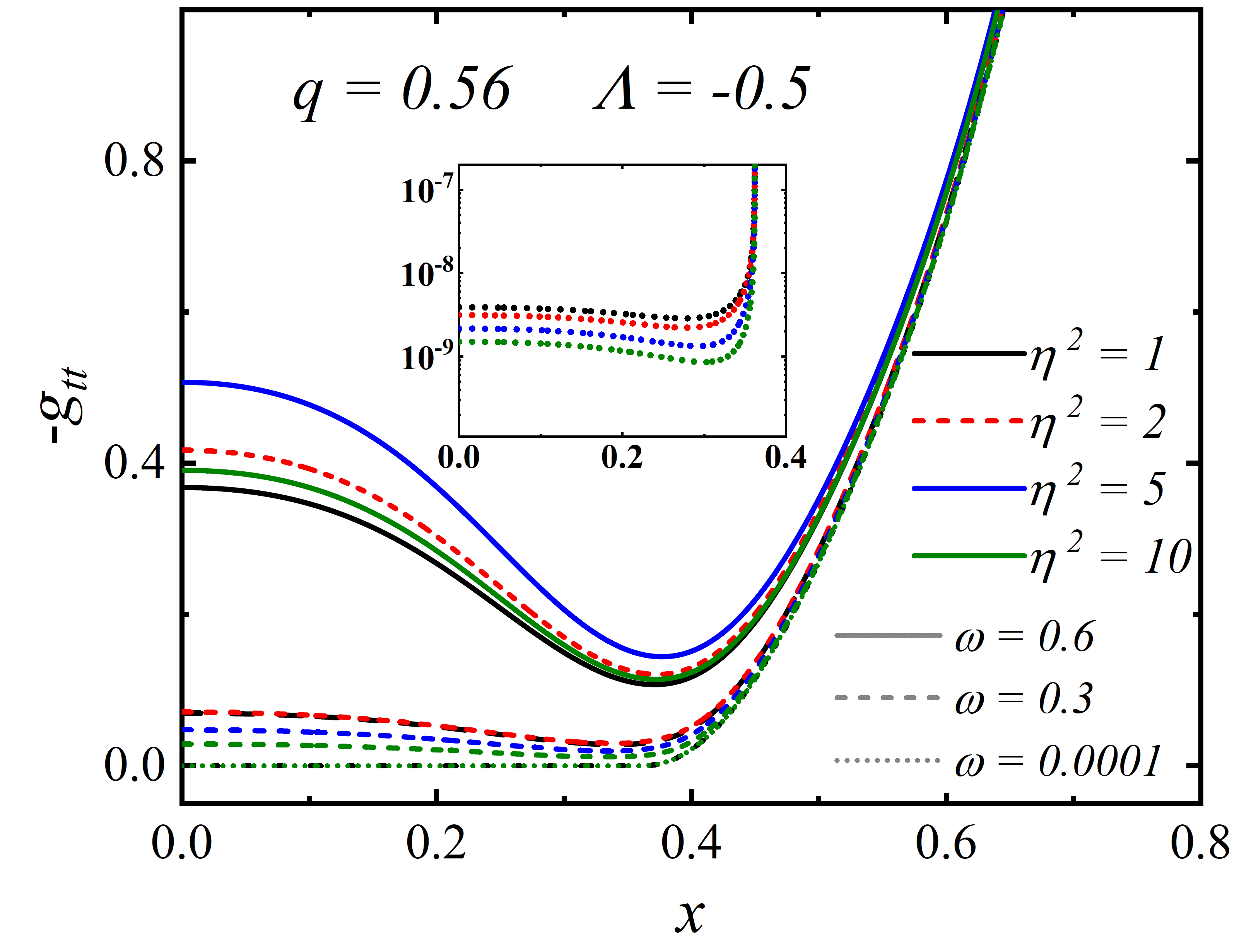}}
\subfigure{\includegraphics[width=0.49\textwidth]{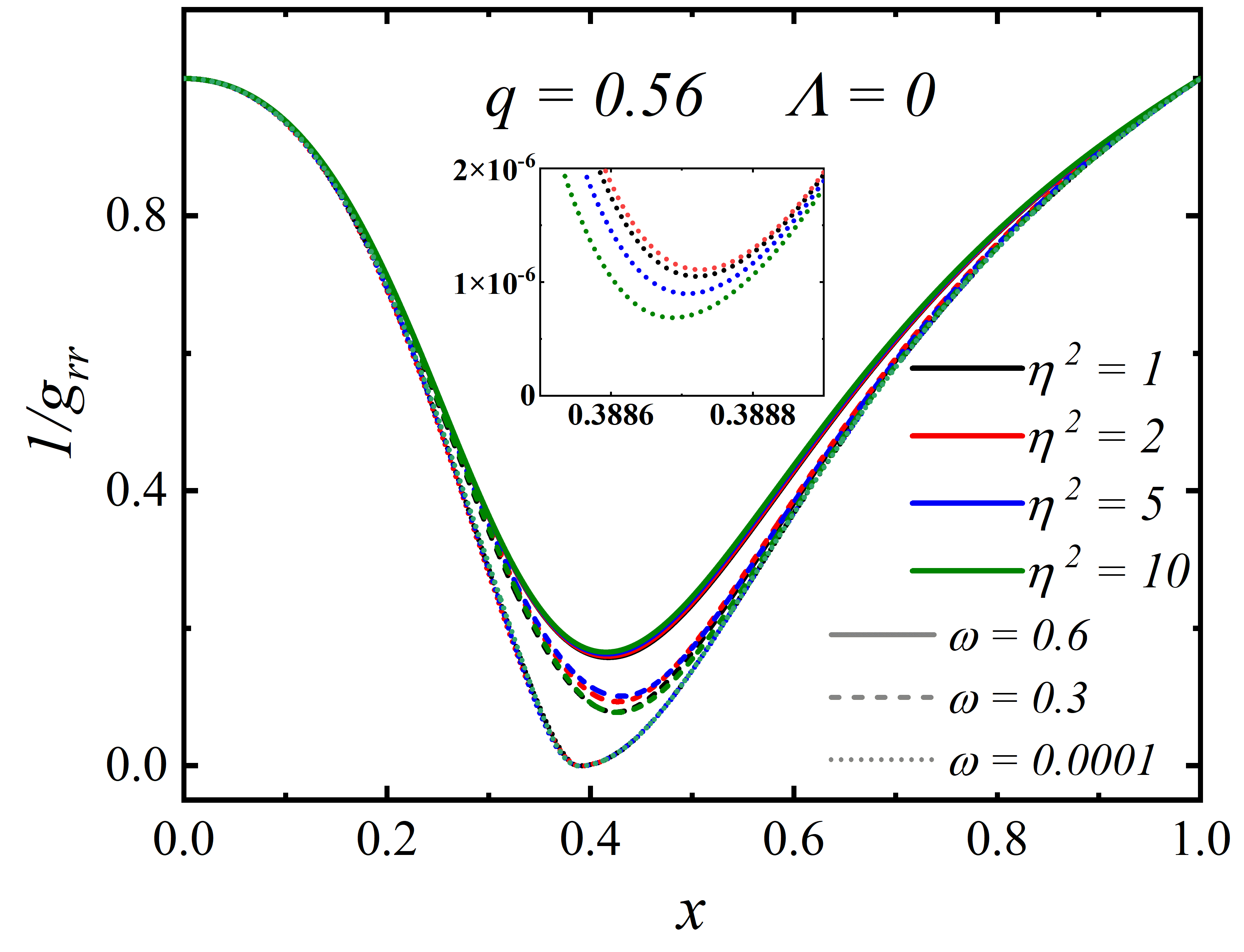}}
\subfigure{\includegraphics[width=0.49\textwidth]{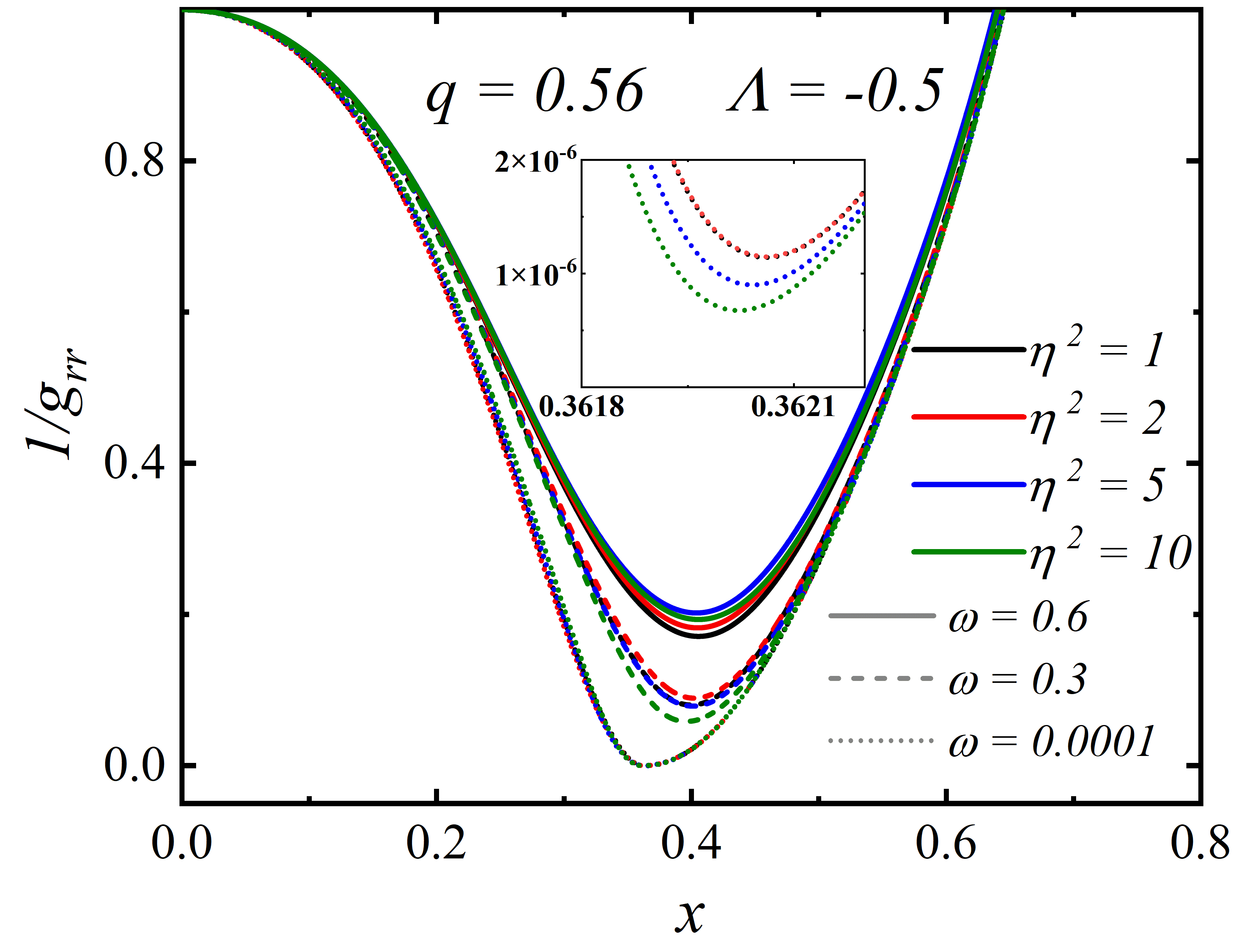}}
\caption{The radial distribution of the metric components with $\eta^2 = \{1,2,5,10\}$ and $\omega =\{ 0.6, 0.3, 0.0001\}$ at $q = 0.56$. The left panels corresponds to $\Lambda = 0$ , the right graphs corresponds to $\Lambda = -0.5$.
  }\label{fig.8}
\end{figure}

Moreover, when $\omega \to 0$, $-g_{tt}=N\,\sigma^2$ in the region $r < r_c$ approaches zero but have not vanished (less than $10^{-8}$ for both $\Lambda=0$ and $\Lambda = -0.5$). And $1/g_{rr}=N$ at $x_c$ is on the order of $10^{-6}$ (see Fig. \ref{fig.8}). Thus, from the perspective of a distant observer, objects move extremely slowly near $x = x_c$, as if ``frozen". The solution of  $\omega \to 0$ is therefore referred to as  ``frozen solitonic Hayward boson stars (FSHBSs)." Given that these characteristics resemble the event horizon of a black hole, we refer to the spherical surface represented by $x = x_c$ as the ``critical horizon." 

\begin{figure}[!htbp]
\centering
\subfigure{\includegraphics[width=0.49\textwidth]{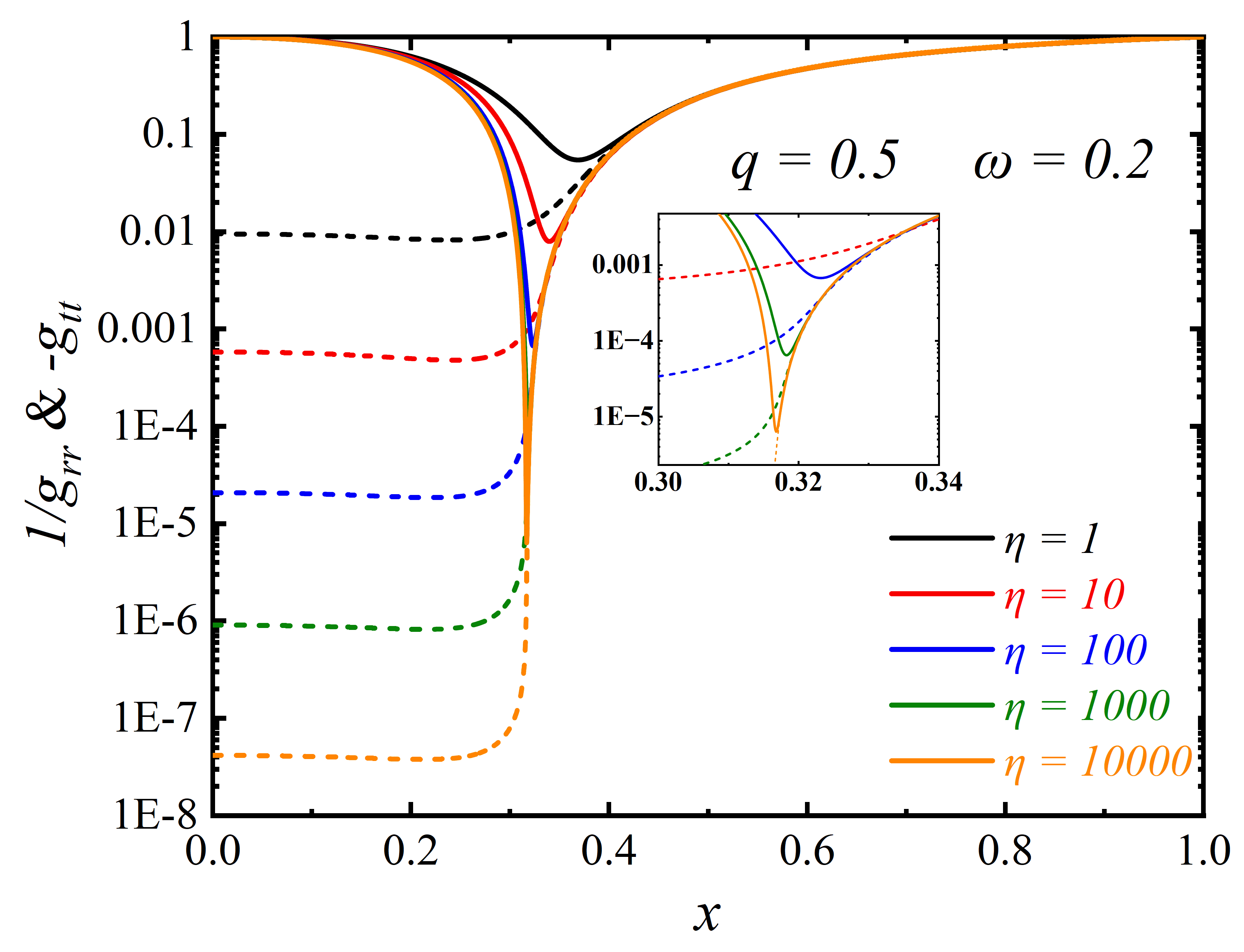}}
\subfigure{\includegraphics[width=0.49\textwidth]{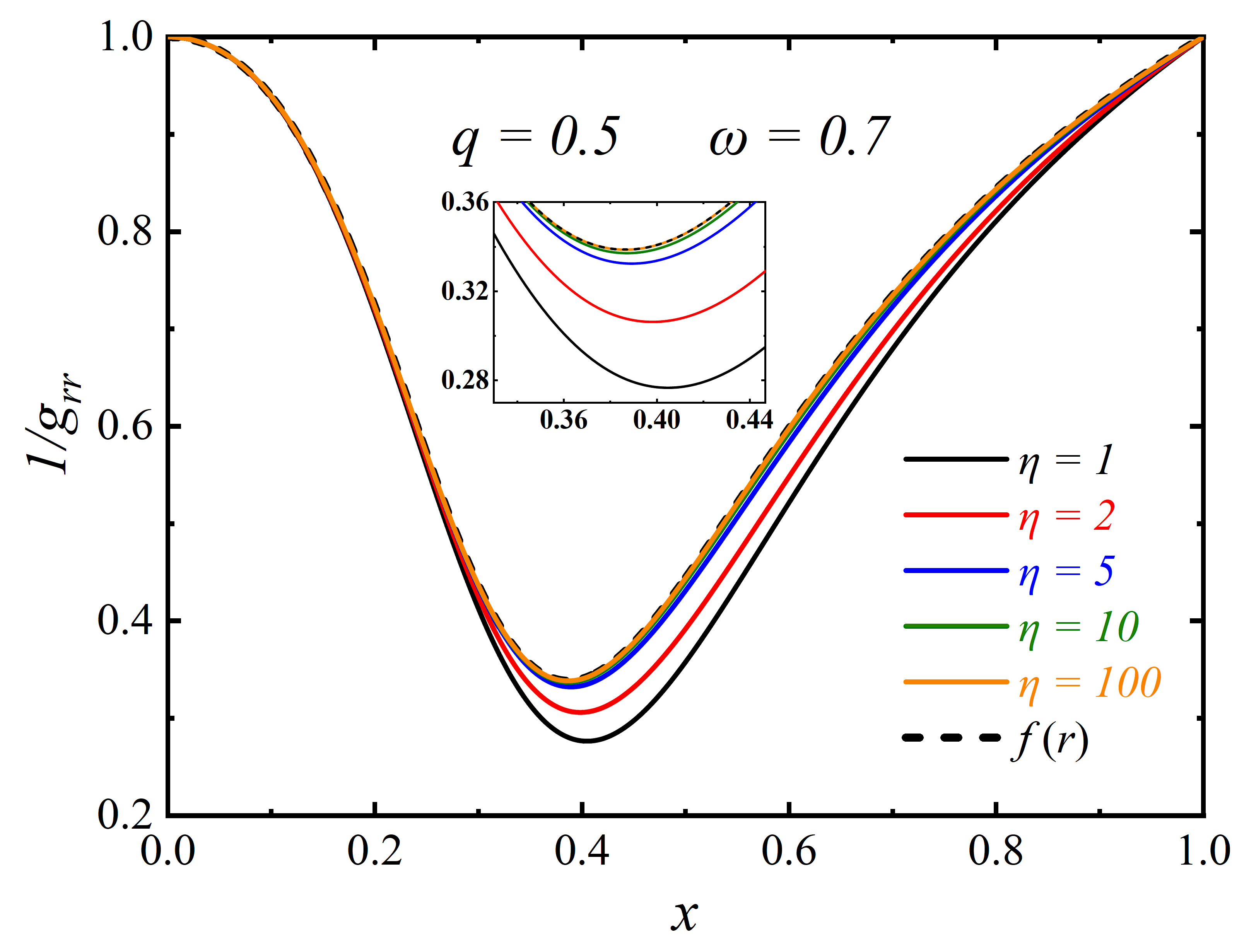}}
\subfigure{\includegraphics[width=0.49\textwidth]{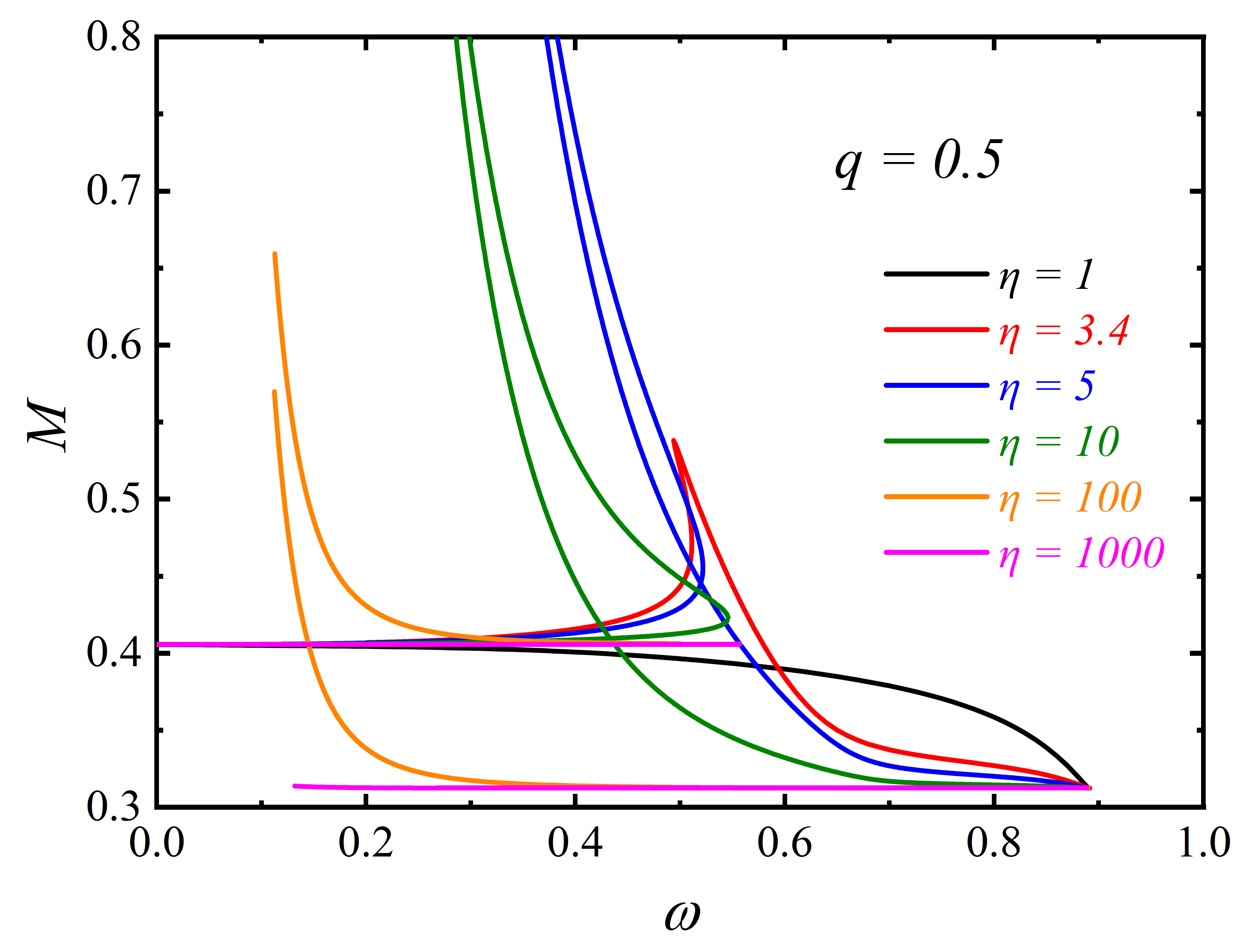}}
\subfigure{\includegraphics[width=0.49\textwidth]{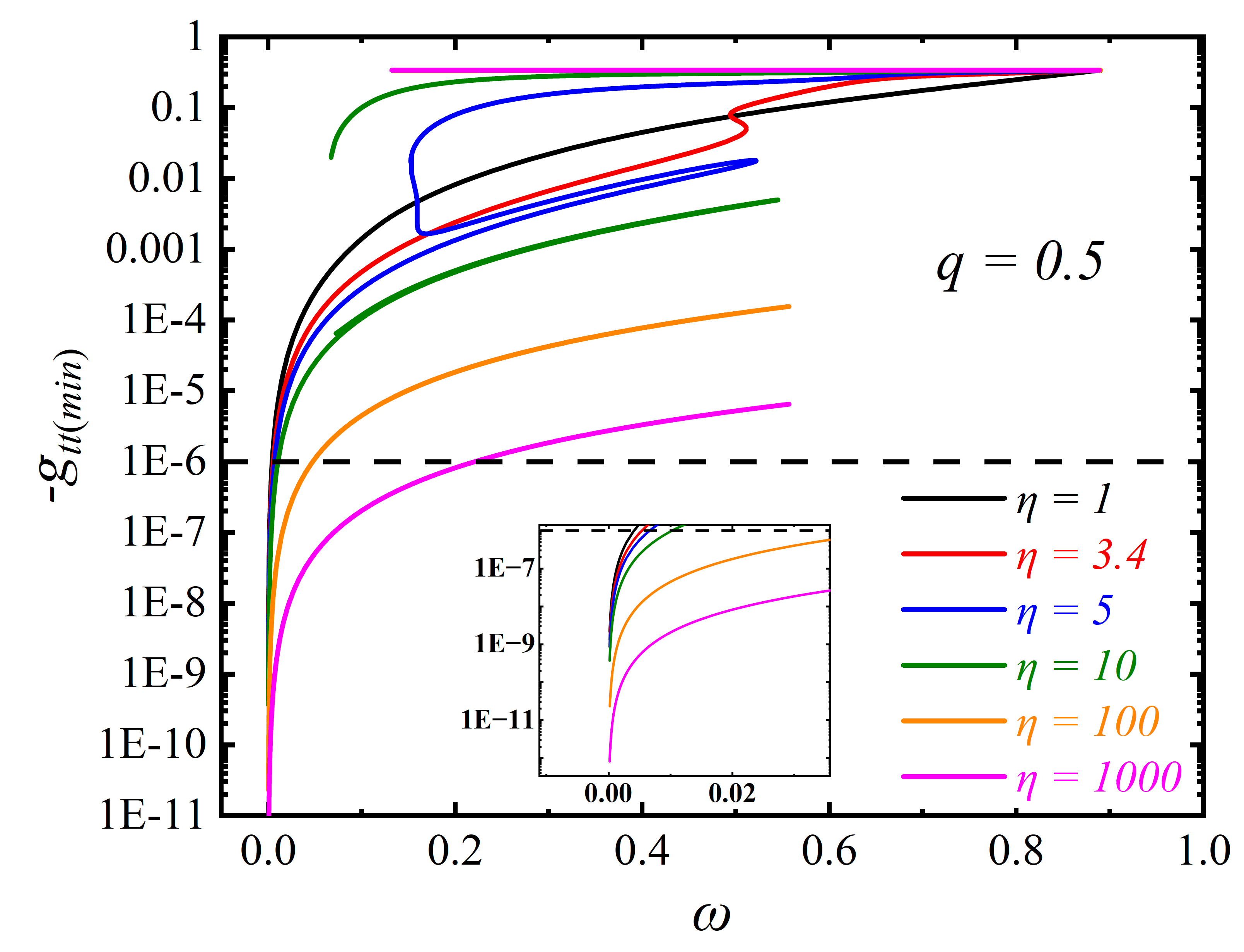}}
\caption{Top left panel: the radial distribution of the metric components $1/g_{rr}\,\&\, -g_{tt}$ for different $\eta$ with $\omega=0.2$. Top right panel: the radial distribution of the metric components $1/g_{rr}$ for different $\eta$ with $\omega=0.7$. The black dashed line represents $f(r)$. Bottom left panel: the ADM mass $M$ as a function of $\omega$ for different $\eta$. Bottom right panel: $-g_{tt(\min)}$, the minimum value of $-g_{tt}$, as a function of $\omega$ for different $\eta$. For reference, a black dashed line indicates the level of $10^{-6}$. The magnetic charge and cosmological constant are fixed at $q = 0.5$ and
$\Lambda = 0$, respectively.}\label{fig.9}
\end{figure}

\begin{figure}[!htbp]
\centering
\subfigure{\includegraphics[width=0.49\textwidth]{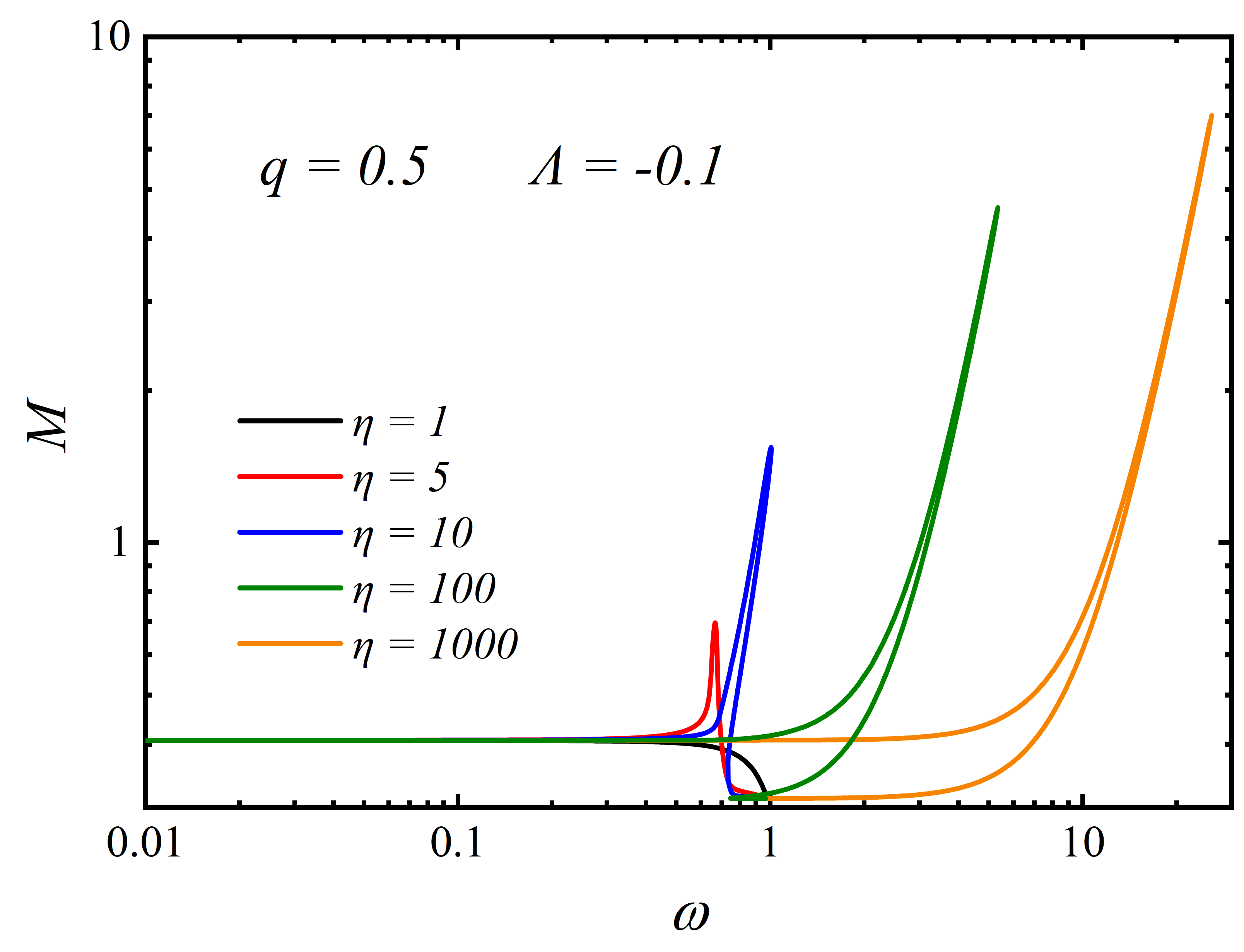}}
\subfigure{\includegraphics[width=0.49\textwidth]{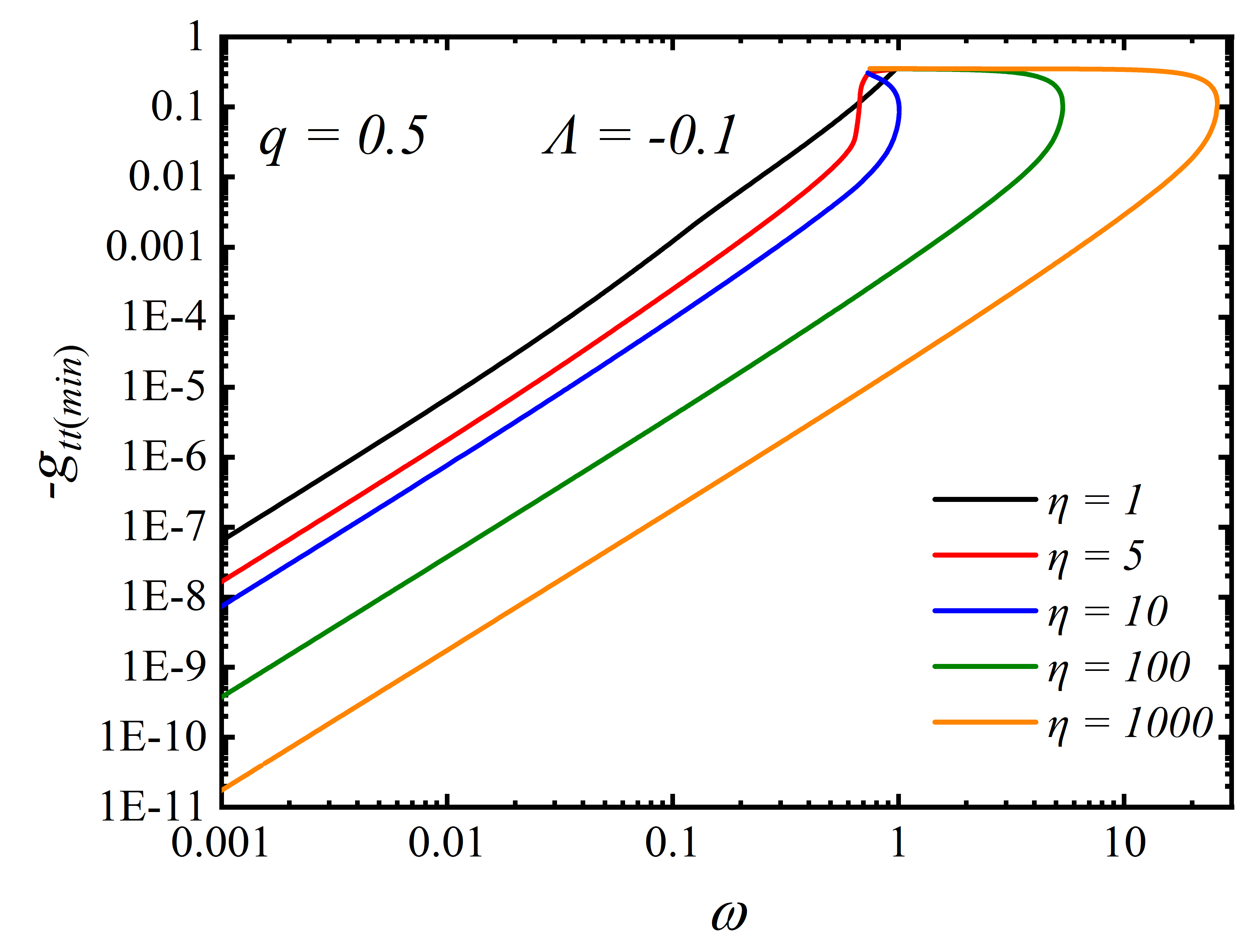}}
\caption{Left panel: the ADM mass $M$ as a function of $\omega$ for different $\eta$ with $q=0.5,\,\Lambda=-0.1$. Right panel: $-g_{tt(\min)}$ as a function of $\omega$ for different $\eta$ with $q=0.5,\,\Lambda=-0.1$.
  }\label{fig.10}
\end{figure}

When the frequency takes different values, increasing $\eta$ may yield significantly different solutions. We illustrate this scenario in Fig. \ref{fig.9} for $\Lambda=0$ and Fig. \ref{fig.10} for $\Lambda=-0.1$. In Fig. \ref{fig.9}, the top left panel corresponds to $\omega=0.2$. As $\eta$ increases, the minimum values of $1/g_{rr}$ and $-g_{tt}$ decrease. However, for the top right panel with $\omega=0.7$. As $\eta$ increase, $1/g_{rr}$ becomes increasingly close to $f(r)$. This indicates that the solution at this point is close to the pure Hayward solution. The two bottom panels reveal the underlying reason. From the bottom left panel, it can be observed that an increase in $\eta$ causes the $M-\omega$ curve to form multiple branches (due to computational difficulties, we have not provided the complete branch structure). The bottom right panel displays the variation of the minimum value of $-g_{tt}$~(denoted as $-g_{tt(\min)}$) with $\omega$ for different $\eta$ values, further illustrating that high and low frequencies belong to distinct branches.

Then, we discuss the case of $\Lambda=-0.1$ in Fig. \ref{fig.10}. Unlike the $\Lambda=0$ scenario, the introduction of a negative cosmological constant leads to a significant increase in $\omega_{\text{max}}$ as $\eta$ grows. Furthermore, when $\eta$ is large (e.g., $\eta=1000$), the $M-\omega$ curve exhibits two branches at high frequencies. These two branches merge at $\omega_{\text{max}}$. The corresponding $-g_{tt(\min)}-\omega$ curve is shown in the right panel, demonstrating properties similar to those observed in the $\Lambda=0$ case.

\begin{figure}[!htbp]
\centering
\subfigure{\includegraphics[width=0.49\textwidth]{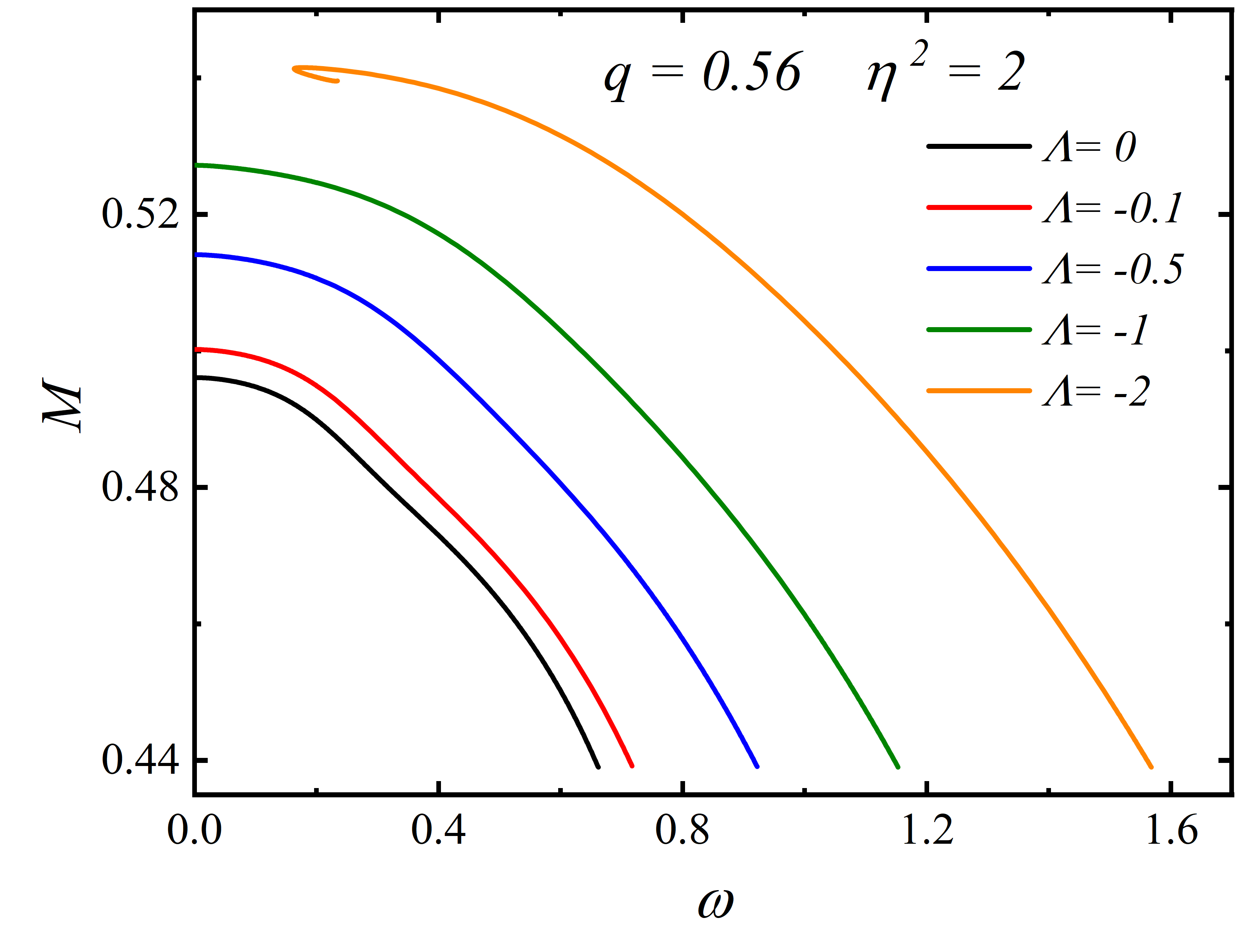}}
\subfigure{\includegraphics[width=0.49\textwidth]{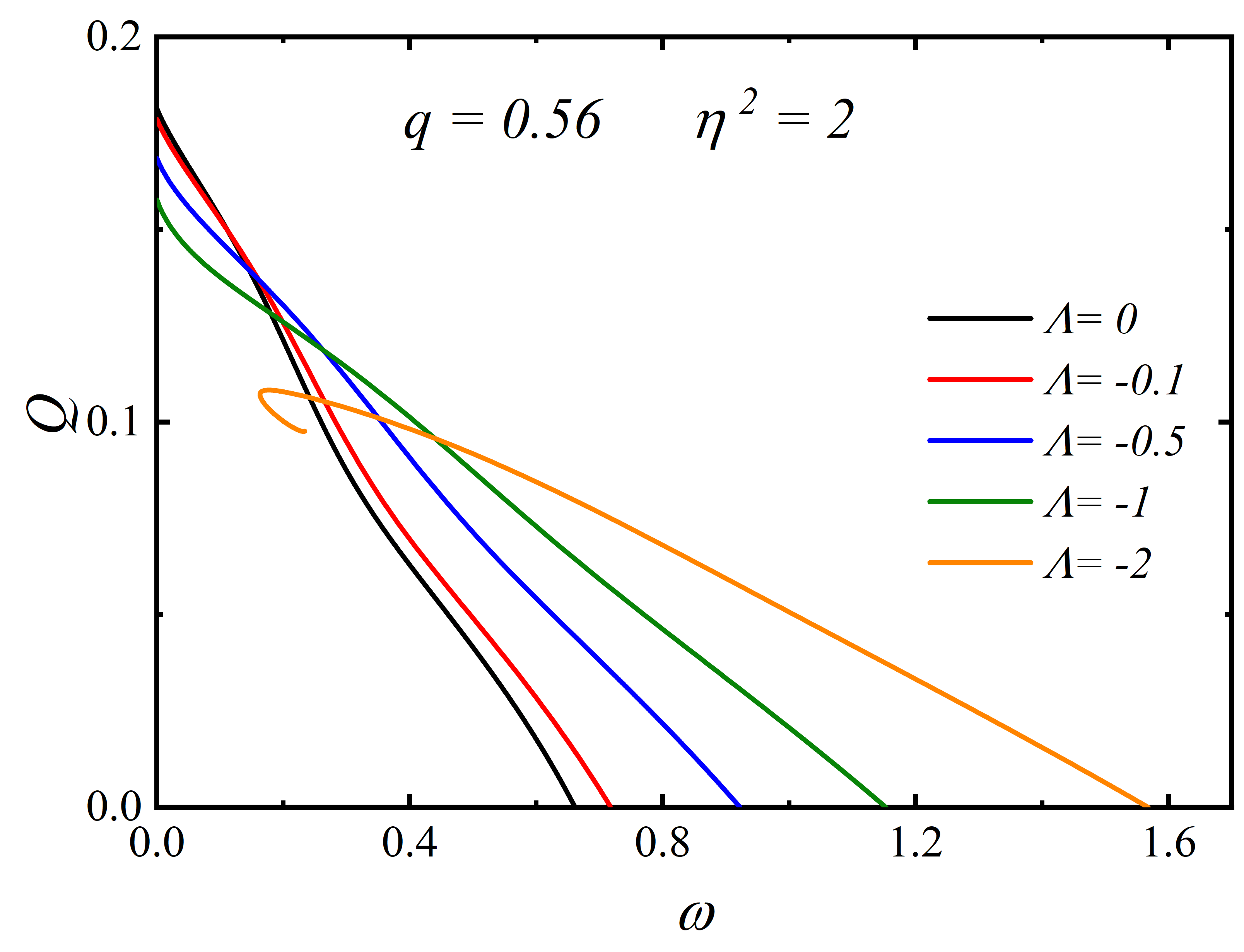}}
\subfigure{\includegraphics[width=0.49\textwidth]{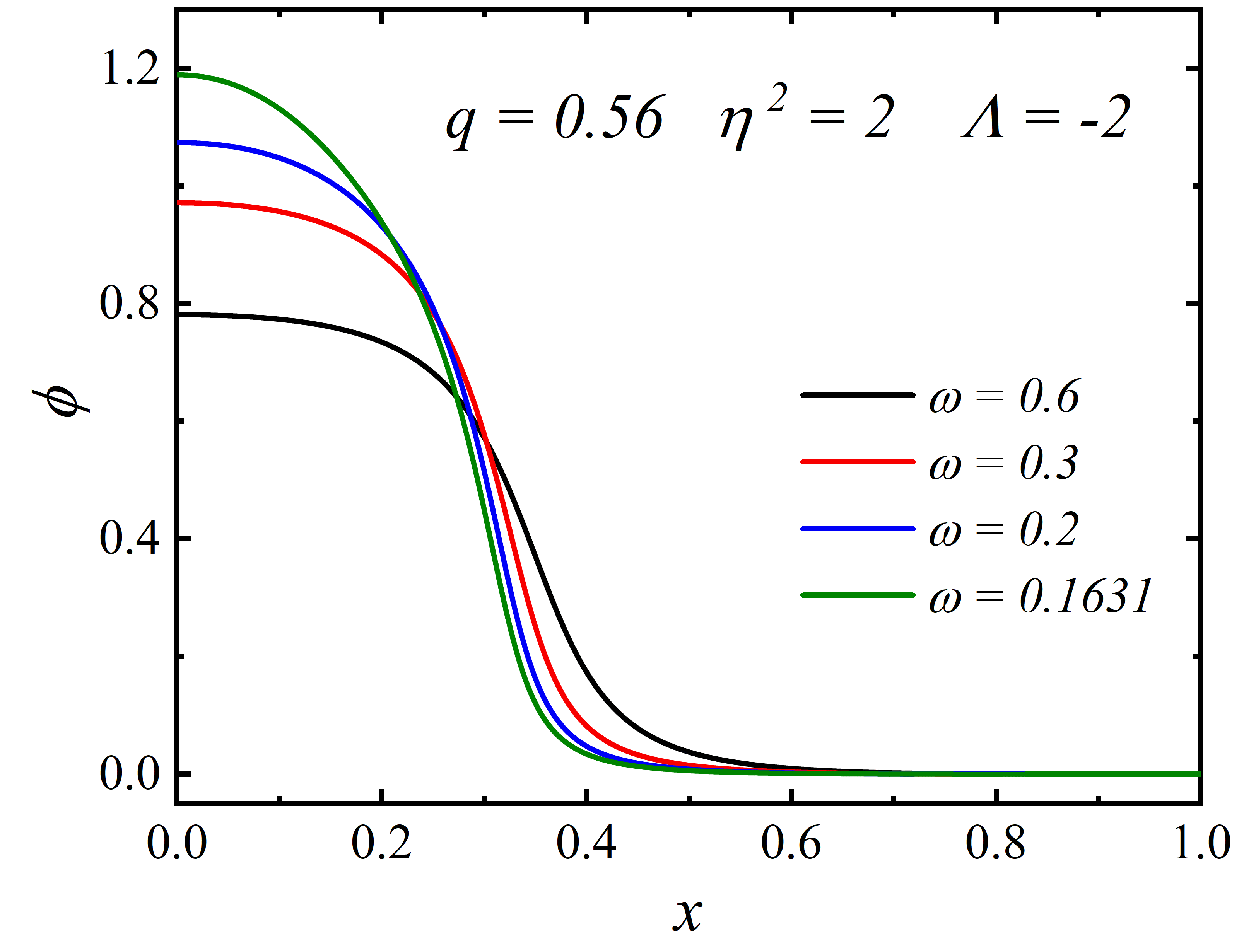}}
\subfigure{\includegraphics[width=0.49\textwidth]{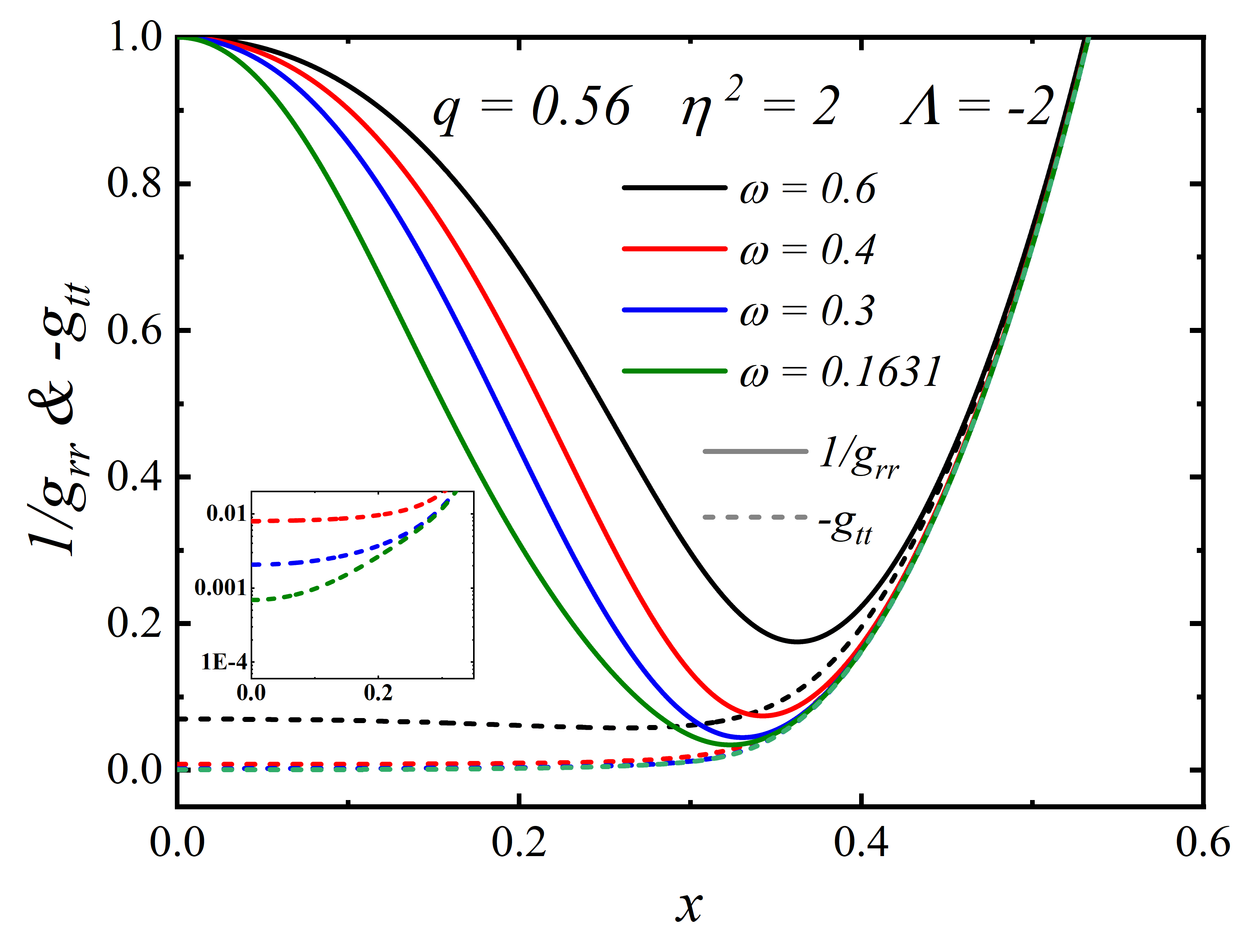}}
\caption{Top panels: the $M$ and $Q$ as a function of $\omega$ for different $\Lambda$. Bottom panels: the radial distribution of the scalar field $\phi$, the metric functions $g_{tt}$ and $1/g_{rr}$ with different frequencies. The magnetic and coupling parameter are fixed to $q=0.56$ and $\eta^2 = 2$.
  }\label{fig.11}
\end{figure}

To investigate the influence of the cosmological constant on the solutions, we display the ADM mass $M$ and Noether charge $Q$ for different values of $\Lambda$ when $q=0.56$ and $\eta^2=2$ in top two panels of Fig. \ref{fig.11}. As $\Lambda$ decreases, the maximum value of the ADM mass $M_{\max}$ gradually increases. This differs from the case of $q < q_c$ in Fig. \ref{fig.2}. In contrast, the maximum value of the Noether charge $Q_{\max}$ decreases with decreasing $\Lambda$, which is consistent with the behavior observed when $q<q_c$. In addition, a decrease in $\Lambda$ causes the curves to redevelop a spiral structure. 

Since a decreasing cosmological constant can cause the extreme solution with $\omega\to0$ to vanish, a natural question that arises is whether this would affect the frozen behavior. In bottom panels of Fig. \ref{fig.11}, we present the field function and metric functions for $\Lambda=-2$. The frequency can only be reduced to $\omega=0.1631$. Due to the slower decay of the complex scalar field, the value of $r_c$ becomes difficult to determine here, and $-g_{tt}$ is more than 0.0001, so the freezing behavior of SHBSs vanishes. Combined with Table \ref{table1}, a smaller $\Lambda$ requires a larger $q$ to maintain the frozen state.

 \begin{figure}[!htbp]
\centering
\subfigure{\includegraphics[width=0.49\textwidth]{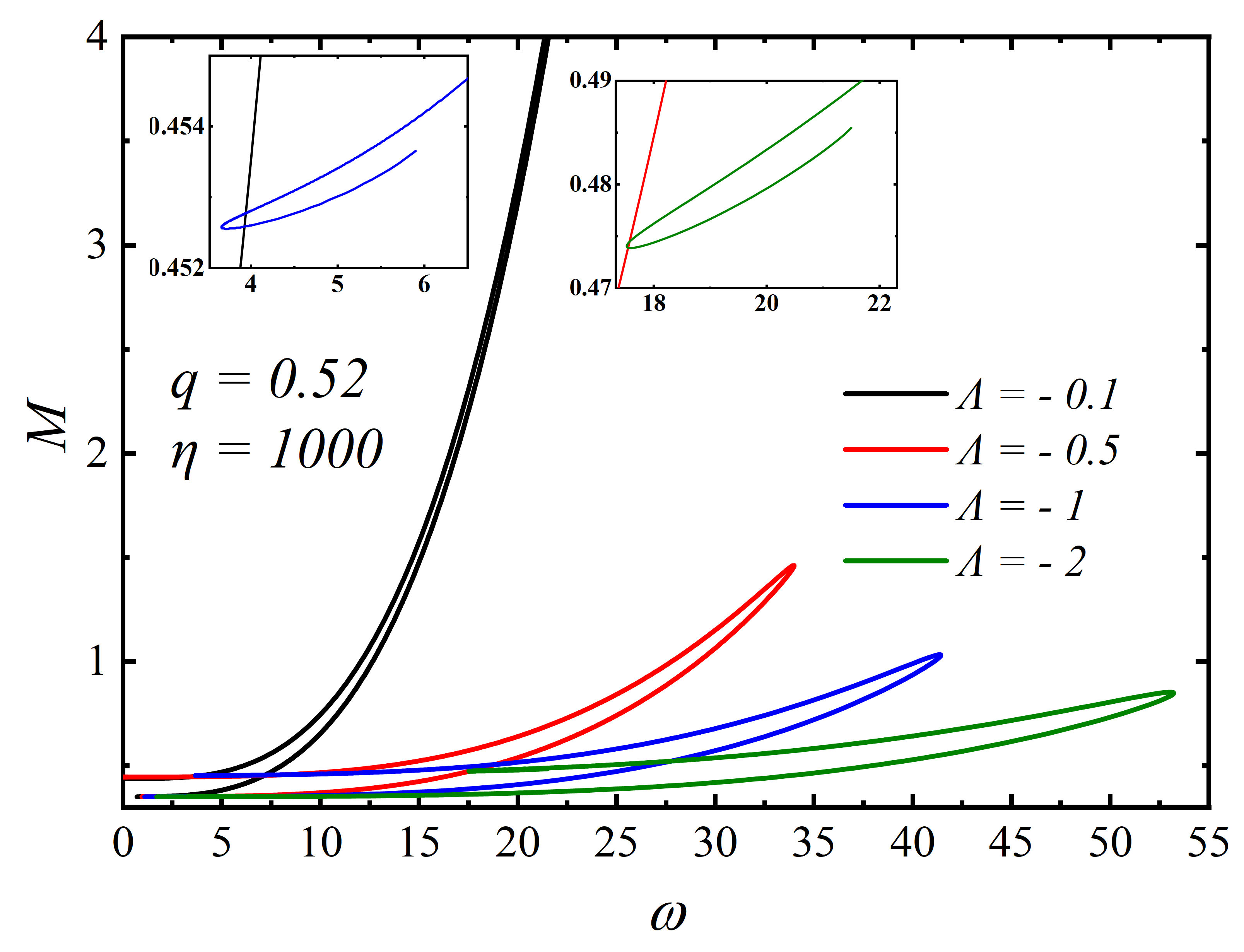}}
\subfigure{\includegraphics[width=0.49\textwidth]{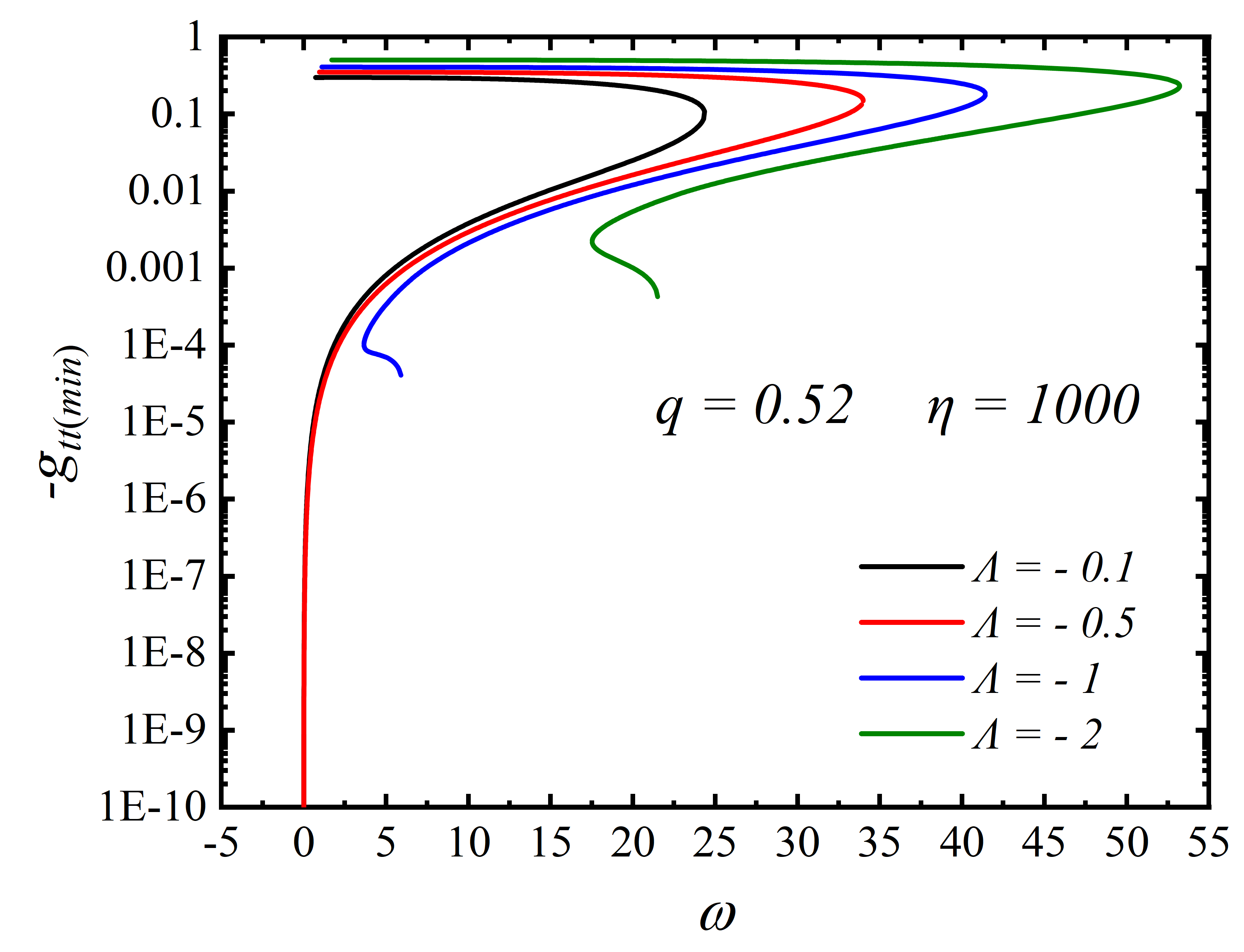}}
\caption{Left panel: the ADM mass $M$ as a function of $\omega$ for different $\Lambda$. Right bottom panel: $-g_{tt(\min)}$ as a function of $\omega$ for different $\Lambda$. The magnetic and coupling parameter are fixed to $q=0.52$ and $\eta = 1000$.
  }\label{fig.12}
\end{figure}

In Fig. \ref{fig.12}, we examine the influence of reducing the cosmological constant on the solutions for $\eta=1000$. Without loss of generality, we set $q=0.52$. The $M-\omega$ curve is shown in the left panel. As $\Lambda$ decreases, the curve gradually flattens: $M_{\text{max}}$ decreases while $\omega_{\text{max}}$ increases. For solutions with $\Lambda=-1$ and $\Lambda=-2$, the solutions with $\omega \to 0$ vanish, and the curves develop  spiral structures (see insets). Meanwhile, from the right panel, $-g_{tt(\text{min})}$ no longer approaches zero infinitely, indicating the breakdown of the ``frozen" behavior.

\begin{figure}[!htbp]
\centering
\subfigure{\includegraphics[width=0.49\textwidth]{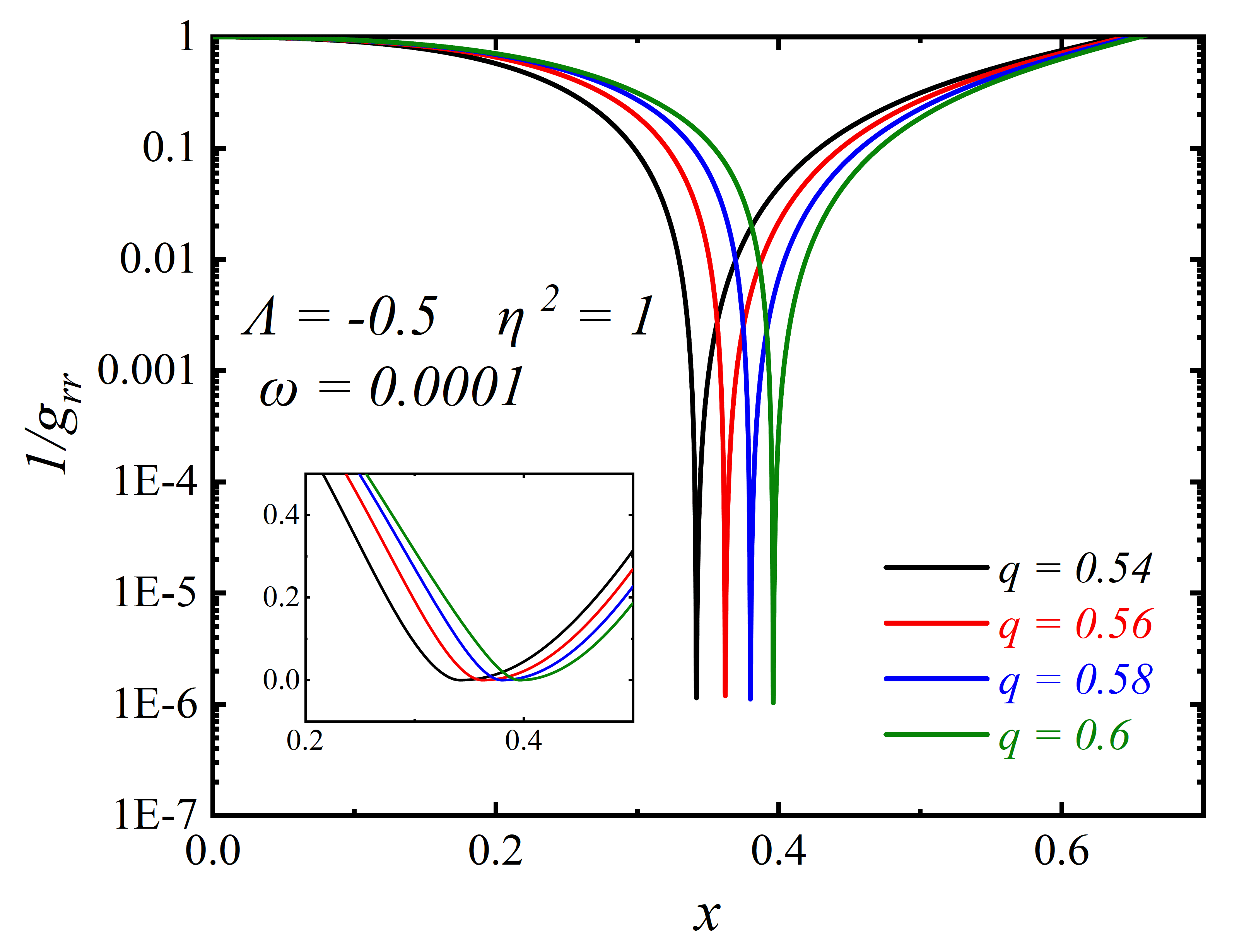}}
\subfigure{\includegraphics[width=0.49\textwidth]{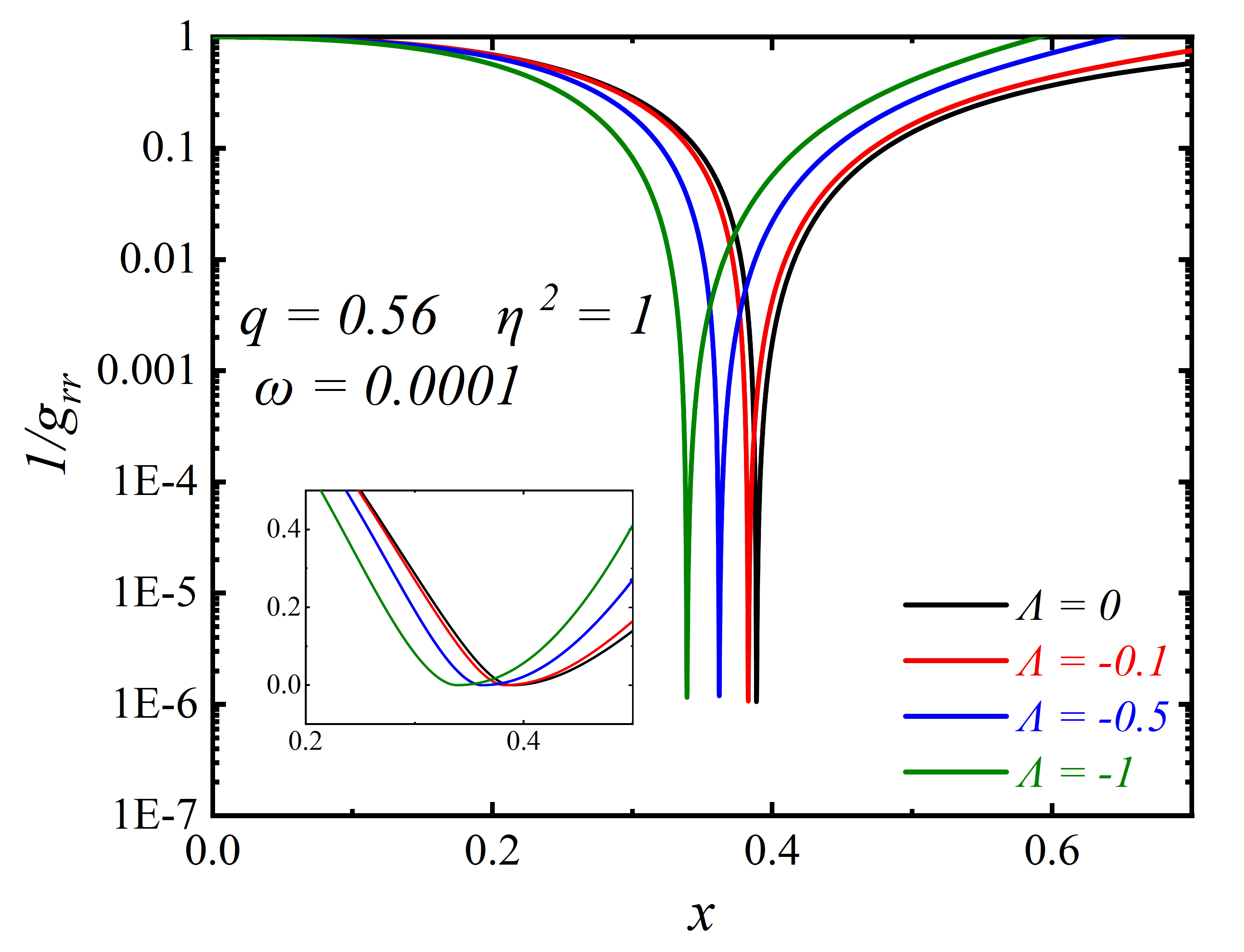}}
\caption{The radial distribution of the metric components for different parameters with $\omega=0.0001$.
  }\label{fig.13}
\end{figure}

For frozen Bardeen boson stars (FBBSs) and Bardeen Dirac stars (FBDSs), increasing the magnetic charge $q$ leads to an increase in the critical radius $x_c$~\cite{Huang:2025css,Zhang:2024ljd}. To verify whether this scenario applies to FSHBSs, we present our findings in Fig. \ref{fig.13}. As can be seen from the left and right panels, the critical radius $x_c$ decreases with the reduction of both $q$ and $\Lambda$ (see inset panel).

\begin{figure}[!htbp]
\centering
\includegraphics[width=0.49\textwidth]{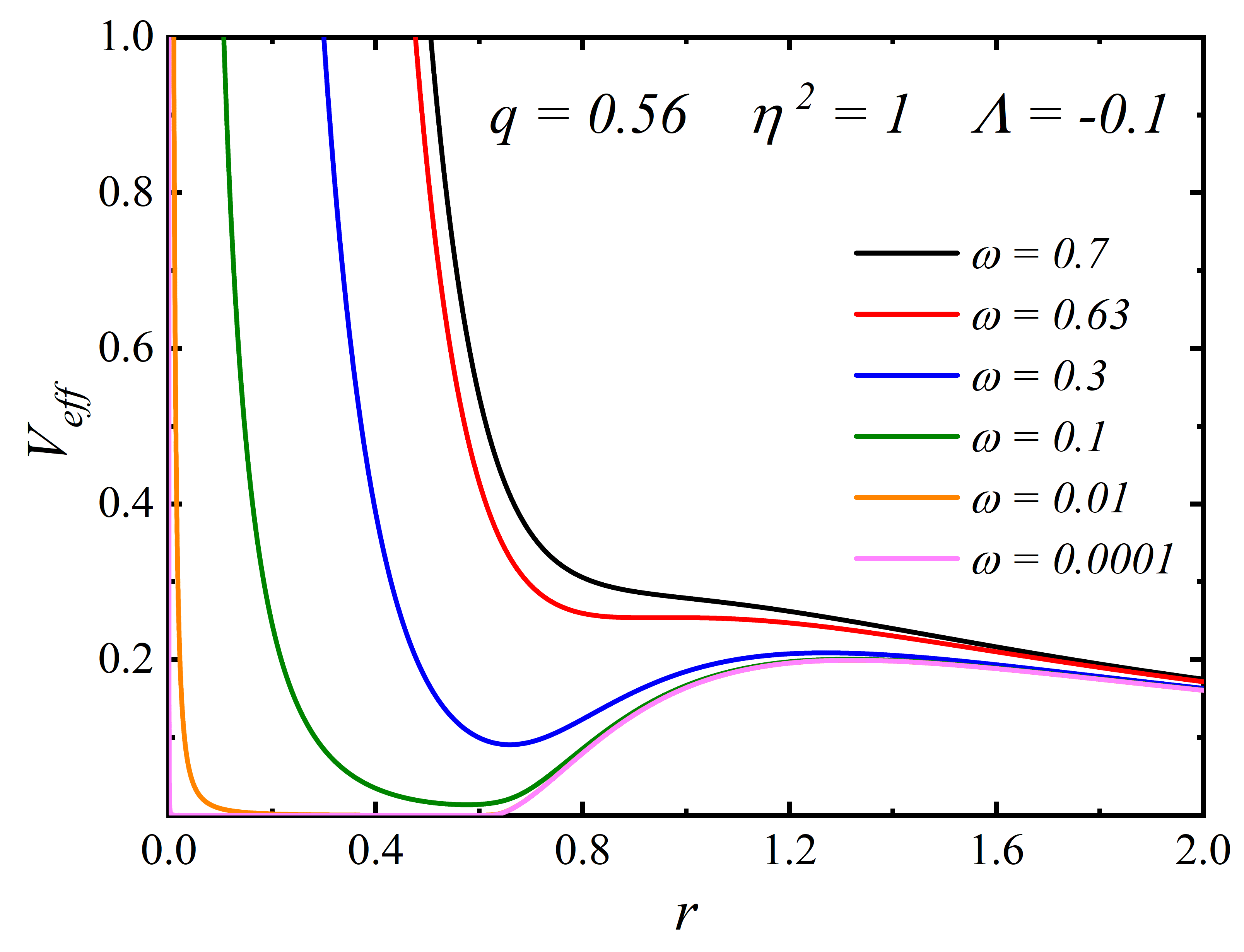}
\includegraphics[width=0.49\textwidth]{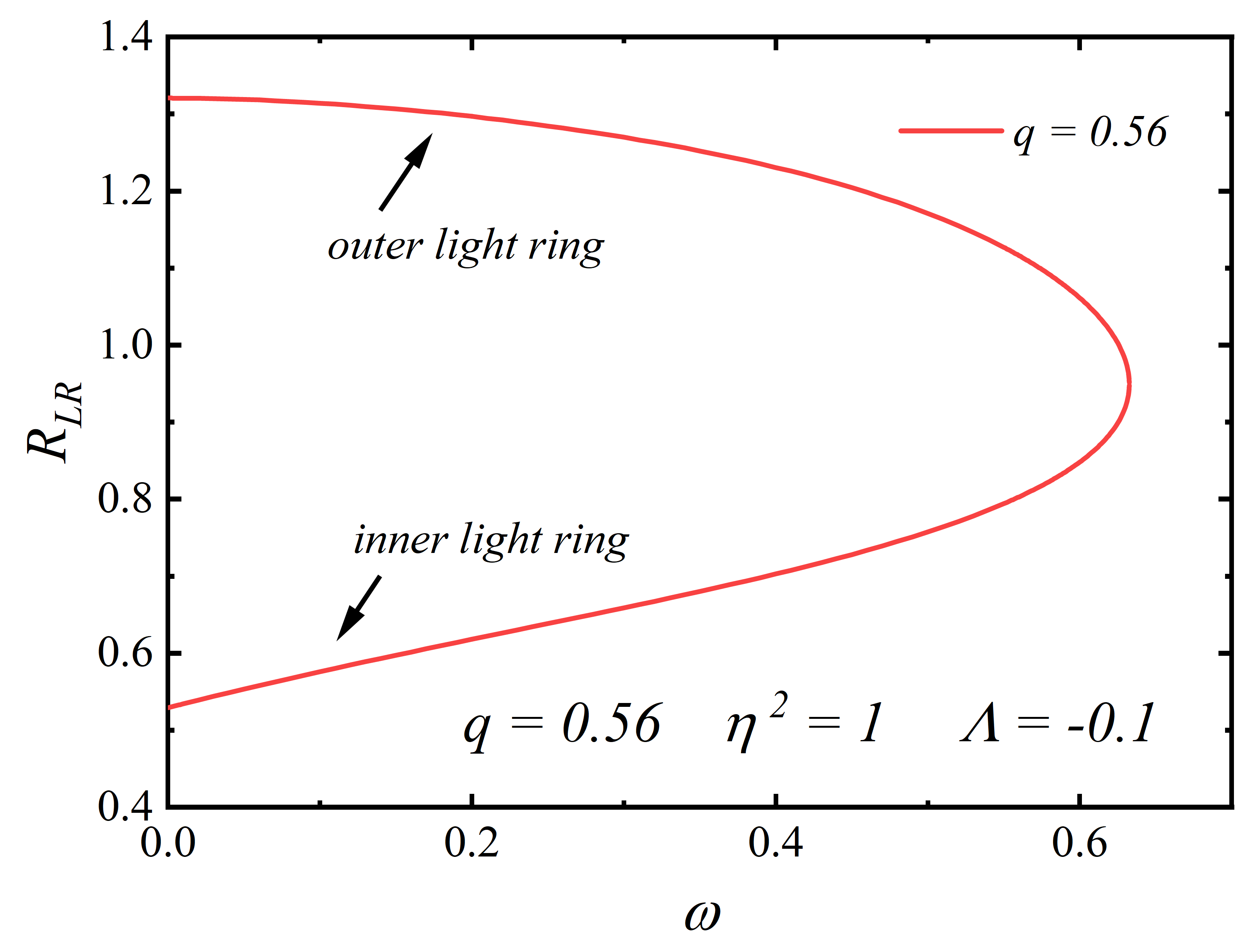}
\caption{Left panel: Effective potential $V_{eff}$ of HBSs as a function of radius $r$ with different frequencies $\omega$. Right panel: The position of light ring $R_{LR}$ as a function of frequency $\omega$ for $q=0.56$. The coupling parameter and cosmological constant are fixed to $\eta^2=1$ and $\Lambda=-0.1$, respectively.
  }\label{fig.14}
\end{figure}

Next, we will discuss the light rings of SHBSs for $q \geq q_c$. In the left panel of Fig. \ref{fig.14}, we present the effective potential $V_{eff}$ as a function of $r$ with different frequencies $\omega$ for $q=0.56$, $\eta^2 = 1$ and  $\Lambda = -0.1 $. The right panel shows the relationship between the radius of the light ring $R_{LR}$ and the frequency $\omega$ for $q=0.56$. We can see that as $\omega$ increases, the inner and outer rings gradually approach each other, merge into a single ring around $\omega \approx 0.630$, and eventually disappear.

\begin{figure}[!htbp]
\centering
\includegraphics[width=0.49\textwidth]{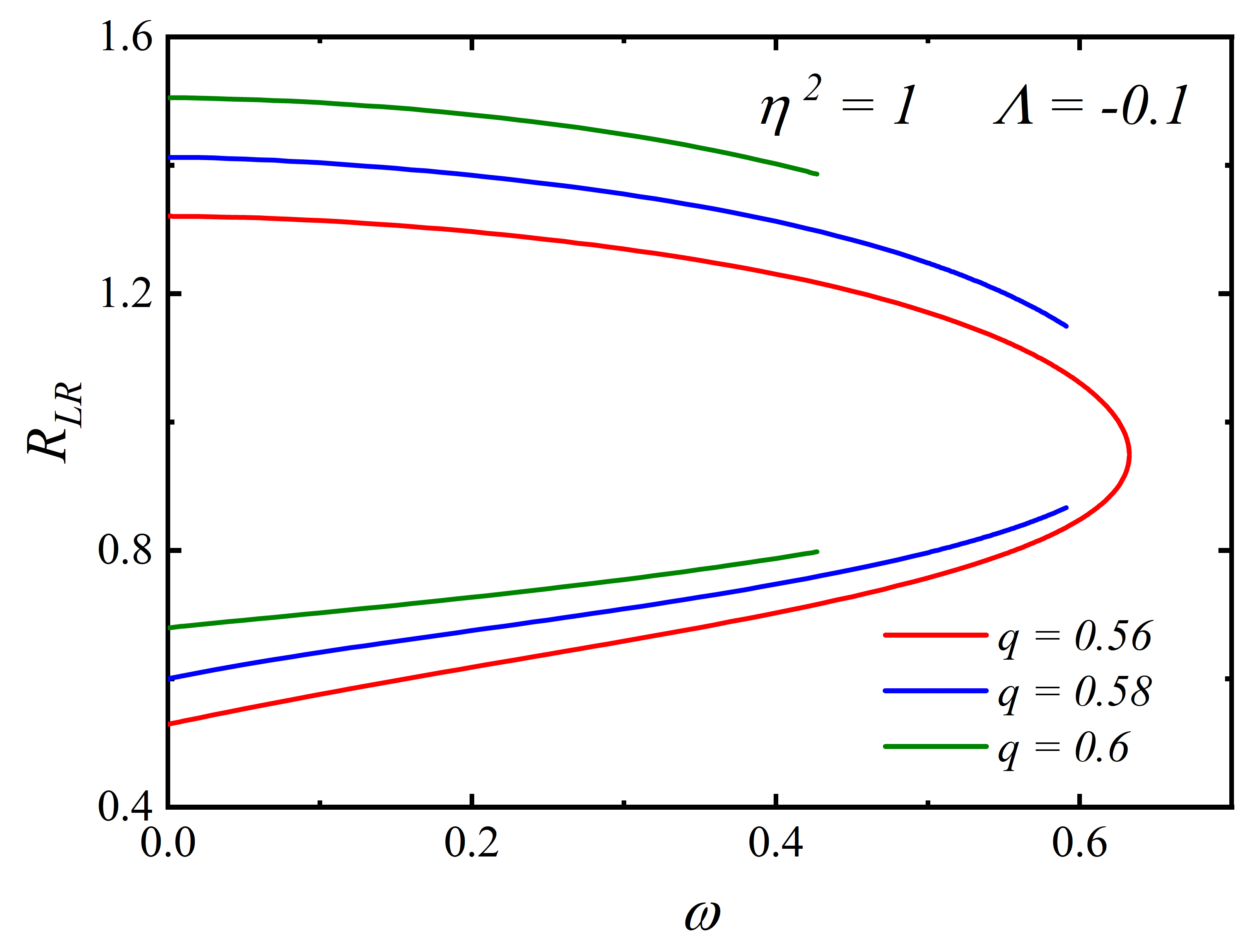}
\includegraphics[width=0.49\textwidth]{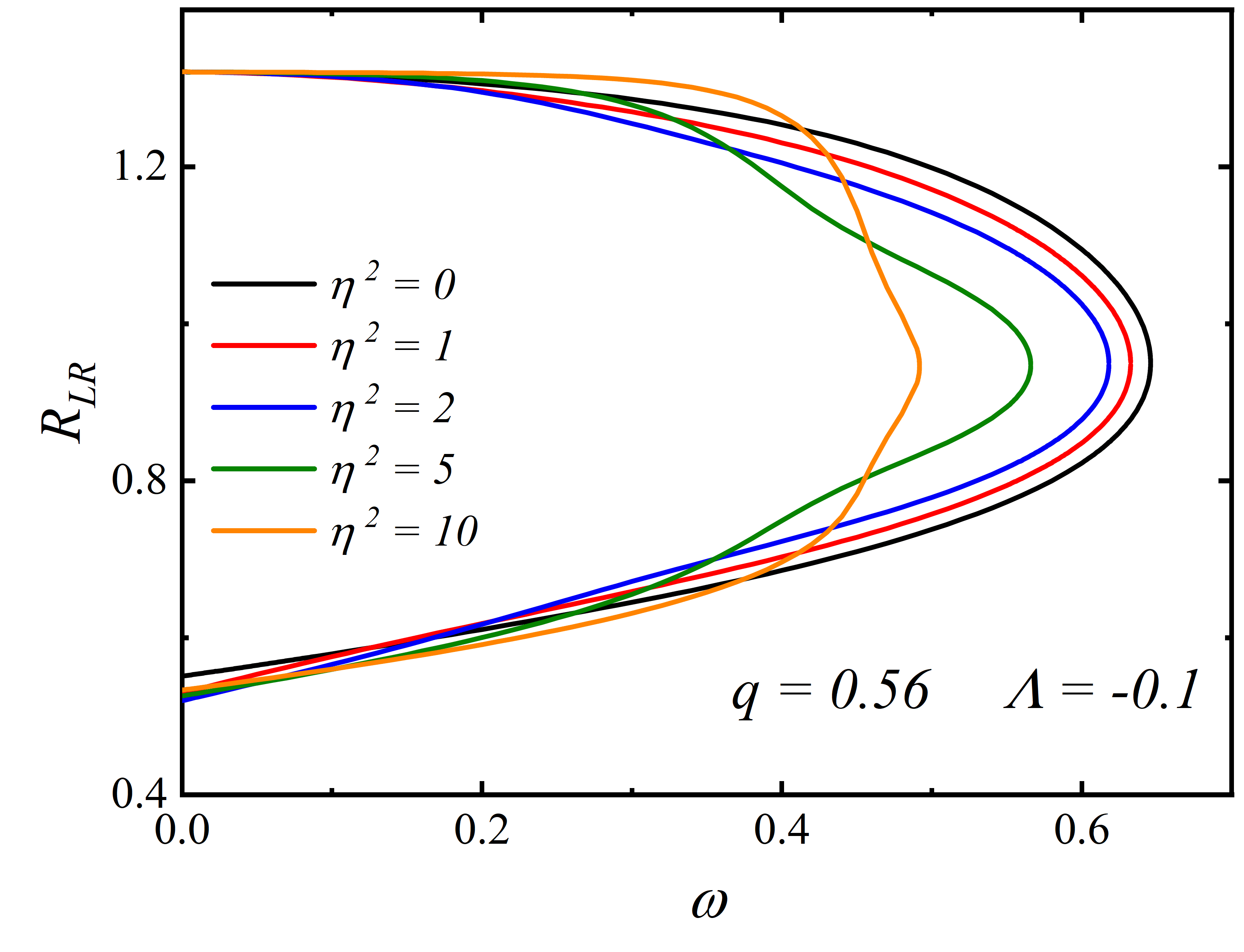}
\includegraphics[width=0.49\textwidth]{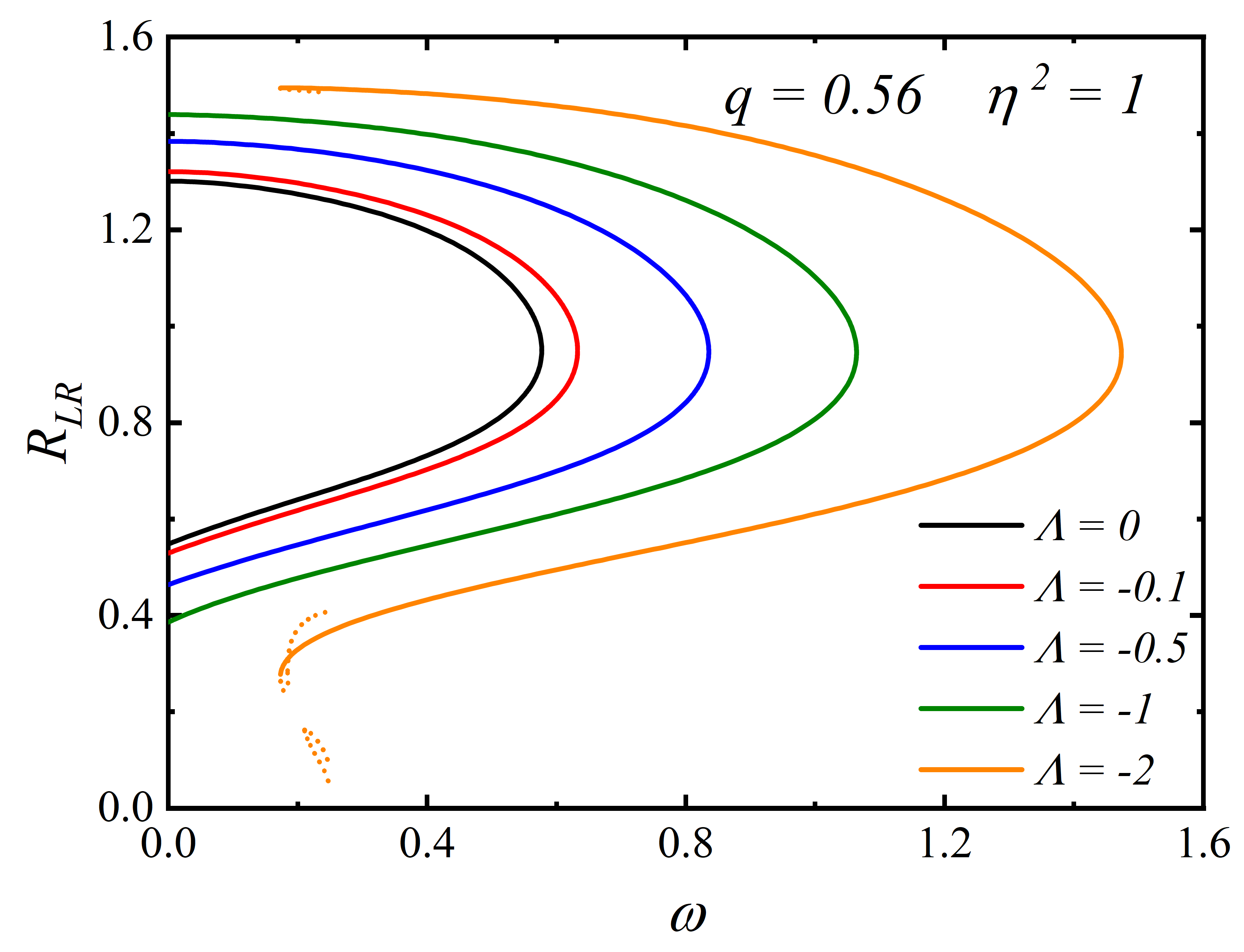}
\caption{The position of light ring $R_{LR}$ as a function of frequency $\omega$ for different magnetic charge $q$, coupling parameter $\eta$ and cosmological constant $\Lambda$.
  }\label{fig.15}
\end{figure}

In Fig.\ \ref{fig.15}, we study the effects of parameters $q$, $\eta^2$, and $\Lambda$ on light ring distributions for $q \geq q_c$. From the top left panel, it can be observed that as $q$ increases, the two light rings do not merge (blue line and green line). Moreover, the minimum separation between the inner and outer light rings grows, and their frequency range narrows. Moreover, as shown in the top right panel, increasing $\eta^2$ reduces the frequency range that possesses light ring solutions. However, unlike $q$, the inner and outer light rings can still degenerate into one at different values of $\eta^2$. The bottom panel demonstrates that reducing $\Lambda$ enables the existence of light rings across a broader frequency range before the extreme solution of $\omega \to 0$ disappears. As the cosmological constant decreases, the extreme solution disappears (see the orange curve for $\Lambda = -2$), while the light ring solutions of the second branch appears. 

\section{CONCLUSION}\label{sec5}
In this work, we have constructed solitonic Hayward boson stars (SHBSs) in AdS spacetime, formed by the Einstein gravity coupled with  a nonlinear electromagnetic field and a complex scalar field with soliton potential. We find that after introducing the complex scalar field, the magnetic charge $q$ is constrained within a specific range. Within this range, no black hole solutions were found.

Similar to Ref.~\cite{Wang:2023tdz,Huang:2023fnt,Yue:2023sep,Chen:2024bfj}, we find that there exists a critical magnetic charge $q_c$: when $q<q_c$, the frequency $\omega$ cannot approach zero, manifesting as a spiral structure in the $M-\omega$ curve. When $q\geq q_c$, $\omega$ can indefinitely approach zero.  In particular, in the extreme limit of $\omega\to0$, the field functions and the energy density decay rapidly beyond $r_c$. The metric components $g_{tt}$ and $1/g_{rr}$ approach zero at the critical horizon, we found the FSHBSs solutions. However, a continued decrease in $\Lambda$ causes the $M-\omega$ curve to redevelop a spiral structure and also disrupts the ``frozen" phenomenon (further studies on this effect can be found in~\cite{Zhang:2024ljd,Zhao:2025yhy}). Moreover, the value of $q_c$ is not fixed but is determined collectively by $\eta$ and $\Lambda$. $q_c$ does not vary monotonically with $\eta$, it exhibits a maximum near $\eta = 1$. However, when $\eta$ is fixed, $q_c$ varies monotonically with $\Lambda$: the smaller $\Lambda$ is ($\Lambda \leq 0$), the larger $q_c$ becomes.

Unlike the magnetic charge and cosmological constant, the coupling coefficient $\eta$ hardly alters the ADM mass of the FHBSs. Axion boson stars also exhibit similar properties~\cite{Chen:2024bfj}. We extend \cite{Zhao:2025hdg} to the large $\eta$ regime and find that at low frequencies, increasing $\eta$ leads to a significant decrease in both $1/g_{rr}$ and $-g_{tt}$. In contrast, at high frequencies, increasing $\eta$ yields the pure Hayward solution.

The light rings of SHBSs always appear in pairs. The inner light ring is stable, while the outer ring is unstable. We analyzed the influence of different parameters ($q$, $\eta $, $\Lambda$) on them. It is worth noting that we found an additional pair of light rings in the second branch of the solution.

Future research could further investigate the evolutionary stability of SHBSs in dynamical spacetimes. Given the growing evidence that AdS boson stars provide the gravitational dual of the lowest-energy states in the large-charge sector of three-dimensional CFT~\cite{delaFuente:2020yua,Liu:2020uaz}, it is particularly intriguing to investigate whether SHBSs exhibit similar features. Additionally, extending the current model to higher-dimensional spacetime is one of our future research objectives.

\section*{Acknowledgements}

This work is supported by the National Natural Science Foundation of China (Grant
No. 12275110 and No. 12247101) and the National Key Research and Development Program of
China (Grant No. 2022YFC2204101 and 2020YFC2201503).

\end{document}